\documentclass[numberedappendix, appendixfloats]{emulateapj}
\usepackage{aas_macros}
\usepackage{graphicx}
\usepackage{amssymb}
\usepackage{subfigure}
\usepackage{rotating}

\slugcomment{}

\shorttitle{Pop~III galaxy signature}
\shortauthors{Rydberg et al.}

\def\Abell209{Abell209-994}
\def\MACS1931{MACS1931-777}
\def\RXJ1347{RXJ1347-1951}

\newcommand{\macst}{MACS0416-1828}
\newcommand{\macstt}{MACS0647-610}

\newcommand{\SFT}{SD}

\begin{document}

\title{A search for Population III galaxies in CLASH. I. Singly-imaged candidates at high redshift}

\author{Claes-Erik Rydberg\altaffilmark{1}$^*$, Erik Zackrisson\altaffilmark{1, 2}, Adi Zitrin\altaffilmark{3, 4}, Lucia Guaita\altaffilmark{1}, Jens Melinder\altaffilmark{1}, Saghar Asadi\altaffilmark{1}, Juan Gonzalez\altaffilmark{1}, G$\ddot{\mbox{o}}$ran $\ddot{\mbox{O}}$stlin\altaffilmark{1} \& Tina Str$\ddot{\mbox{o}}$m\altaffilmark{1}}
\altaffiltext{*}{E-mail: claes-erik.rydberg@astro.su.se}
\altaffiltext{1}{The Oskar Klein Centre, Department of Astronomy, Albanova, Stockholm University, SE-106 91 Stockholm, Sweden}
\altaffiltext{2}{Department of Physics and Astronomy, Uppsala University, Box 515, SE-751 20 Uppsala, Sweden}
\altaffiltext{3}{Cahill Center for Astronomy and Astrophysics, California Institute of Technology, MS 249-17, Pasadena, CA 91125, USA}
\altaffiltext{4}{Hubble Fellow}

\begin{abstract}

Population~III galaxies are predicted to exist at high redshifts and may be rendered sufficiently bright for detection with current telescopes when gravitationally lensed by a foreground galaxy cluster. Population~III galaxies that exhibit strong Ly$\alpha$ emission should furthermore be identifiable from broadband photometry because of their unusual colors. Here, we report on a search for such objects at $z\gtrsim 6$ in the imaging data from the Cluster Lensing And Supernova survey with Hubble (CLASH), covering 25 galaxy clusters in 16 filters. Our selection algorithm returns five singly-imaged candidates with Ly$\alpha$-like color signatures, for which ground-based spectroscopy with current 8--10 m class telescopes should be able to test the predicted strength of the Ly$\alpha$ line. None of these five objects have been included in previous CLASH compilations of high-redshift galaxy candidates. However, when large grids of spectral synthesis models are applied to the study of these objects, we find that only two of these candidates are significantly better fitted by Population~III models than by more mundane, low-metallicity stellar populations.

\end{abstract}


\section{Introduction}
\label{introduction}

The first generation of stars -- the chemically pristine population~III (hereafter Pop~III) -- likely started forming in small numbers within $10^5$--$10^6\ \mathrm{M}_{\odot}$ dark matter halos at redshifts $z\gtrsim 30$ \citep[e.g.][]{2004ARA&A..42...79B}. Current simulations predict that the characteristic masses of these metal-free stars were high \citep[$\gtrsim 10\ \mathrm{M}_{\odot}$;][]{2011Sci...334.1250H,2014ApJ...781...60H}, with very short lifetimes as a result \citep[$\lesssim 20$~Myr; ][]{2002A&A...382...28S}. While this star formation mode may have continued in chemically unenriched pockets of gas down to much lower redshifts \citep{2007MNRAS.382..945T,2010MNRAS.404.1425J,2011Sci...334.1245F, 2013ApJ...773...83X}, Pop~III stars that form in isolation or in very small star clusters will likely remain undetectable even with the next generation of telescopes. The upcoming James Webb Space Telescope (JWST) will, for instance, not be able to detect individual Pop III stars, even when gravitationally lensed by a foreground galaxy cluster, unless their characteristic masses are $\gtrsim 300~\mathrm{M}_{\odot}$ or the cosmic star formation rate in Pop~III stars is an order of magnitude higher than suggested by current simulations \citep{2013MNRAS.429.3658R}.  Alternative ways to probe the properties of Pop~III stars include searches for Pop~III supernovae \citep[e.g.][]{2013ApJ...778...17W, 2013arXiv1312.6330W}, Pop~III gamma-ray bursts \citep[][]{2011A&A...533A..32D}, low-mass Pop~III stars surviving to the present day \citep[e.g.][]{2012A&A...542A..51C}, Pop~III chemical enrichment signatures \citep[e.g.][]{2013RvMP...85..809K} and Pop~III galaxies \citep[e.g.][]{2010ApJ...716L.190S}. 

Pop~III galaxies may be envisioned in two different forms -- pure Pop~III galaxies and hybrids. Pure Pop~III galaxies may have formed at $z\lesssim 15$ within $\sim 10^7$--$10^8\ \mathrm{M}_{\odot}$ dark matter halos that have remained chemically pristine \citep[e.g.][]{2009MNRAS.399...37J,2010ApJ...716L.190S,2012MNRAS.427.2212Z}, whereas hybrid Pop~III galaxies have significant metal content, but are still able to form Pop~III stars in sufficient numbers to give rise to tell-tale observational signatures due to incomplete chemical mixing or infall of zero-metallicity gas. The expected properties of both types of objects remain highly uncertain. Estimates of the likely star formation efficiencies in pure Pop~III galaxies differ greatly \citep[e.g.][]{2012MNRAS.426.1159S, 2013ApJ...772..106M}, and while simulations indicate that many metal-enriched galaxies may contain some zero-metallicity stars, the Pop~III mass fractions in such objects are predicted to be very low \citep[e.g.][]{2011MNRAS.414..847S}. 

Pop~III galaxies may, in principle, be distinguished from more mundane objects based on a number of spectral features, like the strengths of the Lyman-$\alpha$ (Ly$\alpha$) and He\,\textsc{ii}$\lambda$1640 emission lines, the Lyman  ``bump'', the slope of the ultraviolet (UV) continuum and the lack of metal lines like [O\,\textsc{iii}]$\lambda$5007 \citep[e.g.][]{2003A&A...397..527S,2010MNRAS.401.1325I,2010A&A...523A..64R,2011MNRAS.415.2920I,2011ApJ...740...13Z, 2014arXiv1411.6628Z}. Pop~III-like signatures of this type have already been reported for galaxies at low to intermediate redshifts \citep[e.g.][]{2002ApJ...565L..71M, 2003ApJ...596..797F,2008IAUS..255...75D,2011MNRAS.411.2336I,2013A&A...556A..68C}, but the nature of these objects remain unclear, as they are all too bright to be easily reconciled with the Pop~III objects seen in current simulations. Unfortunately, many of these diagnostics become increasingly difficult to use at higher redshifts, where most pure Pop~III galaxies are expected to exist \citep{2012MNRAS.427.2212Z}. The Lyman bump would be absorbed by the largely neutral intergalactic medium (IGM), the [O\,\textsc{iii}]$\lambda$5007 line gets redshifted out of range of current spectrographs, and the He\,\textsc{ii} line falls below current spectroscopic detection thresholds for Pop~III galaxies in the relevant mass range. 

Pure Pop~III galaxies may be detectable in deep Hubble Space Telescope images when gravitationally lensed by a foreground galaxy cluster, provided that their star formation efficiencies are sufficiently high \citep{2012MNRAS.427.2212Z}. While anomalously blue UV continuum slopes, due to the presence of Pop~III stars, are in principle detectable in multiband imaging surveys,  this requires extremely high Lyman continuum (LyC) escape fractions, with even harsher star formation efficiency requirements as a result \citep{2013ApJ...777...39Z}. In this paper, we will therefore focus on the use of the Ly$\alpha$ line to search for high-redshift Pop~III galaxy candidates.   

Even though Pop~III stars can drive the Ly$\alpha$ emission-line equivalent width, $\mathrm{EW}(\mathrm{Ly}\alpha)$, to very high values \citep[][]{2010A&A...523A..64R}, a substantial fraction of the Ly$\alpha$ photons emitted from objects in the reionization epoch ($z\gtrsim 7$) is likely to be absorbed by the neutral intergalactic medium \citep[e.g.][]{2010Natur.464..562H, 2011ApJ...730....8H, 2014MNRAS.444.2114J}. One Pop~III candidate with strong Ly$\alpha$ emission at $z\approx 6.5$ has already been reported by \citet{2012ApJ...761...85K}, but no similar objects have been spectroscopically confirmed at higher redshifts. Simulations involving galactic outflows, patchy reionization and source clustering \citep[e.g.][]{2010MNRAS.402.1449D, 2011MNRAS.414.2139D, 2012MNRAS.424.2193J} nonetheless suggest that significant Ly$\alpha$ transmission may occur along fortunate sightlines. As argued by \citet{2011MNRAS.418L.104Z}, reionization-epoch objects with very strong Ly$\alpha$ emission can potentially be identified from multiband photometry data because of their unusual colors, and this signature may arise even if as little as $1\%$ of the stellar mass in these objects is in the form of Pop~III stars. This technique is similar in spirit to the one used by \citet{2014ApJ...784...58S} to identify lensed high-redshift galaxies with prominent [O\,\textsc{iii}] and H$\beta$ emission lines. Here, we present a search for anomalous Ly$\alpha$ emitters in the imaging data from the Cluster Lensing And Supernova survey with Hubble (CLASH) survey \citep{2012ApJS..199...25P}, covering 25 galaxy clusters in 16 broadband filters. 

The paper is organized as follows: We describe the observational data from the CLASH program in Section~\ref{sec:observationaldata}. In Section~\ref{sec:models}, we present our models for Pop~III and comparison galaxies, and our color-color diagram selection. We then explain the statistical methods we have used, $\chi^2$ and cross-validation fitting, as well as our search procedure, in Section~\ref{sec:statisticalmethods}. Section~\ref{sec:results} contains the results, data and description of the five objects we have discovered. In Section~\ref{sec:gravitationallensing}, we shortly describe the gravitational lensing models used. We present the dependence of the results on different physical parameters in Section~\ref{sec:physicalproperties}, and Section~\ref{sec:summaryconclusion} contain summary and conclusion. Our cosmological model uses the parameters $H_0 = 67.3$, $\Omega_\mathrm{M}=0.308$, and $\Omega_{\Lambda}=0.692$. The cosmological parameters are derived from Planck, WP, highL, and BAO data in \citet{2014A&A...571A..16P}.

\section{Observational data}
\label{sec:observationaldata}

This paper is based on data from CLASH \citep{2012ApJS..199...25P}, a 524-orbit multi-cycle treasury program that uses gravitational lensing effects to study the mass distribution in 25 galaxy clusters. One of the scientific goals of this survey is to detect and characterize gravitationally lensed high-redshift galaxies, and five of the target clusters -- MACS0416, MACS0647, MACS0717, MACS1149, and MACS2129 -- are selected primarily on the basis of being strong gravitational lenses.

\begin{figure}
    \begin{center}
        \includegraphics[width = 8 cm]{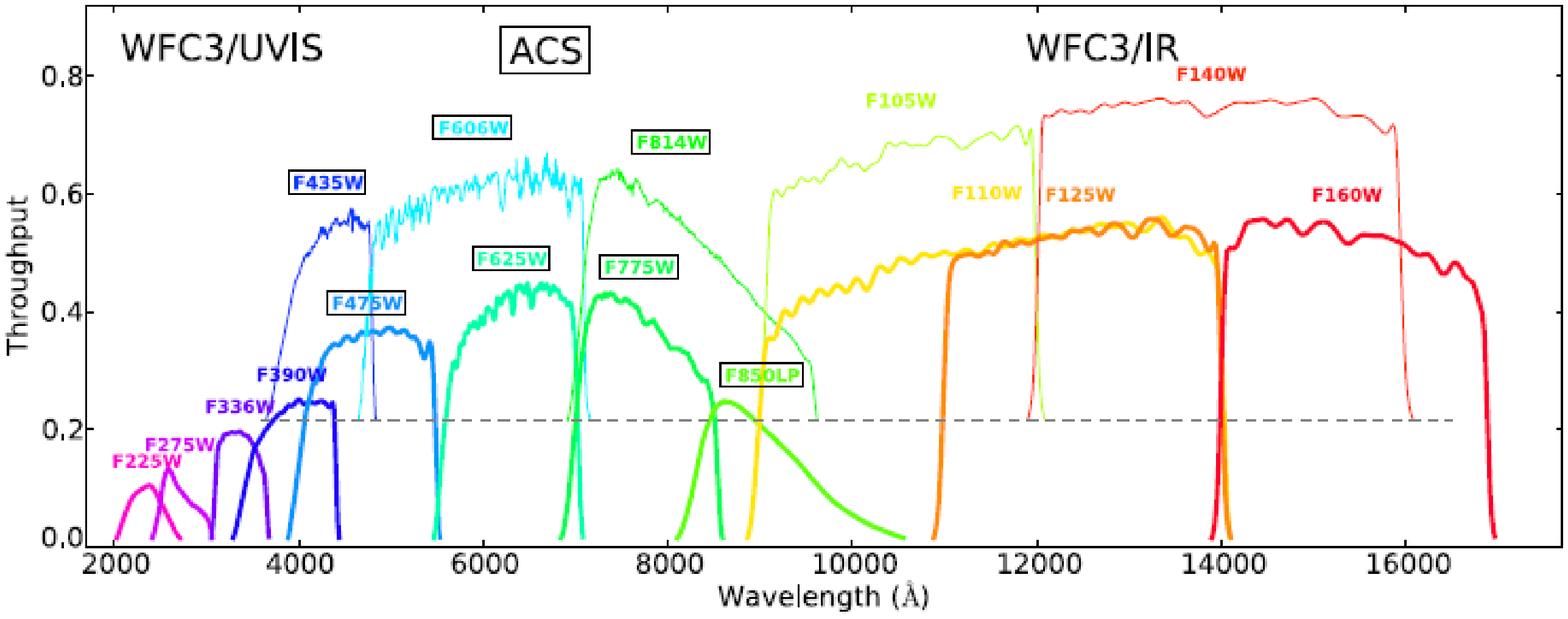}
        \caption{The passbands used in the CLASH survey. As can be seen, they cover a broad wavelength range, from 2,000 to 17,000~$\mathrm{\AA}$. Figure courtesy of D. Coe.}
        \label{fig:clashpassbands}
    \end{center}
\end{figure}

Each CLASH cluster is observed in 16 different (but partly overlapping) passbands using three different cameras. The WFC3/UVIS camera covers the ultraviolet part of the spectrum with four passbands: F225W, F275W, F336W, and F390W, ranging between 2,000 and 4,500~$\mathrm{\AA}$. The ACS/WFC camera covers the optical part of the spectrum using 7 filters: F435W, F475W, F606W, F625W, F775W, F814W, and F850LP. For some clusters, archived earlier observations are also available in the F555W filter. The wavelength coverage of ACS/WFC is approximately 3,500 to 10,500~$\mathrm{\AA}$. Finally, there is an infrared camera WFC3/IR that observes in five passbands, F105W, F110W, F125W, F140W, and F160W at wavelengths from 9,000 to 17,000~$\mathrm{\AA}$. Figure~\ref{fig:clashpassbands} shows the transmission profiles of the various CLASH filters.

The 5$\sigma$ detection limit of CLASH varies from 26.4 to 27.8, depending on the filter used. The point spread function (PSF) has a FWHM of $\sim 0.17''$ for the WFC3/IR and $\sim 0.1''$ for the other two cameras. AB-magnitudes are used throughout the paper.

All CLASH clusters have also been covered by a series of programs with the Spitzer Space Telescope (Spitzer). Data is available\footnote{http://sha.ipac.caltech.edu/applications/Spitzer/SHA/} for all clusters in the 3.6 and 4.5~micron filters of the Infra Red Array Camera \citep[IRAC;][]{2004ApJS..154...10F}, and for some clusters in the 5.8 and 8.0~micron filters as well. The Spitzer data used in this paper were observations in all four IRAC filters for the clusters Abell209 and RXJ1347 while for the clusters MACS0416, MACS0647, and MACS1931 observations in only 3.6 and 4.5~micron filters were available.

\subsection{Official catalogs}
\label{sec:officialcatalogs}

We have used the photometric data from the official CLASH catalogs \citep{2012ApJS..199...25P} both in the search for high-redshift objects and in the detailed analysis of the resulting candidates. The images on which the catalogs are based have been reduced by the CLASH team and are publicly available as FITS files on the CLASH homepage. The SExtractor software \citep{1996A&AS..117..393B} has been used to identify potential objects, resulting in 25 catalogs with a total of 53,030 entries. Photometric redshifts are derived using two different codes, BPZ \citep{2000ApJ...536..571B, 2004ApJS..150....1B, 2006AJ....132..926C} and \textsc{Le Phare} \citep{2006A&A...457..841I}. The catalogs are made publicly available on the CLASH homepage\footnote{http://archive.stsci.edu/prepds/clash/}.

When the data were reduced, no procedure for correcting bias-stripping and charge transfer efficiency (CTE) degradation effects \citep{2010PASP..122.1035A} was available for the WFC3/UVIS filters. Neither have we performed any correction. This leads to the possible contamination by cosmic rays boosting the observed flux or causing spurious detections. Since we are basically searching for objects with non-detections in these filters, this contamination merely removes possible candidates. Therefore, the high-redshift candidates identified should be safe.

\subsection{Consistency check}
\label{sec:consistencycheck}

The official catalogs are automatically produced and serve a more general purpose than the search for high-redshift galaxies. Therefore, we also generated our own source catalogs, optimizing the configuration parameters for our goals. Appendix~\ref{sec:guaitacatalog} describes the procedure. Since we found consistency in detecting sources following our approach and considering the official CLASH catalog, we will refer to the official catalog throughout the rest of the paper.

The consistency check described above, as well as the official catalogs, are automatically produced, once the parameters for each run of SExtractor is set. To confirm the usefulness of the official catalogs to investigate the spectral energy distributions (SED) of the objects, we have generated a small catalog for the five objects found through the search process (Section~\ref{sec:colorcolordiagrams} and \ref{sec:searchprocess}). The magnitudes have been measured isophotally with SExtractor, as shown in Appendix~\ref{sec:melindercatalog}. The resulting errors are larger due to the use of bigger apertures but the magnitudes are consistent with the official catalogs.

\section{Models}
\label{sec:models}

Since to date there are no confirmed Pop~III galaxies observed, model spectra based on spectral synthesis codes are required for the selection and analysis of candidates. A combination of synthetic and empirical spectra for non-pop III objects will also be used for reference purposes. Both sets of spectral templates will be fitted to the CLASH catalogs in order to single out the best Pop~III galaxy candidates. 

In this section, we give a brief description of the Pop~III and $\mathrm{Z}>0$ Yggdrasil galaxy models we will use, as well as other models of more mundane objects. The treatment of Ly$\alpha$-line absorption is also outlined.

\subsection{Yggdrasil}
\label{sec:modelsyggdrasil}

The Yggdrasil population synthesis code is described in detail in \citet{2011ApJ...740...13Z}, so only a very brief overview is presented here. In codes of this type, the galaxy spectra are modeled by summing up the flux contributions from all luminous constituents assumed to be present within galaxies. In this paper, we will assume that the stellar spectrum is produced by either metal-enriched (Pop I/II) or zero-metallicity (Pop~III) stars, although more exotic objects such as dark stars \citep{2009NJPh...11j5014F, 2010ApJ...717..257Z} in principle also could be included. The flux contribution from the photoionized interstellar medium is then computed self-consistently from this stellar mixture using the photoionization code \textsc{Cloudy} \citep{1998PASP..110..761F} and added to the total.

The grid of Yggdrasil models used in this paper uses the following set of parameters:

\begin{description}

\item[Covering factor ($f_\mathrm{cov}$)] This parameter controls the relative impact of nebular emission on the emerging spectrum and represents, in physical terms, the fraction of the stellar population covered by photoionized gas. In the model grid used in this paper, three different $f_\mathrm{cov}$ values are tested: 0.0 (no nebular emission), 0.5 (intermediate case) and 1.0 (maximal nebular contribution). Photoionized gas around a galaxy (high $f_\mathrm{cov}$) implies a strong nebular continuum with an SED which is redder than the underlying stars as well as emission lines.
\item[Initial mass function (IMF)] The IMF regulates the distribution of stellar masses at birth (the zero age main sequence). For galaxies with metallicity $\mathrm{Z} > 0$, we use the standard Kroupa IMF \citep{2001MNRAS.322..231K}. Since the IMF for Pop~III stars is largely unknown, three different IMF options are considered for these objects: the Kroupa IMF, a top-heavy, log-normal distributions peaking at mass 10~M$_{\odot}$, and a top-heavy, power-law distribution with a characteristic mass of 100~M$_{\odot}$.
\item[Starburst duration (\SFT{})] We use models of starburst galaxies since these will be the most luminous, and thus the most likely to be detected. The models include instantaneous burst (all stars are formed at the same time) and \SFT{}s of 10~Myr, 30~Myr, and 100~Myr with constant star formation rate. For Pop~III galaxies, we only present the results from the instantaneous burst models while we use all four \SFT{}s for $\mathrm{Z}>0$ galaxies. The results from the models with longer \SFT{} is nearly redundant for Pop~III galaxies since we are dealing with very young galaxies (according to our Yggdrasil fits). The models with longer \SFT{}s constantly form new stars, and as the SED in the rest-frame ultraviolet/optical is mostly dependent on young massive stars, the resulting model spectrum will be similar to that of an instantaneous burst model at young age.
\item[Age] This is a very important parameter, especially for the models with a top-heavy IMF. Massive stars age fast and die young. This means the galaxy spectrum could experience huge variations in just a few million years for models with a top-heavy IMF. We use a lower limit of 1~Myr, since ages even lower is highly unlikely as this would imply an implausible star formation rate. We also remove models with an age above the age of the universe at the redshift used.
\item[Metallicity] In this article we will use Yggdrasil zero-metallicity models to represent Pop~III galaxies. To distinguish Pop~III galaxies from metal-enriched galaxies, we use the Yggdrasil models with Z~=~0.0004, 0.004, 0.008, and 0.02 for comparison. We also use other empirical and model templates, as described in Section~\ref{sec:mundanealternatives}.
\item[Lyman-$\alpha$ escape fraction $f_{\mathrm{Ly\alpha}}$] Galaxies surrounded by gas (i.e. $f_\mathrm{cov}>0$) ionizes its surroundings and emits a potentially strong Ly$\alpha$ line. The parameter $f_{\mathrm{Ly\alpha}}$ represents the fraction of Ly$\alpha$ photons escaping both the ISM and the IGM without getting absorbed or scattered. See Section~\ref{sec:lyaline} for a thorough explanation.

\end{description}

\subsection{Mundane alternatives}
\label{sec:mundanealternatives}

In order to identify robust Pop~III galaxy candidates, it is important to compare the quality of the photometric fits (Section~\ref{sec:statisticalmethods}) resulting from Pop III models to those produced by more mundane galaxies. In the literature there are both pure synthetically built galaxy models and pure empirical comparison spectra, but also hybrids of the two. In addition to comparing with Z~$>$~0 Yggdrasil models, we employ two of the grids of comparison spectra used by \textsc{Le Phare}:

\begin{description}

\item[Gissel] This is a set of synthetic model templates from \citet{2003MNRAS.344.1000B}. The models include different metallicities, aging, and extinction.
\item[CWW, Kinney] A set of hybrid models building on UV observations of nearby galaxies by \citet{1980ApJS...43..393C, 1996ApJ...467...38K, 1999MNRAS.310..540A}. The galaxy spectra are also extrapolated into the infrared. Some spectra are also included from the Kinney atlas.

\end{description}

\subsection{The Lyman-$\alpha$ line}
\label{sec:lyaline}

Pop~III stars have a very high effective temperature of up to 100,000~K. This implies that huge amounts of hydrogen ionizing photons are emitted. For galaxies surrounded by a gas, i.e. $f_\mathrm{cov}>0$, the gas gets photoionized. Through recombination the photoionized gas emits photons at the Ly$\alpha$ wavelength ($\lambda_{\mathrm{Ly\alpha}}=1,216~\mathrm{\AA}$) as well as at other radiation levels. For most Pop~III galaxy models, the Ly$\alpha$-line is by far the strongest rest-frame emission line.

However, the Ly$\alpha$ line is a highly resonant line. This means that the Ly$\alpha$-photons are easily absorbed by neutral gas, both in the interstellar medium (ISM) and the intergalactic medium (IGM). If absorbed, the photons will quickly be re-emitted in a new direction, i.e. they scatter. In the ISM the photons are easily absorbed by dust while scattering. Pop~III galaxies are believed to have very small amounts of dust, or no dust at all, since dust need metals to form. This could imply that a significant fraction of Ly$\alpha$ photons will eventually exit the ISM by scattering. If absorbed in the IGM, they will simply scatter out of the line of sight. We will use the parameter $f_\mathrm{Ly\alpha}$ to denote the combined fraction of Ly$\alpha$ photons escaping from both the ISM and the IGM. We impose an upper limit $f_\mathrm{Ly\alpha} < 0.5$, as simulations indicate a higher escape fraction to be implausible for galaxies in the reionization epoch \citep{2011MNRAS.414.2139D}.

For objects at $z<6$, the radiation generally passes several neutral clouds. In each cloud the radiation suffers Ly$\alpha$ absorption, as lower rest-frame wavelengths are redshifted to the Ly$\alpha$ wavelength. The result is several absorption lines (the number increasing with redshift) in the spectrum called the Ly$\alpha$-forest. To model this, we use \citet{1995ApJ...441...18M} that also takes into account higher order Lyman absorbers.

When $z>6$, the neutral hydrogen reaches a density that essentially absorbs all radiation with wavelength below that of the Ly$\alpha$ line. This is called the Gunn–-Peterson trough \citep{1965ApJ...142.1633G}, and to simulate this feature, all flux in model spectra below $\lambda_{\mathrm{Ly\alpha}}$ are set to zero when modeling objects at $z>6$. This effect forms the basis for photometric redshift estimates at these redshifts. Since all flux below the $\lambda_{\mathrm{Ly\alpha}}$ is zero, at $z>6$, and since $\lambda_{\mathrm{Ly\alpha}}$ is redshifted, the result is that all flux below $(1+z) \lambda_{\mathrm{Ly\alpha}}$ for an observed object at $z$ is zero. The observational effect when observing in several filters at different wavelengths is that filters with transmission below $(1+z) \lambda_{\mathrm{Ly\alpha}}$ contain no flux, while filters with transmission above $(1+z) \lambda_{\mathrm{Ly\alpha}}$ contain flux. By inspecting what filters contain flux, an estimate of the object's redshift can be attained. This is the Ly$\alpha$-break technique, similar to the Lyman-break technique which uses the Lyman limit ($912~\mathrm{\AA}$) at $z < 6$ in the same manner. In general an object is called a drop-out in the filter bordering to this Ly$\alpha$ induced limit, and such a galaxy is called a Lyman-break galaxy (LBG). The Ly$\alpha$-break technique can be improved by using $\chi^2$ or cross-validation fitting to models, as outlined in Section~\ref{sec:statisticalmethods}.

The optical depth due to Ly$\alpha$ scattering is broadened by the natural width of the line as well as the velocity distribution, giving rise to the Ly$\alpha$ damping wing \citep{1998ApJ...501...15M, 2009fflr.book.....S}. This scattering is accounted for by the $f_{\mathrm{Ly}\alpha}$ parameter unless the Ly$\alpha$ line is outside the filter containing the scatter. However, we have estimated the possible effect of this and found it to be negligible for our purposes.

\subsection{Color-color diagrams}
\label{sec:colorcolordiagrams}

\begin{figure*}[htp]
  \begin{center}
	  \subfigure[$z=6.5-8.0$]{\label{fig:colorcolorlowz}\includegraphics[scale=0.25]{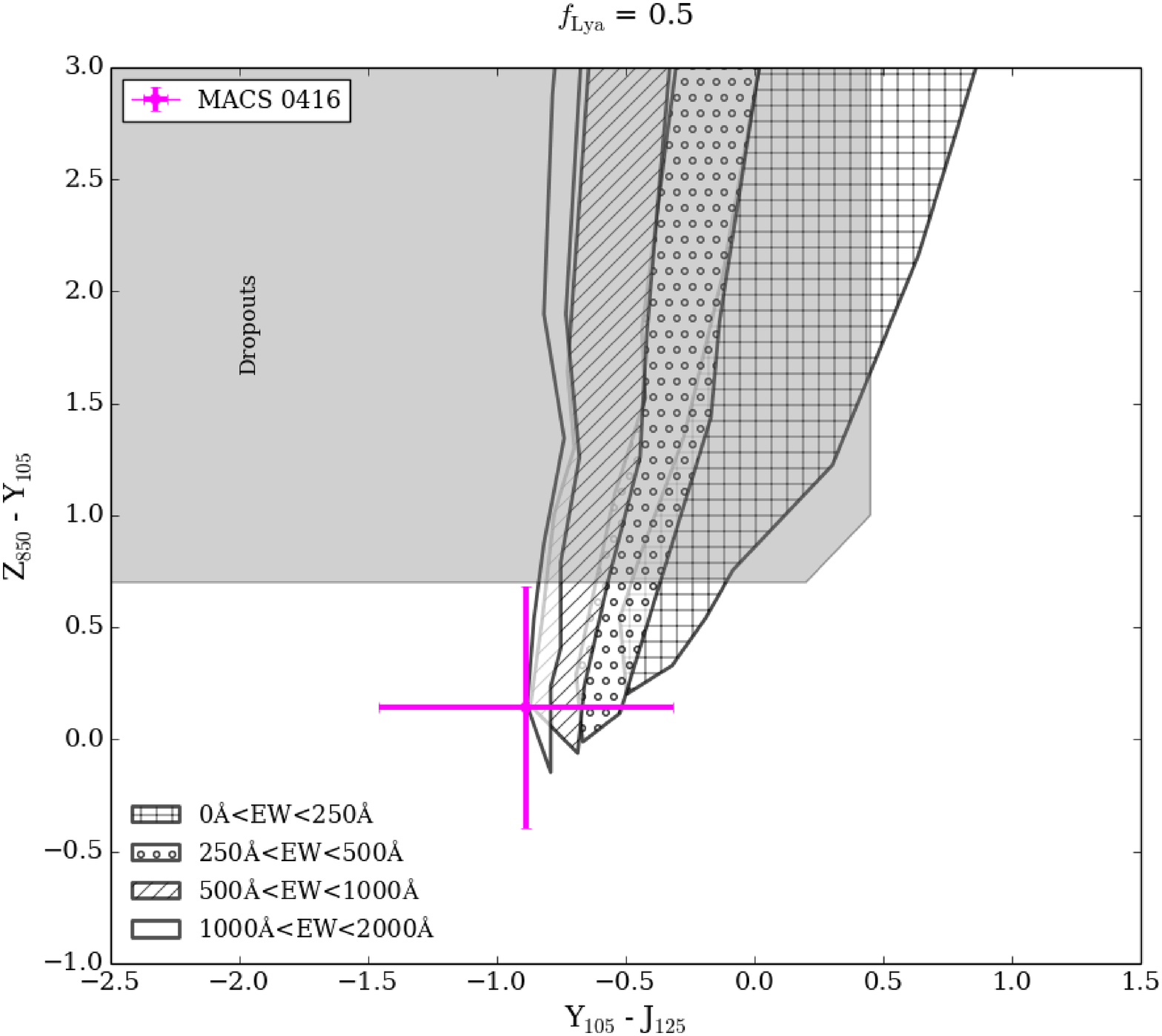}}
		\subfigure[$z=7.0-9.0$]{\label{fig:colorcolorhighz}\includegraphics[scale=0.25]{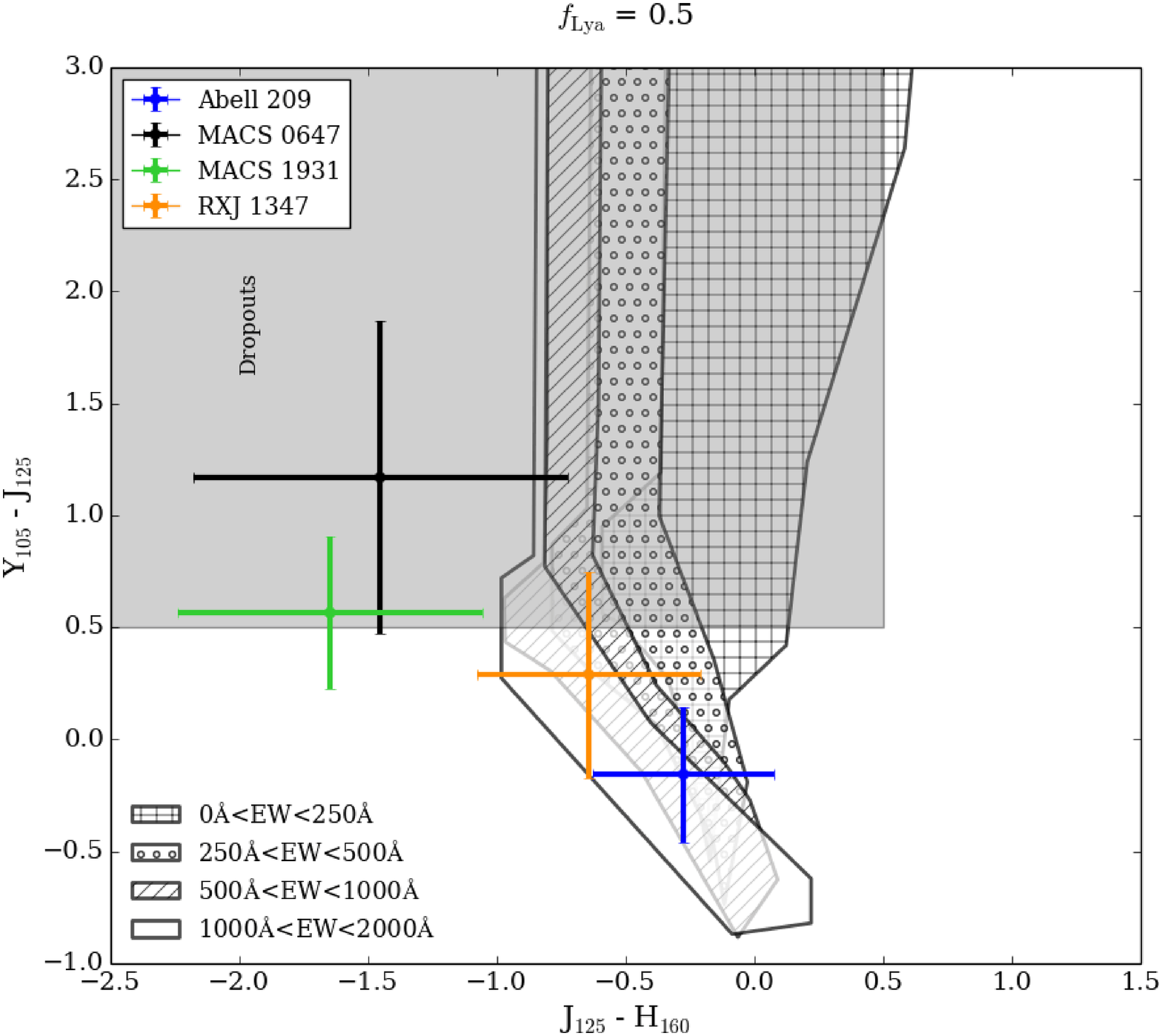}}
  \end{center}
  \caption{\footnotesize
Color-color diagrams for two different redshift intervals. The filters are selected to single out drop-outs with high Ly$\alpha$-emission. The different areas show in what regions in the diagrams models of Pop~III galaxies with $f_\mathrm{Ly\alpha}=0.5$ are located depending on the rest-frame EW(Ly$\alpha$) of the model. The areas are constructed using pure model expectations, hence observational errors are not modeled. The light gray areas correspond to the drop-out criterias used in \citet{2011ApJ...737...90B} to select $z \approx 7$ and $z \approx 8$ LBGs, respectively. High-redshift Pop~I/II galaxy candidates generally appear to the right or within the $0 < \mathrm{EW} < 250~\mathrm{\AA}$ Pop~III areas displayed, but still within the light gray area \citep[see][]{2011MNRAS.418L.104Z}. Our objects are marked with bars corresponding to the observational errors. In the color-color diagram for the interval $z=6.5-8.0$ we have one object, found behind the galaxy cluster MACS0416. As can be seen, the object is on the border of the area for Pop~III galaxies, with error bars covering areas both within the Pop~III region and outside. For the four objects in the interval $z=7.0-9.0$, we see that two of these (behind the clusters Abell209 and RXJ1347) are well positioned within the Pop~III region, also having a large fraction of their error bars within the region. The other two objects, behind MACS0647 and MACS1931, appear outside the Pop~III region displayed.
}
  \label{fig:colorcolordiagram}
\end{figure*}

Color-color diagrams can be used to identify drop-outs at different redshifts \citep[e.g.][]{2011ApJ...737...90B}. Moreover, these diagrams can serve as a diagnostic of Pop III galaxies with strong Ly$\alpha$ emission, as these may appear in regions of suitably selected region not occupied by Pop I/II galaxies \citep{2011MNRAS.418L.104Z}. In Figure \ref{fig:colorcolordiagram}, we display two color-color diagrams set-up to identify Pop~III galaxies in the redshift ranges $z=6.5-8.0$ (a) and $z=7.0-9.0$ (b). The Pop~III models used all have $f_\mathrm{Ly\alpha}=0.5$. However, the Ly$\alpha$ equivalent width (EW) varies, depending on the intrinsic luminosity of the Ly$\alpha$ line and the luminosity and form of the continuum. Various regions are distinguished with different patterns for different EW ranges. The regions are constructed using pure model expectation, hence observational errors are not modeled. The light gray areas correspond to the drop-out criteria used in \citet{2011ApJ...737...90B} to select $z \approx 7$ and $z \approx 8$ LBGs, respectively. High-redshift Pop~I/II galaxy candidates generally appear to the right or within the $0 < \mathrm{EW} < 250~\mathrm{\AA}$ Pop~III areas displayed, but still within the light gray area, as discussed by \citet{2011MNRAS.418L.104Z}. We have also confirmed this by comparing to the high-redshift candidates in \citet{2014ApJ...792...76B}. Compared to the Pop~III regions with high Ly$\alpha$ EW, a normal Pop~II stellar population with a Salpeter IMF, has a maximum predicted equivalent width of Ly$\alpha$ $200-400~\mathrm{\AA}$ in the rest frame \citep{1993ApJ...415..580C, 2003A&A...397..527S, 2010A&A...523A..64R}. The objects discussed in this paper are included with error bars indicating the observational errors. The $z=6.5-8.0$ diagnostic diagram includes one object behind the MACS0416 cluster. As can be seen, the object is on the border of the area for Pop~III galaxies, with error bars covering areas both within the Pop~III region and outside. In the $z=7.0-9.0$ diagram, we have two objects (located behind the clusters Abell209 and RXJ1347), well positioned within the Pop~III region. Both have a large fraction of their error bars within the Pop~III region. We also have two other objects, behind MACS0647 and MACS1931, appearing outside the Pop~III region displayed. We will henceforth refer to these objects as \Abell209{}, \macst{}, \macstt{}, \MACS1931{}, and \RXJ1347{}. The number after the galaxy cluster name indicates their row number in the public CLASH catalog.

Color-color diagrams are useful diagnostics, but in this case account only for three filters each, hence they are not perfect. Observational errors, for example, may cause other objects to spuriously appear in the Pop~III region. Other types of high-redshift Ly$\alpha$-emitters (LAE) may also mimic this color signature. An observed high equivalent width does not necessarily imply the presence of Pop~III stars, but can be produced in a number of other scenarios. The fact that Ly$\alpha$ resonantly scatters on the neutral hydrogen atoms in the ISM of galaxies, means the Ly$\alpha$ photons follow different paths through galaxies than the continuum photons. Depending on the structure, dust content and kinematics of the ISM, the Ly$\alpha$ line may be either enhanced or suppressed compared to the continuum \citep{1991ApJ...370L..85N, 2013ApJ...766..124L, 2014A&A...562A..52D}. The sensitivity of Ly$\alpha$ on the ISM physics may also result in highly anisotropic Ly$\alpha$ emission, which has been found both in simulations \citep{2009ApJ...704.1640L, 2012A&A...546A.111V} and observations of starburst galaxies in the local universe \citep{2005A&A...438...71H, 2014ApJ...782....6H, 2009AJ....138..923O}. Moreover, accreting black holes produce a very hard ionizing spectrum \citep{2001ApJ...556...87H} that can produce a similar Pop~III signature. Finally Ly$\alpha$ and continuum can also be emitted from neutral cooling gas, so-called cooling radiation \citep{2000ApJ...537L...5H, 2009ApJ...690...82D, 2012MNRAS.423..344R}. Hence, Pop~III candidates need to be assessed with care, with the above effects kept in mind.

Statistical SED fitting using all filters, like $\chi^2$ and cross-validation techniques, could result in a more reliable identification of objects than color-color diagrams. These methods are sometimes able to find acceptable fits to Pop~III models for objects outside the Pop~III regions of color-color diagrams. For instance, our two objects outside the Pop~III region have an acceptable Pop~III fit when considering all available filters. They also have colors causing them to appear to the left of even the Pop~III models with the most extreme EW we have displayed. Since mundane objects generally appear to the right of the Pop~III regions, this means the objects have extreme colors even for Pop~III objects. In the same way as observational errors could masquerade LAEs as Pop~III galaxies, observational errors could make Pop~III galaxies appear outside the Pop~III region. These difficulties with identifying objects are of course augmented when the observational uncertainty is high, and the faint objects we are working with naturally have large errors. We will discuss the robustness of results more thorough in Section~\ref{sec:results}.

\section{Statistical methods}
\label{sec:statisticalmethods}

To search the CLASH data, and to provide various statistical tools as well as FITS image handling, we have developed a program called Observational data scanner (ODS). ODS is a .net application for Windows written in C\#. ODS use CLASH data as well as data loaded from other sources linked to the CLASH data (such as, for example, results from $\chi^2$ fits with other programs). 

\subsection{$\chi^2$ fitting}

To constrain the photometric redshifts and compare different models, we first use the code \textsc{Le Phare} \citep{2006A&A...457..841I}. The code performs a $\chi^2$ fit of observed to model fluxes weighted by the observational uncertainty. The fit is obtained through minimization defined by:

\begin{center}

$\chi^2 = \sum_i{\left[ \frac{F_{obs, i} - s F_{mod, i}}{\sigma_{obs, i}} \right]^2}$

\end{center}

where $F_{obs, i}$, $\sigma_{obs, i}$ are the observed fluxes with uncertainty and $F_{mod, i}$ is the comparison model flux in filter i. The parameter $s$ is the scaling factor used to minimize $\chi^2$. We calculate $s$ to minimize $\chi^2$ through use of the analytical formula:

\begin{center}

$s = \sum_i{\left[ \frac{F_{obs, i} \times F_{mod, i}}{\sigma_{obs, i}^2} \right]} / \sum_i{\left[ \frac{F_{mod, i}^2}{\sigma_{obs, i}^2} \right]}$

\end{center}

The resulting $\chi^2$ value increases with the number of filters. Since we need to compare fits with different number of valid data points, we deduce a reduced\footnote{The number of degrees of freedom is not strictly the number of filters with valid data. The factor $s$ used for optimization removes one degree of freedom. The Yggdrasil model have several parameters correlating in complicated ways (for example, the $f_{\mathrm{Ly}\alpha}$ parameter only matters when $f_{\mathrm{cov}} > 0$), removing an indeterminate number of degrees of freedom. Both these effects are ignored which means it is not a proper reduced $\chi^2$ we calculate. We use it anyway since it is only used as a part of the search procedure and carefully scrutinize the results afterwards.} $\chi^2$ through dividing $\chi^2$ with the number of filters. The usefulness and limitations of reduced $\chi^2$ has been discussed by \citet{2010arXiv1012.3754A} and should be used with some care. We use it as a complementary search criteria to our color-color diagrams.

\subsection{Cross-validation fitting}
\label{sec:crossvalidation}

Cross-validation \citep[see ][]{Singh1981} is a statistical method using $\chi^2$ minimization, regression or other similar statistical tools. Cross-validation provides a measure on the effectiveness of a model. In general, a portion of the data is withheld from the calculation when using cross-validation. The withheld data is later used for evaluation purposes. We have used a special case of cross-validation called k-fold cross-validation resampling, with k equal to the number of filters. k-fold cross-validation means that, instead of calculating a single $\chi^2$ fit, one fit is calculated for each filter (a fold), each time with the observation in one filter removed. The removed filter's observational value is then compared to the $\chi^2$ minimization implied value from the remaining filters to derive a measure on the model's effectiveness.

The deviations of the predicted removed observation from the actual observation are used with their uncertainty to construct a measure of the accuracy akin to $\chi^2$. We assume a normal distribution of the deviation and use a geometric mean to arrive at a measure corresponding to a reduced $\chi^2$ value, which we refer to as k-fold accuracy. The measure is between 0 and 0.39 (maximum value for a normalized Gaussian distribution), where 0.39 implies a perfect fit to data. This measure, which is closely correlated with reduced $\chi^2$, will be used as an approximation for probability calculations of redshift, $f_{\mathrm{Ly\alpha}}$ etc. A disadvantage with cross-validation, compared to $\chi^2$, is that the scale factor (used to infer total stellar mass) is not as well defined, since each fold contributes one, albeit similar, scale factor. We will use the average of the scale factors from each k-fold as an approximation of the scale factor.

When compared, the results from our $\chi^2$ minimization and cross-validation optimization are very similar. The best fit models are the same for four of our objects. Only for \macst{} does the best fit model change and the $\chi^2$ minimization find an optimum solution slightly younger and at $z=6.9$ instead of at $z=6.8$ for cross-validation. We have compared the resulting fits of the two methods on our Pop~III galaxy models. By recalculating the cross-validation as the number of standard deviations that the fitted optimum corresponds to (using a normal distribution), we get something comparable to the minimum $\chi^2$ value. Generally the $\chi^2$ was larger but the optimum values proved to be similar. Even though related, it is reassuring to have two measures confirming the results.

\textsc{Le Phare} is unable to compute cross-validation, so for this purpose, we use ODS. Since ODS is far too slow to compute cross-validation for the whole CLASH dataset we use it as a tool to more closely analyze objects found through our color-color diagrams and \textsc{Le Phare} $\chi^2$-fitting.

\subsection{Secondary optima}

Finding the minimum $\chi^2$ or maximum k-fold accuracy (optima is used as interchangeable name for minimum $\chi^2$/maximum k-fold accuracy) is a good first step, but does by no means guarantee the correct redshift and model. There may be one or several optima that are nearly as good as the best one, but for a completely different model or at a completely different redshift. The optimum could also be extended. An extended optimum has a nearly as good fit as the best one in an extended range of parameter values, rendering the optimum position uncertain.

\textsc{Le Phare} provides means to find secondary optima and/or saving $\chi^2$ for different models. We will, however, use ODS for this analysis as it can produce both $\chi^2$ and cross-validation results. This aids us in calculating probability distributions for redshifts, thus identifying secondary optima and low-redshift interlopers. The probability distribution also allows us to compare the fit for different model parameters and $f_\mathrm{Ly\alpha}$. Finally, this provides us with a method to compare our Pop~III models to fits of more mundane objects (Section~\ref{sec:mundanealternatives}).

\subsection{Search process}
\label{sec:searchprocess}

We ran \textsc{Le Phare} for each observation in CLASH, using the models in Yggdrasil in the redshift range 0 to 12. In the fitting, we regarded all $\mathrm{S/N}<1$ detections as non-detections and converted them to non-detections. In \textsc{Le Phare}, the implementation of non-detections exclude models that imply more flux in a filter with a non-detection than the upper limit given by the non-detection (determined by 1$\sigma$ in our case). I.e. a model is first fitted using filters with detections. The fitted model is then compared to filters with non-detections and is removed if it implies more flux in any of these than their non-detection flux limit. The optimization resulted in a list containing $\chi^2$, redshift, scale factor, and models for the minimum $\chi^2$s. The results were then linked to the CLASH observations in the ODS and used for scanning. The criteria of the scan were set according to our color-color diagrams, including the Pop~III regions with surroundings. Two scans were performed, matching the two redshift windows for the color-color diagrams. Criteria were also set on reduced $\chi^2 < 2$ and estimated redshift to correspond to the redshift intervals of the color-color diagrams. The full width at half maximum (FWHM) provided for each object in the CLASH catalogs were compared to the PSF, to ensure that only resolved objects remained (thus, there is some risk of missing unresolved high-redshift candidates), eliminating contamination from low-redshift stars. We then carefully analyzed the results through examination of the FITS images to remove obvious optical artifacts. This resulted in a list of eighteen potential candidates.

The ODS has a more flexible code for $\chi^2$ minimization and is also able to run cross-validation fitting. It provides results for every combination of models and redshifts, as well as several ways of handling non-detections. The drawback compared to \textsc{Le Phare} is the speed. We ran ODS $\chi^2$ and cross-validation fitting for the eighteen objects achieved in the first scan. We also used ODS to fit the objects to mundane alternatives see Section~\ref{sec:mundanealternatives}. We excluded objects with cross-validation fits to mundane models good enough to rival those of Pop~III models. To exclude low-redshift interlopers, we also removed objects possessing lower-redshift solution competitive with its high redshift solution, no matter if the low-redshift solution stems either from Pop~III models or our comparison models. This resulted in the five objects we present.

In recent years, it has been pointed out that extreme [O\,\textsc{iii}]-emitters at $z\approx 1$--2 may represent an important interloper population in photometrically selected samples of $z>6$ galaxies \citep[e.g.][]{2011ApJ...743..121A, 2013ApJ...765L...2B, 2013ApJ...773L..14C}. While the overall SEDs of such objects are markedly different from the Ly$\alpha$-emitting Pop~III galaxies we consider in this paper, the two types of objects may -- over certain redshift intervals -- display somewhat similar CLASH colors for very high [O\,\textsc{iii}] equivalent widths \citep{2014arXiv1412.7909H}. Since the Yggdrasil grid of metal-enriched stellar populations contains objects with rest-frame [O\,\textsc{iii}] equivalent widths of up to EW([O\,\textsc{iii}])$\approx 4000~\mathrm{\AA}$ \citep[similar to the most extreme objects discussed by][]{2014arXiv1412.7909H}, this interloper class has already been rejected by the fitting procedure for the Pop~III galaxy candidates that are singled out as the most promising ones.

\section{Results}
\label{sec:results}

Through the process described in Section~\ref{sec:searchprocess}, we have discovered five interesting objects when searching the CLASH data. These objects, when fitted to Pop~III galaxy templates, have photometric redshifts in the range 6.8--8.8. Table~\ref{tab:RT1} shows coordinates and photometric data for each object. AB-magnitudes with errors for the seven filters covering the longest wavelength ranges in the CLASH survey are listed for each object. The data from CLASH filters covering lower-wavelengths have also been used in the analysis but consists of non-detections or very weak detections (plausibly spurious detections), and are not presented. Table~\ref{tab:RT1} also displays derived values, using Pop~III templates, our photometric redshift, magnification, restframe EW(Ly$\alpha$) as well as mass estimates. The reduced $\chi^2$ is supplied as a familiar measure of the quality of the fit. Thumbnail images of the five objects are displayed in Figure~\ref{fig:RF1}. The same seven filters as in Table~\ref{tab:RT1} are used, showing the Ly$\alpha$-break for each of the objects. None of the objects have been discussed in the literature before, nor did any appear in the compilation by \citet{2014ApJ...792...76B}. \Abell209{}, \macst{}, and \RXJ1347{} may have been overlooked because they appear outside the drop-out regions normally used to identify $z \approx 7$ or $z \approx 8$ objects, illustrated in Figure~\ref{fig:colorcolordiagram}. The other objects are located within the drop-out regions. However, their photometric magnitudes imply a very blue spectrum which mundane models reproduce poorly. The photometric redshift estimates in the official catalogs using \citet{2000ApJ...536..571B} have huge error bars with lower limits $z_{\mathrm{l}}<1$. This implies that the objects are not very well fit by mundane object spectra. With the exception of \RXJ1347{}, the magnifications (see Section~\ref{sec:gravitationallensing}) are low ($\mu<2$), akin to Pop~III galaxy candidates in unlensed fields such as the objects discussed in \citet{2011MNRAS.418L.104Z}.

The P($z$) graphs in this section, and the corresponding graphs in Section~\ref{sec:physicalproperties}, are calculated using the cross-validation described in Section~\ref{sec:crossvalidation}. The results are normalized to the interval [0,~1]. A value of 1 means a model where every removed observation is perfectly reproduced by a fit to the remaining observations, i.e. the model corresponds exactly to the observed data. When calculating P($z$) for a certain grid, the highest cross-validation value for any model in the grid at each $z$ is selected. Hence, a low value at a certain $z$ means there is no model in the whole grid that fits the data well. A high value at a certain $z$ means there is at least one model in the whole grid that fits the data well. We use this as a proxy for a probability distribution with the aim of comparing different grids of models.

\begin{table*}
\caption{
Coordinates, photometry, photometric z, magnification factor, EW(Ly$\alpha$), mass estimate, and reduced $\chi^2$ for the five Pop~III galaxy candidates. The estimates are based on the Yggdrasil Pop~III grid.
}
\label{tab:RT1}
\begin{center}
\begin{tabular}{lccccc}
\hline
\\
Object & \Abell209{} & \macst{} & \macstt{} & \MACS1931{} & \RXJ1347{} \\
\hline
\\
Rectascension & $ 01^{\mathrm{h}}31^{\mathrm{m}}56.94^{\mathrm{s}} $ & $ 04^{\mathrm{h}}16^{\mathrm{m}}10.26^{\mathrm{s}} $ & $ 06^{\mathrm{h}}48^{\mathrm{m}}02.44^{\mathrm{s}} $ & $ 19^{\mathrm{h}}31^{\mathrm{m}}45.44^{\mathrm{s}} $ & $ 13^{\mathrm{h}}47^{\mathrm{m}}30.57^{\mathrm{s}} $ \\
Declination & $ -13^{\circ}36'40~9'' $ & $ -24^{\circ}05'17~9'' $ & $ 70^{\circ}15'28~2'' $ & $ -26^{\circ}34'13~6'' $ & $ -11^{\circ}46'13~7'' $ \\
F814W & $ >28.97  $ & $ >29.02 $& $  >29.97  $ & $ 28.22 \pm 0.68 (1.1\sigma) $ & $ >29.05 $ \\
F850LP & $ >28.39  $ & $ 27.33 \pm 0.34 (2.7\sigma) $ & $ 28.56 \pm 0.72 (1.0\sigma) $ & $ >27.02 $ & $ >28.54 $ \\
F105W & $ 26.79 \pm 0.13 (7.7\sigma) $ & $ 27.19 \pm 0.20 (5.0\sigma) $ & $ 28.54 \pm 0.50 (1.7\sigma) $ & $ 26.67 \pm 0.20 (4.9\sigma) $ & $ 27.55 \pm 0.26 (3.7\sigma) $ \\
F110W & $ 27.13 \pm 0.15 (6.8\sigma) $ & $ 27.41 \pm 0.17 (5.8\sigma) $ & $ 27.75 \pm 0.20 (5.0\sigma) $ & $ 26.89 \pm 0.19 (5.3\sigma) $ & $ 27.58 \pm 0.12 (8.3\sigma) $ \\
F125W & $ 26.95 \pm 0.17 (6.0\sigma) $ & $ 28.08 \pm 0.37 (2.4\sigma) $ & $ 27.37 \pm 0.19 (5.1\sigma) $ & $ 26.10 \pm 0.14 (7.3\sigma) $ & $ 27.26 \pm 0.20 (4.9\sigma) $ \\
F140W & $ 27.22 \pm 0.19 (5.4\sigma) $ & $ 27.90 \pm 0.28 (3.4\sigma) $ & $ 28.13 \pm 0.30 (3.1\sigma) $ & $ 27.14 \pm 0.27 (3.5\sigma) $ & $ 28.33 \pm 0.39 (2.3\sigma) $ \\
F160W & $ 27.22 \pm 0.18 (5.4\sigma) $ & $ 27.87 \pm 0.29 (3.3\sigma) $ & $ 28.83 \pm 0.53 (1.6\sigma) $ & $ 27.75 \pm 0.45 (2.0\sigma) $ & $ 27.91 \pm 0.23 (4.2\sigma) $ \\
$z$ & $ 8.0 $ & $ 6.8 $ & $ 8.8 $ & $ 8.2 $ & $ 8.0 $ \\
$\mu$ & $ 1.5 $ & $ 1.8 $ & $ 1.7 $ & $ 1.5 $ & $ 10 $ \\
EW(Ly$\alpha$) & $ 1,200~\mathrm{\AA} $ & $ 570~\mathrm{\AA} $ & $ 820~\mathrm{\AA} $ & $ 820~\mathrm{\AA} $& $ 640~\mathrm{\AA} $ \\
M & $ 2.2 \times 10^6 \mathrm{M}_{\odot} $ & $ 1.2 \times 10^6 \mathrm{M}_{\odot} $ & $ 1.7 \times 10^6 \mathrm{M}_{\odot} $ & $ 4.8 \times 10^6 \mathrm{M}_{\odot} $ & $ 0.83 \times 10^6 \mathrm{M}_{\odot} $ \\
$\chi^2$ & $ 1.63 $ & $ 0.49 $ & $ 1.46 $ & $ 1.02 $ & $ 0.62 $ \\
\\
\hline
\end{tabular}
\end{center}
\end{table*}

\subsection{The five Pop~III galaxy candidates}
\label{sec:popiiigalaxycandidates}

A more thorough examination of the five objects identified through the search process described in Section~\ref{sec:searchprocess} has also been carried out, using Yggdrasil $\mathrm{Z} > 0$ templates (Section~\ref{sec:modelsyggdrasil}). Even though Pop~III models generally have a better quality of fit than any of the Yggdrasil $\mathrm{Z} > 0$ templates, there are competing low-metallicity fits (except for \MACS1931{} and \RXJ1347{}), as shown in Figure~\ref{fig:RF2}.

For each Pop~III galaxy candidate, the available Spitzer images (Section~\ref{sec:observationaldata}) have also been examined. Low-redshift old and/or dusty objects should be detectable in the longer wavelength filters of Spitzer. However, visual inspection of the Spitzer images at the positions of the objects yield non-detections in all available Spitzer filters, strengthening the case for the high-redshift solutions.

The very thin peak around the maximum of the quality of fit for Pop~III models seen in the case of \RXJ1347{} and to some extent \macstt{} and \Abell209{} is not a numerical artifact. The redshifted Ly$\alpha$ line is at the edge of the filters F125W for \RXJ1347{} and \Abell209{}, and on the edge of F140W for \macstt{}. This means that the throughput of the Ly$\alpha$ radiation in the filter is extremely redshift sensitive. Since the line is simultaneously observed in F105W and F110W for \RXJ1347{} and \Abell209{}, and in 105W, F110W, and F125W for \macstt{}, the parameter $f_{\mathrm{Ly\alpha}}$ is well-constrained and the quality of fit becomes very sensitive to redshift in a small redshift interval. The result is a very narrow peak of better quality of fit. The Yggdrasil $\mathrm{Z} > 0$ templates have lower intrinsic Ly$\alpha$ emission, hence the effect is smaller and only noticeable in \RXJ1347{} as a small local peak around $z=8.0$ in Figure~\ref{fig:RF2}.

\subsubsection{\Abell209{}}

As can be seen in Figure~\ref{fig:RF1}, the object \Abell209{} contains the clearest Ly$\alpha$-break of the five objects we discuss in this paper. It has clear detections (S/N$>5$) in all five IR filters. In Figure~\ref{fig:RF2}, we plot observations versus the best-fit Pop~III galaxy model and the best fitted model from CWW, Kinney and Gissel are compared. The clear Lyman-break is seen and also the steep decline in the flux of observations with wavelength. The strong Ly$\alpha$-line, which enters the filters F105W, F110W and F125W, in the Pop~III model causes it to be in better agreement with the observations compared to Gissel and CWW, Kinney. However, as can be seen in Figure~\ref{fig:RF2}, there are $\mathrm{Z} > 0$ Yggdrasil template fits closely resembling the Pop~III model fits. Only the small peak described in Section~\ref{sec:popiiigalaxycandidates} is not reproduced by the $\mathrm{Z} > 0$ Yggdrasil templates, which instead find a lower photometric redshift solution.

\begin{figure}[t!]
\resizebox{\hsize}{!}{\includegraphics[clip=true]{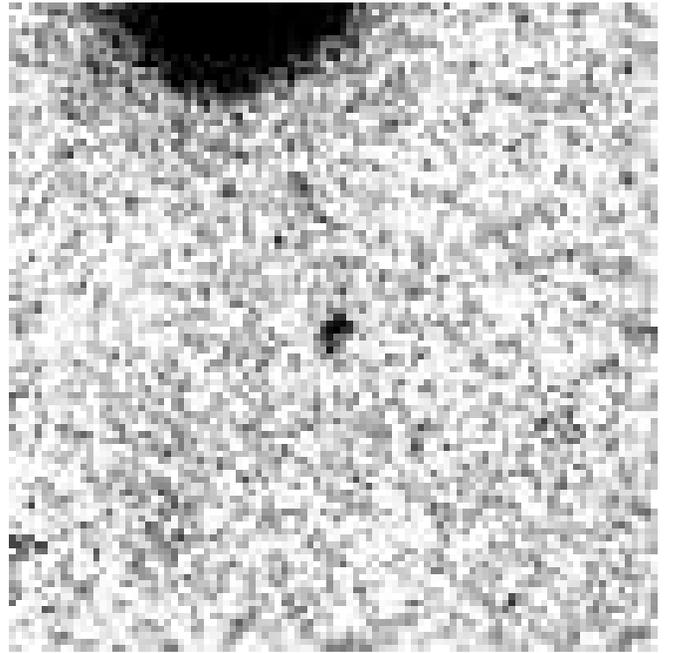}}
\caption{\footnotesize
Wider (6.5$''$ $\times$ 6.5$''$) image of \Abell209{} and its surroundings in F105W. The most nearby object is more than 2$''$ away.
}
\label{fig:RF_a209_3}
\end{figure}

The nearest object to \Abell209{} is situated more than 2$''$ away, as shown in Figure \ref{fig:RF_a209_3}. This is an advantage when considering ground-based spectroscopy, since there will be less interfering radiation from other objects.

\subsubsection{\macst{}}

The Ly$\alpha$-break for \macst{} (see Figure~\ref{fig:RF1}) around F850LP-F105W is not very clear with a 2.7$\sigma$ detection in F850LP. This is also the object with the lowest photometric redshift estimate ($z = 6.8$) of the five objects. The object has just one 5$\sigma$ detection, in the filter F110W. \macst{} also has one $\mathrm{S/N} \approx 4.98$ detection in F105W. As can be seen in Figure~\ref{fig:RF2}, the filters F850LP, F105W and F110W have brighter magnitudes, while the F125W, F140W and F160W filters have fainter magnitudes. This causes a break between F110W and F125W, which suggests a steep spectral slope. The object can also be reproduced with a Ly$\alpha$-emission line entering the three filters with lower wavelength, as in the Pop~III galaxy model. The cross-validation fit to Pop~III galaxy models is excellent, as shown in the P(z) graph in Figure~\ref{fig:RF2}. The fit to $\mathrm{Z} > 0$ Yggdrasil templates are worse but not significantly so. The Gissel grid of models have a good fit around $z=6.2$, and the CWW, Kinney grid has a quite good fit at slightly lower redshift.

\begin{figure}[t!]
\resizebox{\hsize}{!}{\includegraphics[clip=true]{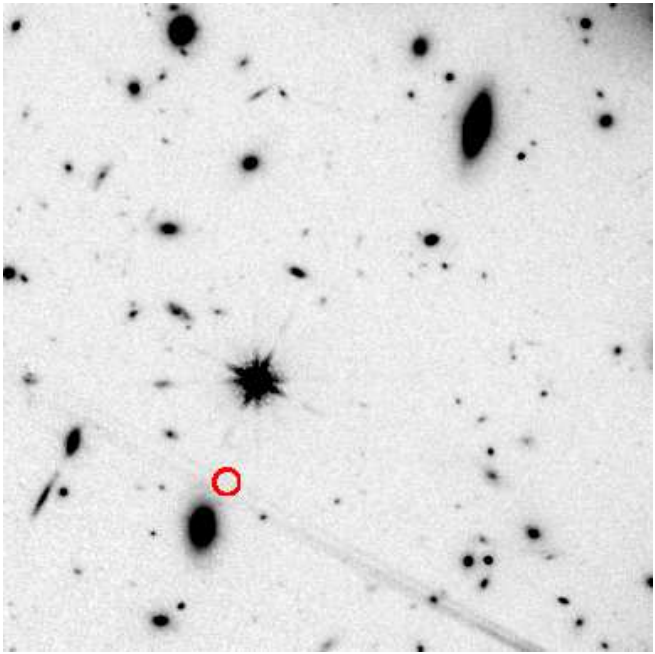}}
\caption{\footnotesize
$50'' \times 50''$ image in F160W of the cluster MACS0416. \macst{} is marked with a circle. As can be seen, there is possible contamination of light from a diffraction spike originating in an object to the lower right of \macst{}.
}
\label{fig:RF_macs0416_3}
\end{figure}

Figure~\ref{fig:RF_macs0416_3} shows an overview of the MACS0416 cluster with the object's location marked. As can be seen, there is possible contamination by light from a diffraction spike to the lower right in the image. Most of the light from the diffraction spike appears to be lower in the image than the object, yet there can be some contamination. This could distort the observed colors.

\subsubsection{\macstt{}}

\begin{figure}[t!]
\resizebox{\hsize}{!}{\includegraphics[clip=true]{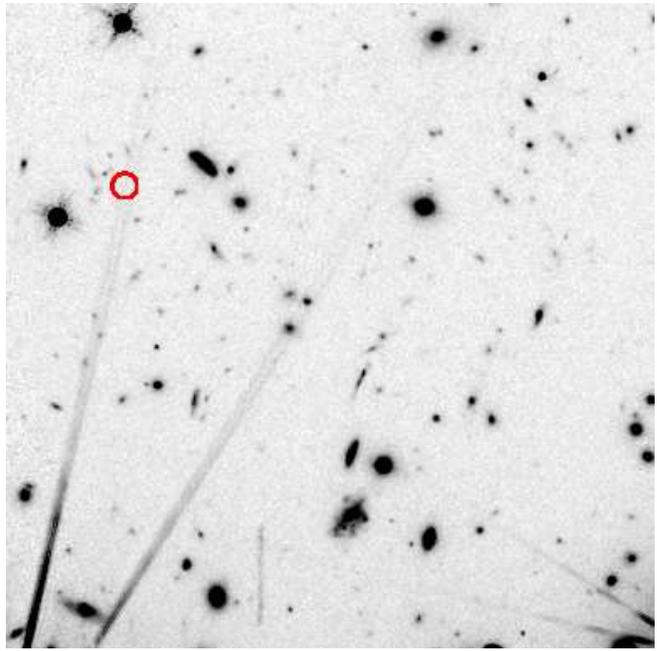}}
\caption{\footnotesize
$50'' \times 50''$ image in F160W of the cluster MACS0647. \macstt{} is marked with a circle. There is plausible light contamination from a diffraction spike from an object below and slightly to the left of \macstt{}.
}
\label{fig:RF_macs0647_3}
\end{figure}

This object, even though something can be discerned in the F105W thumbnail in Figure~\ref{fig:RF1}, seem to have the Ly$\alpha$-break between F105W and F110W. It is also hard to discriminate any coherent morphology for the object, especially when comparing F110W and F125W. F110W contains several smaller detections spread out while F125W has one larger detection but offset to the lower in the image. This raises doubts about the nature of the object. It could even be several smaller objects identified as one by SExtractor. Also, Figure \ref{fig:RF_macs0647_3} shows that the \macstt{} object, similar to the \macst{} object, is plausibly contaminated by light from a diffraction spike. The contamination could possibly explain the morphology as optical artifacts. Examining the images in each filter separately reveals that the diffraction spike in F110W has a completely different angle compared to F125W, F140W, and F160W (where it covers the object). Since this uncontaminated image has a 5$\sigma$ detection the whole object is not likely to be an artifact from the diffraction spike. But closer investigation would be necessary to determine the complete impact of the diffraction spike on the reported fluxes.

If, despite these considerations, considering it as one object, it has two 5$\sigma$ detections in F110W and F125W. For the best-fitting models there is a huge difference in photometric redshifts -- 8.8 for the Pop~III model versus 6.1 and 6.3 for CWW, Kinney and Gissel, respectively, and 8.1 for the Yggdrasil $\mathrm{Z} > 0$ grid. The difference can be understood by investigating the object in Figure~\ref{fig:RF2}. The Pop~III model has a very strong Ly$\alpha$-line entering F105W, F110W and F125W as well as partly into F140W (causing the small peak as described in Section~\ref{sec:popiiigalaxycandidates}). This produces a good fit even though it has hardly any continuum emission in F105W and only half of F110W has continuum flux. The Yggdrasil $\mathrm{Z} > 0$ grid, unable to produce a similar peak at $z=8.8$, produces its optima at $z=8.1$ by having the Ly$\alpha$-line in F105W, F110W and F125W simultaneously and relying on a stronger continuum. The other comparison models need to have more continuum emission instead, causing a lower redshift solution to fit better.

\subsubsection{\MACS1931{}}
\label{sec:results_macs1931}

The object we have discovered behind the cluster MACS1931 has very extreme colors, even for a possible Pop~III galaxy (see Figure~\ref{fig:colorcolordiagram}). In the thumbnail images (Figure~\ref{fig:RF1}), this can be seen as a strong detection in F125W while the other filters have lower fluxes. This object has two 5$\sigma$ detections in F110W and F125W with the detection in F105W falling short at $\mathrm{S/N} \approx 4.9$. The best-fit solutions are shown in Figure~\ref{fig:RF2}, along with the observed fluxes. Whereas Pop~III models do a reasonably good job at explaining the observed SED, comparison models fare much worse and are unable to simultaneously fit the bright flux in the F125W and the lower fluxes in adjacent filters. The object's cross-validation fit, seen in Figure~\ref{fig:RF2}, shows a very clear preference for a Pop~III galaxy solution in the redshift interval $z = 8.0-9.0$. However, caution must be expressed about the nature of this object because of its extreme colors, placing it outside the Pop~III region in the color-color diagram.

\subsubsection{\RXJ1347{}}

\begin{figure*}[h!]
     \begin{center}
     \begin{tabular}{ | c | c | }
     \hline
      Spectrum & P(z) \\
			\hline
			\includegraphics[width=60mm, height=42mm]{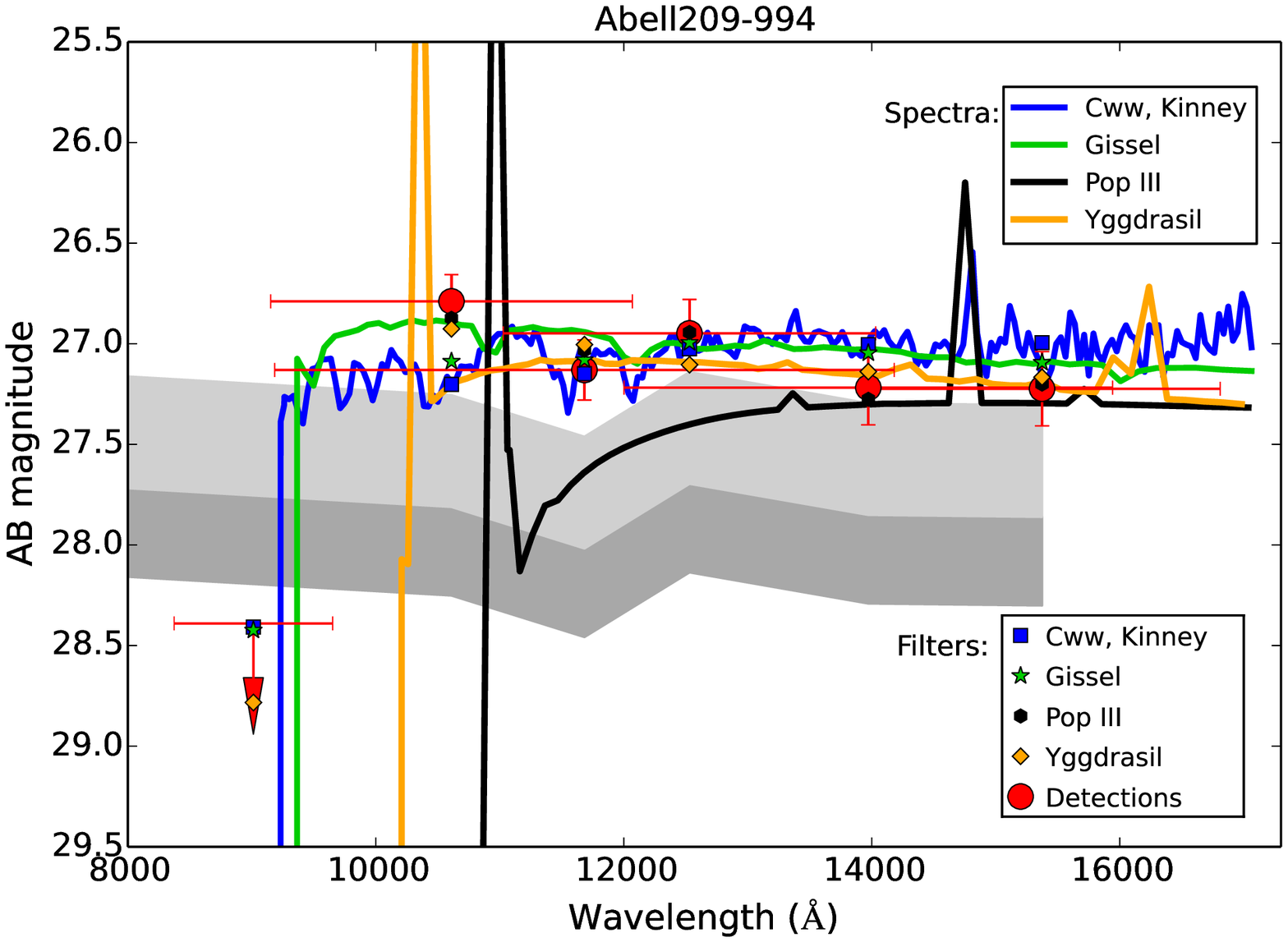}
      & 
			\includegraphics[width=70mm, height=42mm]{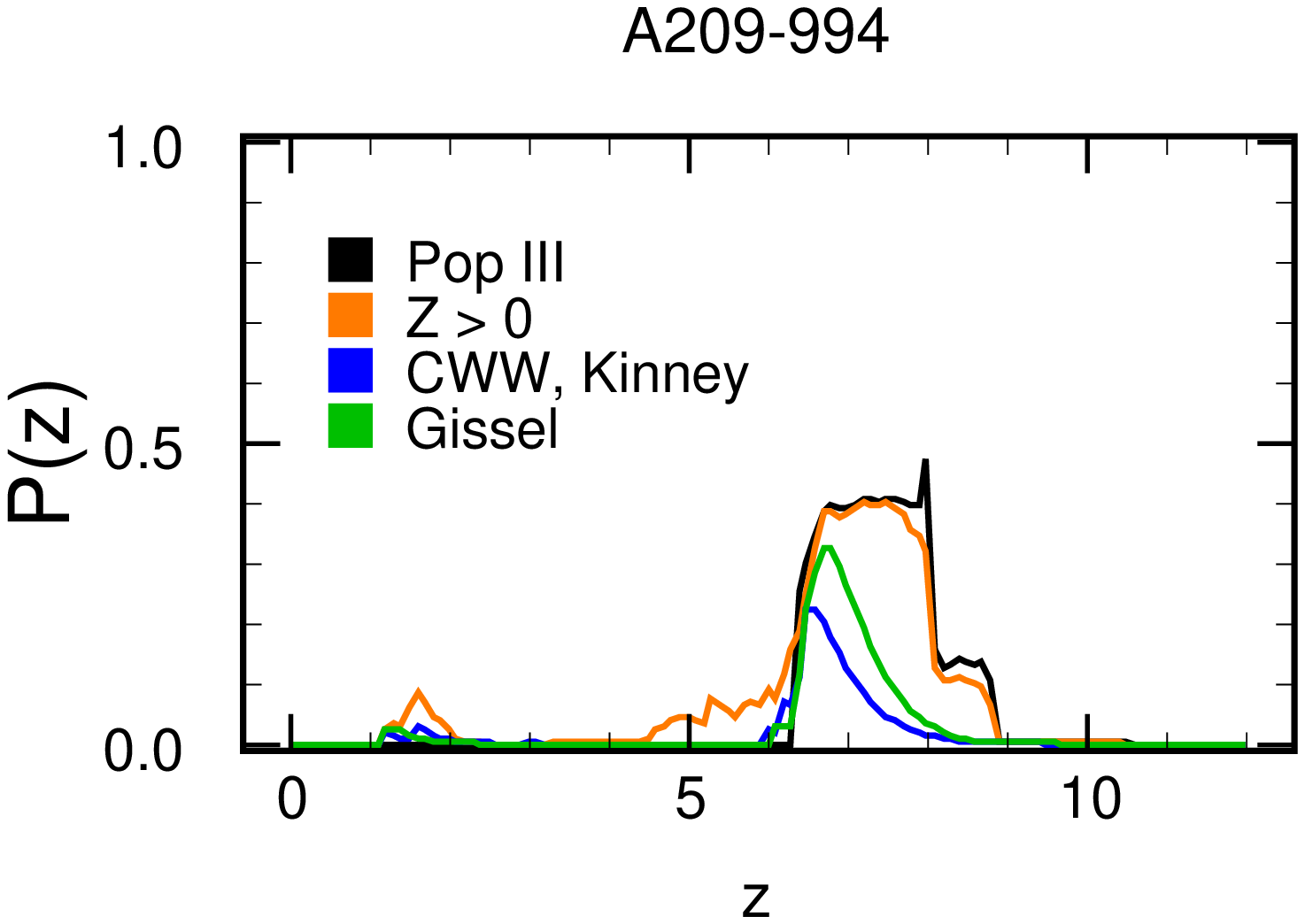}
			\\
			\hline
			\includegraphics[width=60mm, height=42mm]{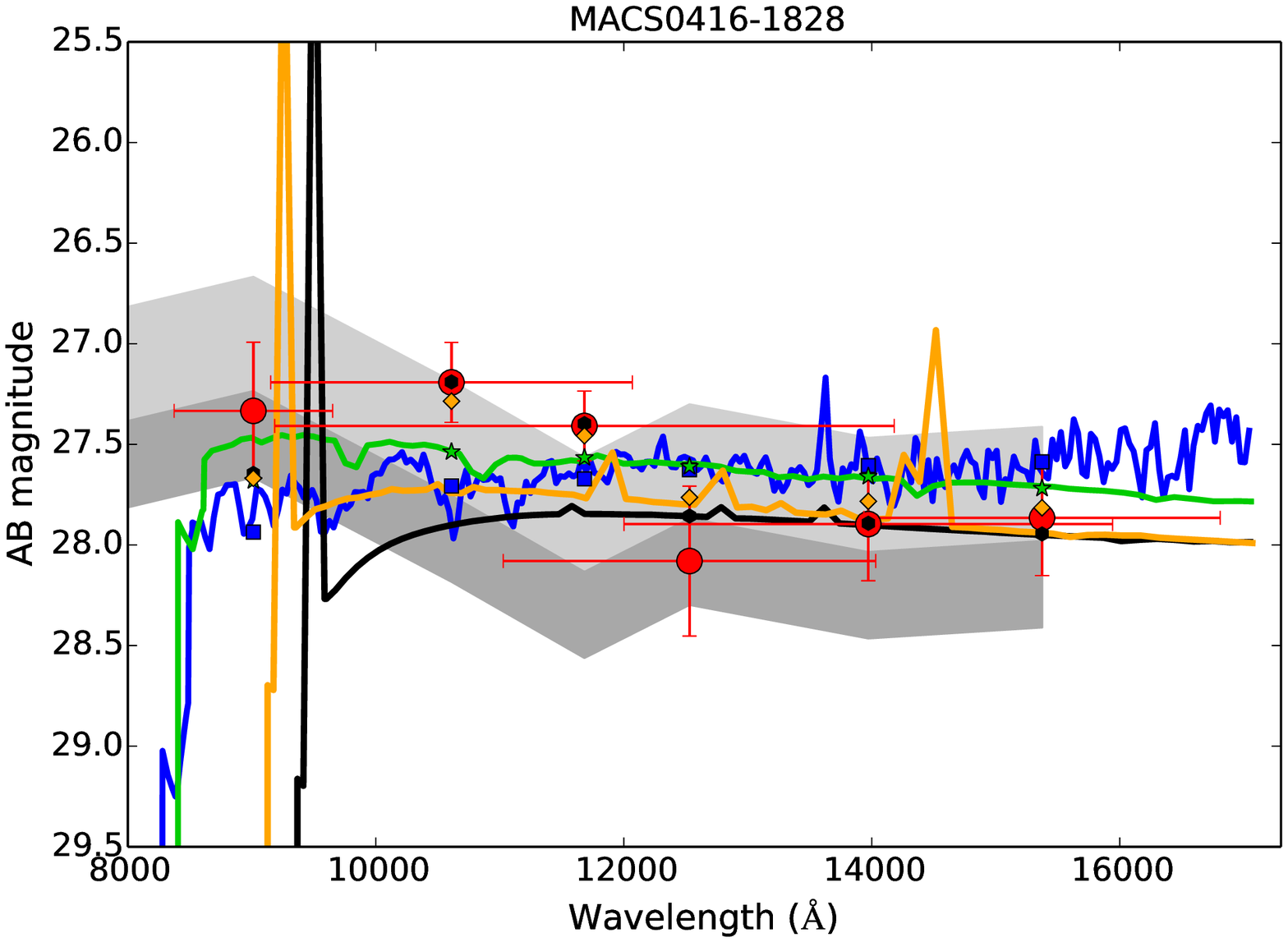}
      & 
			\includegraphics[width=70mm, height=42mm]{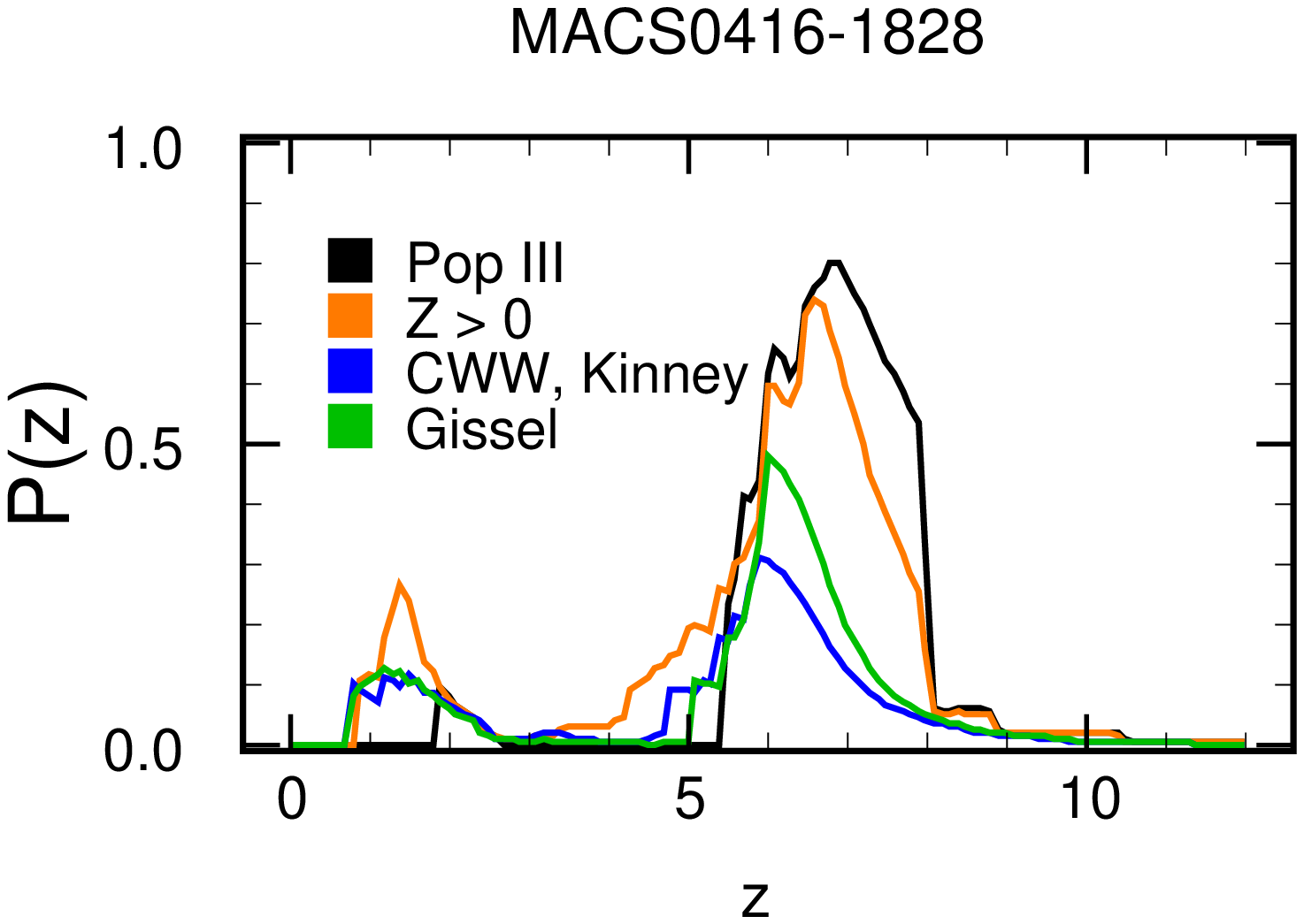}
			\\ 
			\hline
			\includegraphics[width=60mm, height=42mm]{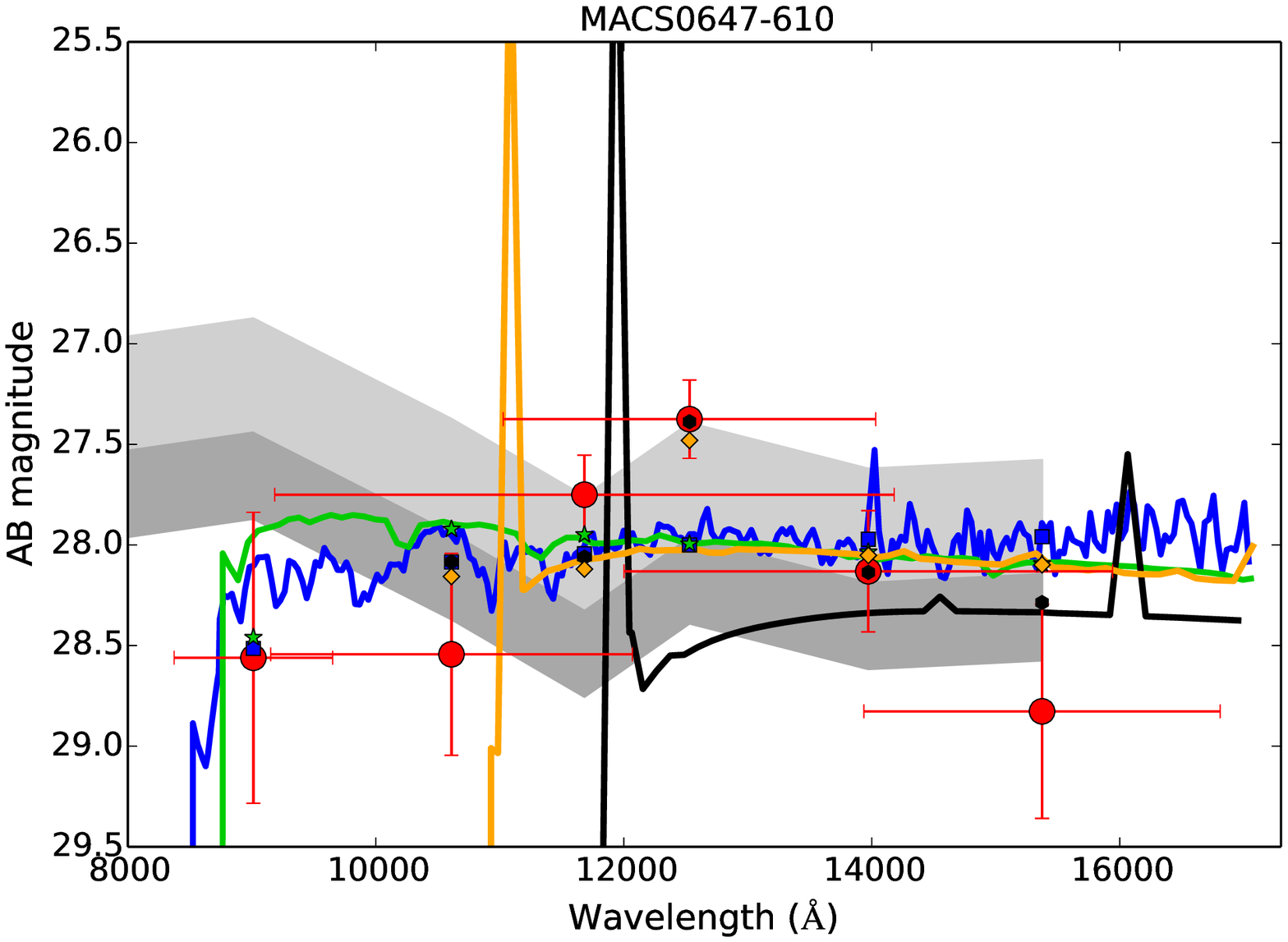}
      & 
			\includegraphics[width=70mm, height=42mm]{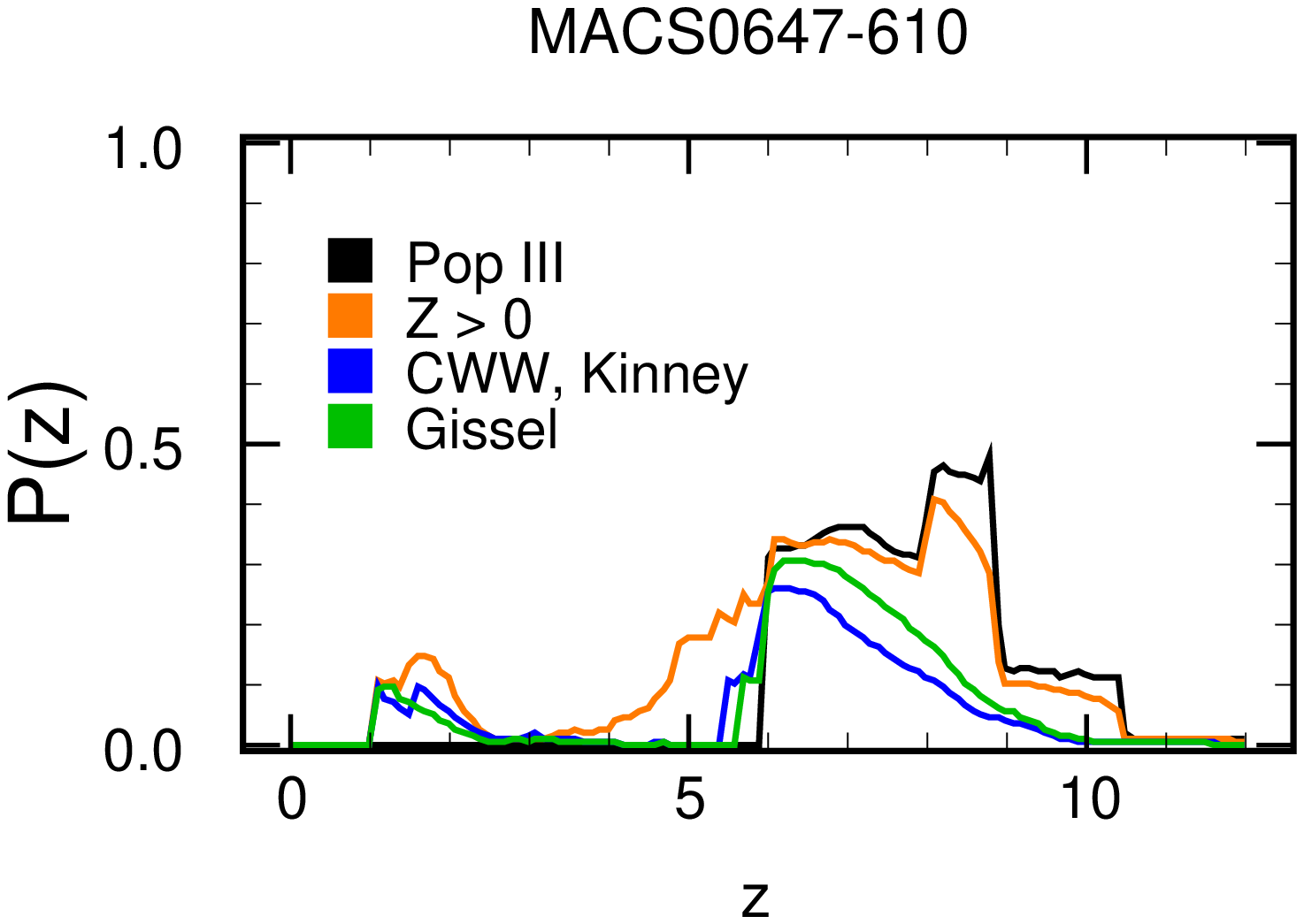}
      \\ 
			\hline
			\includegraphics[width=60mm, height=42mm]{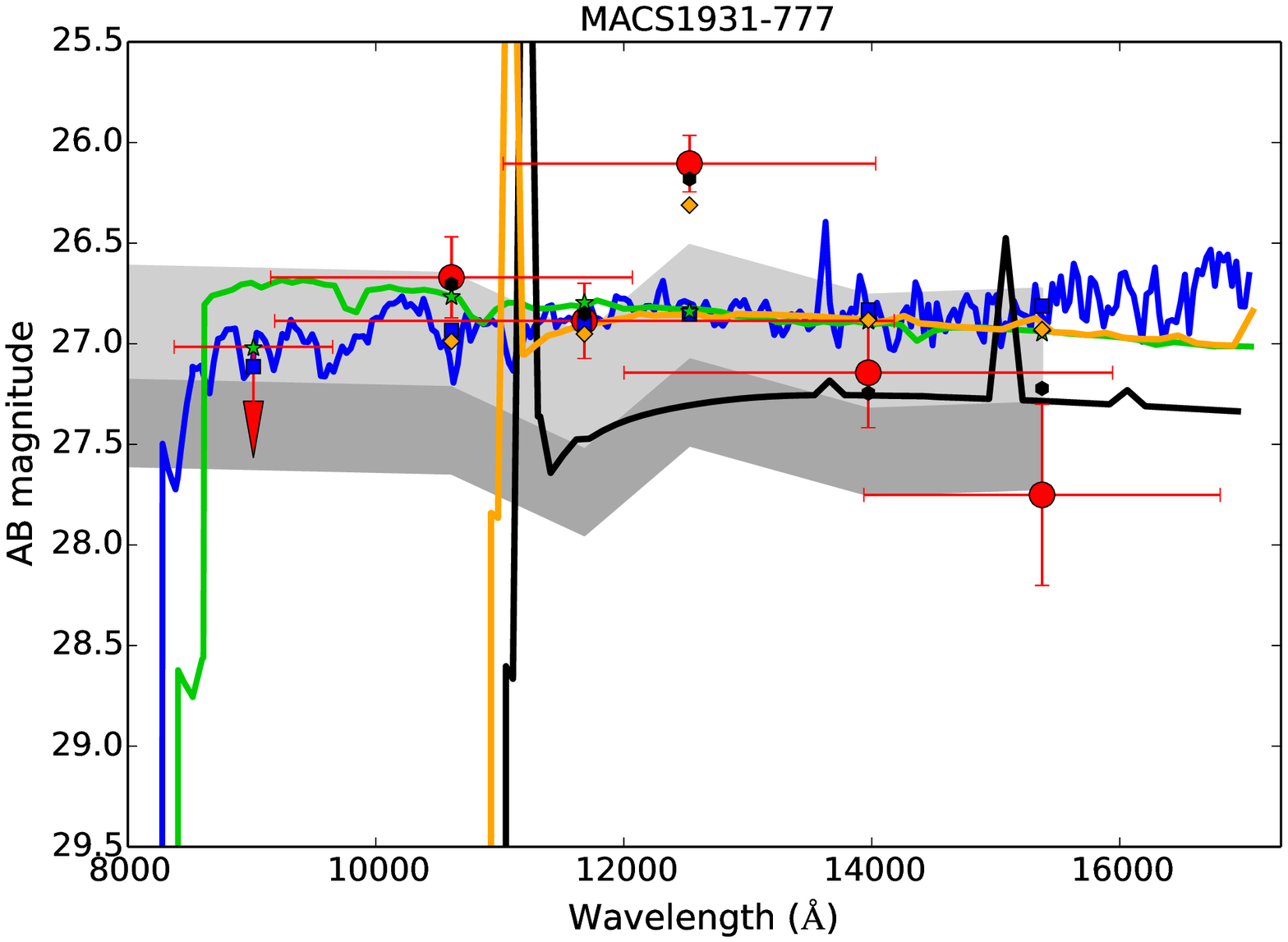}
      & 
			\includegraphics[width=70mm, height=42mm]{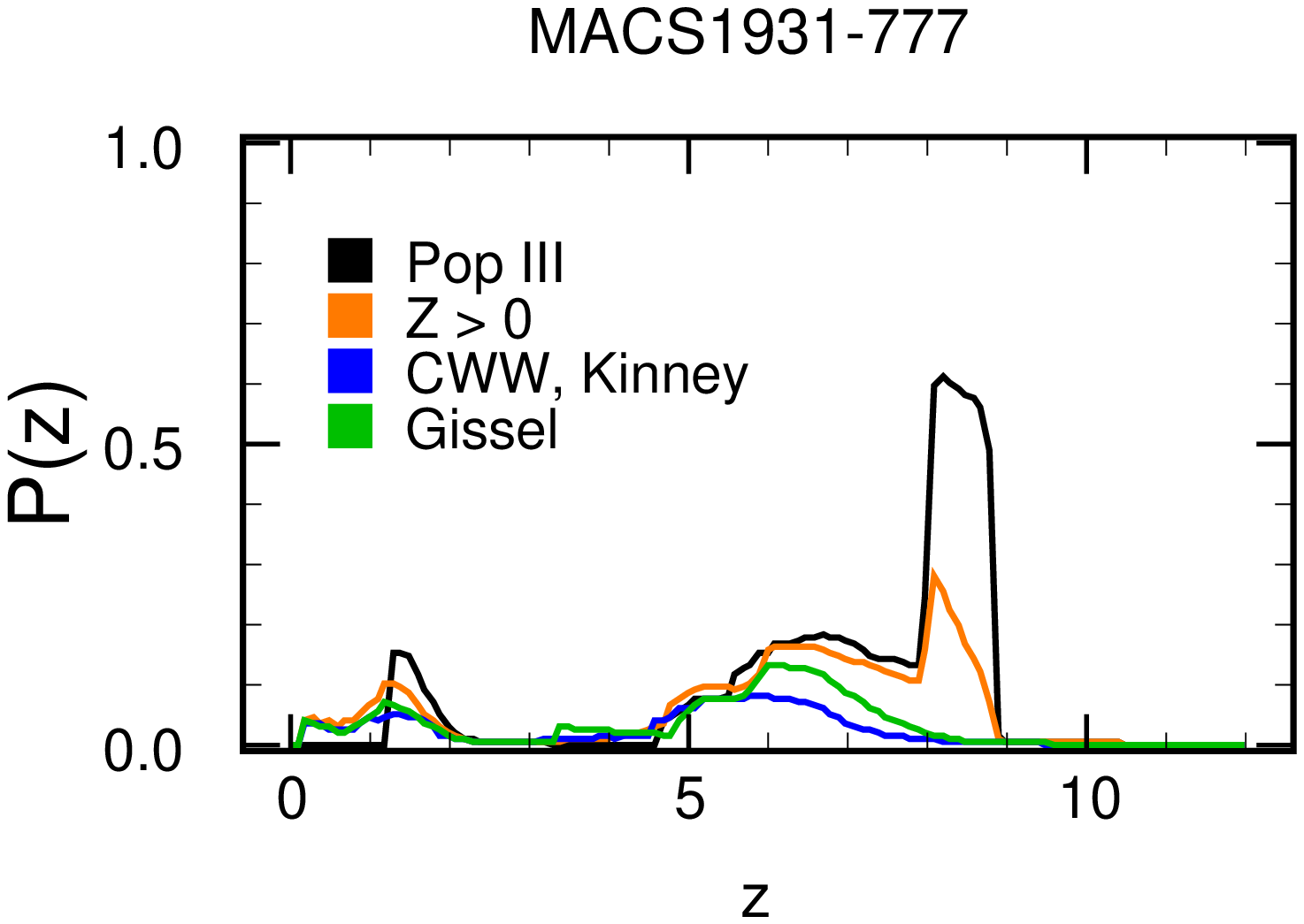}
			\\ 
			\hline
			\includegraphics[width=60mm, height=42mm]{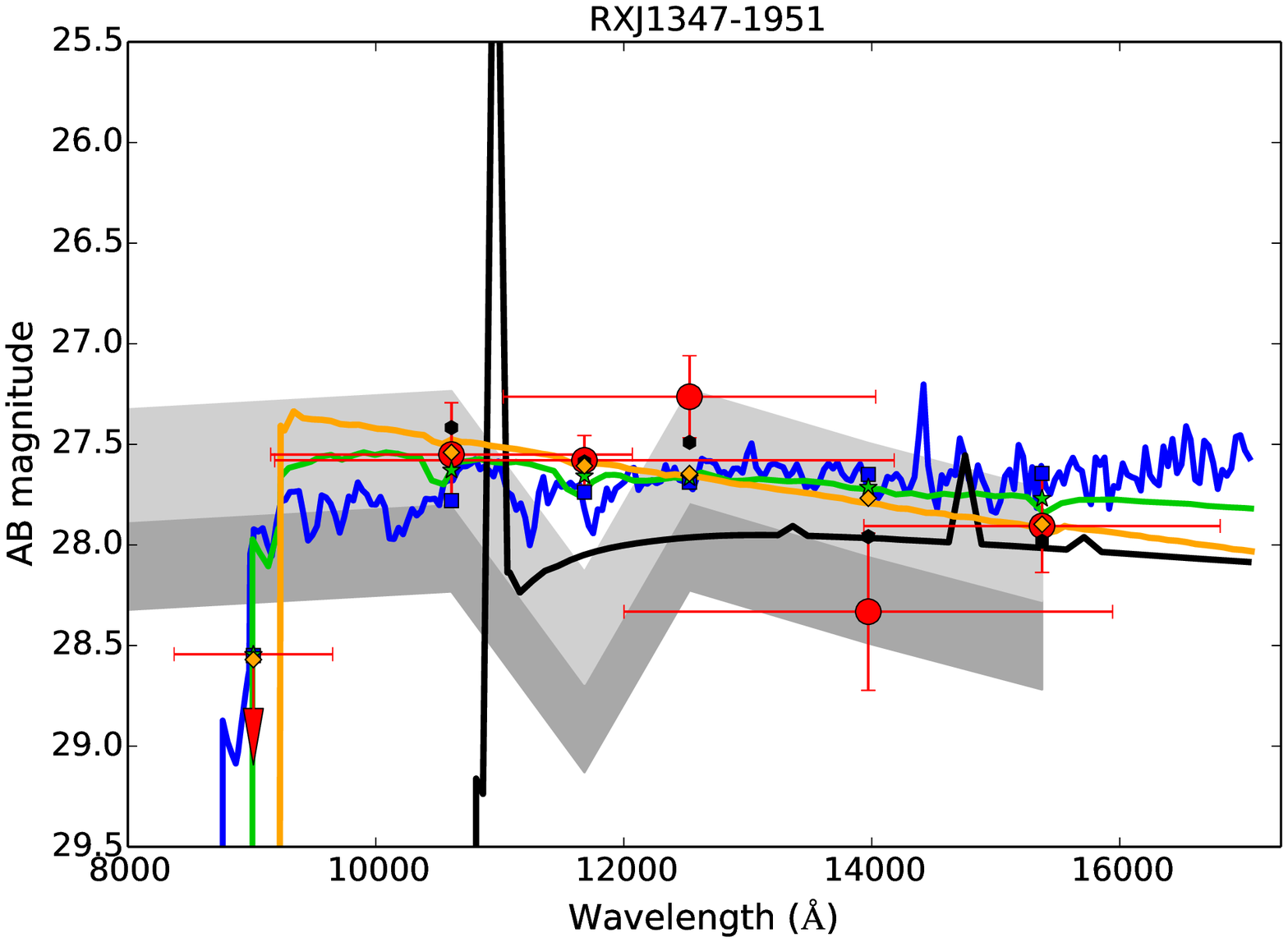}
      & 
			\includegraphics[width=70mm, height=42mm]{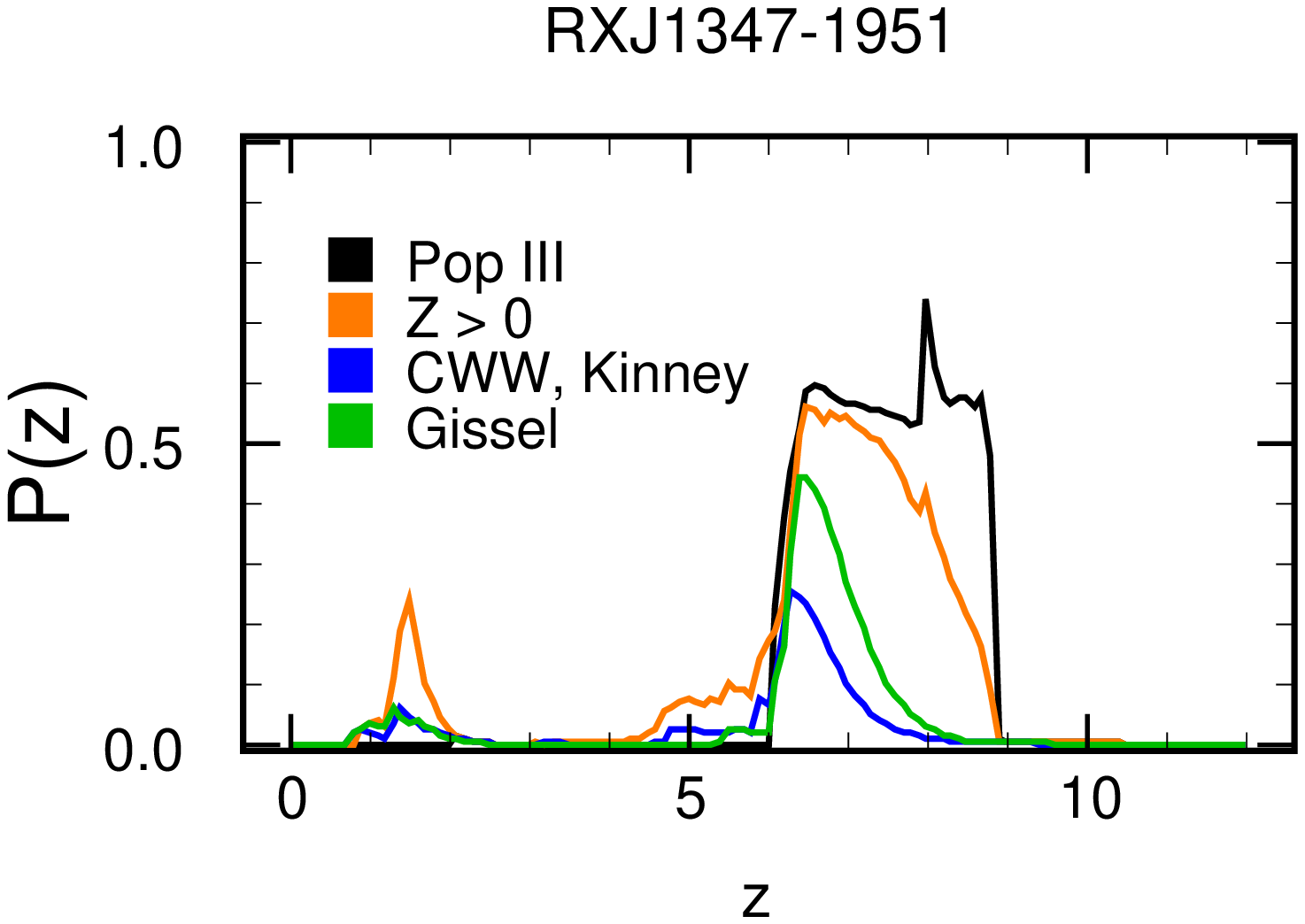}
      \\ 
			\hline
      \end{tabular}
      \caption{\footnotesize
Spectra and cross-validation fits. Each row displays one object. The left column displays observed magnitudes with errors in red versus spectra and integrated fluxes for the best-fitting galaxy model from each grid. The gray areas correspond to different levels of S/N. Values below the gray areas means $\mathrm{S/N}<2$. The dark gray area indicates $2<\mathrm{S/N}<3$ and the light gray area $3<\mathrm{S/N}<5$. Data above the gray areas have $\mathrm{S/N}>5$. The right column contains cross-validation fits for observations of each object with the Gissel, CWW Kinney, Yggdrasil $\mathrm{Z}>0$, and Pop~III galaxy grids. For each redshift, the best fit of any model/parameter selection within each grid is displayed. For \Abell209{}, a clear Ly$\alpha$-break is seen in the observations between F850LP and F105W. \macst{} is bright in filters F105W and F110W relative to the higher wavelength filters, making models with strong Ly$\alpha$ emission reproduce this closer than other comparison models. In \macstt{}, the Pop~III galaxy model has a strong Ly$\alpha$ line entering F105W, as well as F110W and F125W; here the Yggdrasil $\mathrm{Z}>0$ model grid finds a comparably good fit at $z=8.1$. For the \MACS1931{} object, only the Pop~III model manages to fit the observation in F125W, producing a narrow peak of extremely good fits to Pop~III galaxy models at $z \approx 8-9$. For \RXJ1347{}, the Pop~III model also fits the observational data better than the comparison models in an extremely narrow peak.
}
      \label{fig:RF2}
      \end{center}
      \end{figure*}

\begin{figure*}[h!]
     \begin{center}
     \begin{tabular}{ | l | c | c | c | c | c | c | c | }
     \hline
      & F814W & F850LP & F105W & F110W & F125W & F140W & F160W \\
			\hline
			\begin{sideways}
			{\scriptsize \Abell209{}}
			\end{sideways}
			&
      \includegraphics[width=22mm, height=22mm]{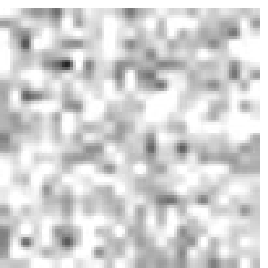}
      & 
			\includegraphics[width=22mm, height=22mm]{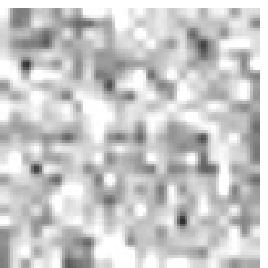}
			& 
			\includegraphics[width=22mm, height=22mm]{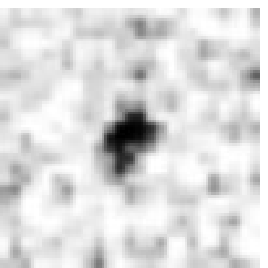}
      & 
			\includegraphics[width=22mm, height=22mm]{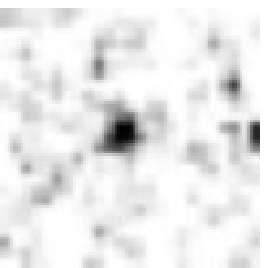}
      & 
			\includegraphics[width=22mm, height=22mm]{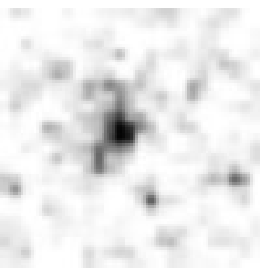}
      & 
			\includegraphics[width=22mm, height=22mm]{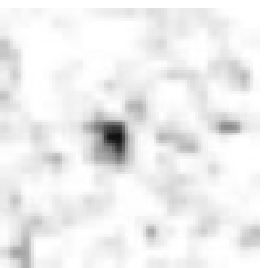}
      & 
			\includegraphics[width=22mm, height=22mm]{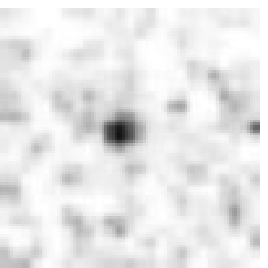}
      \\
			\hline
			\begin{sideways}
			{\scriptsize \macst{}}
			\end{sideways}
			&
      \includegraphics[width=22mm, height=22mm]{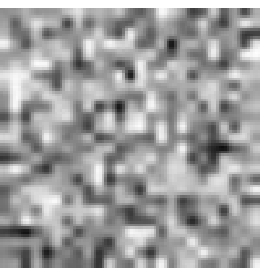}
      & 
			\includegraphics[width=22mm, height=22mm]{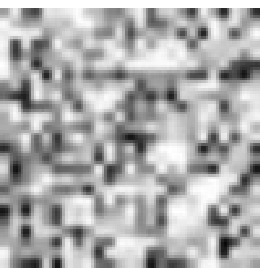}
			& 
			\includegraphics[width=22mm, height=22mm]{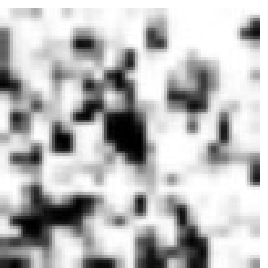}
      & 
			\includegraphics[width=22mm, height=22mm]{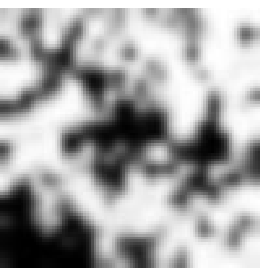}
      & 
			\includegraphics[width=22mm, height=22mm]{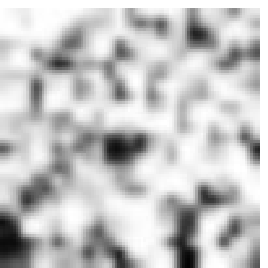}
      & 
			\includegraphics[width=22mm, height=22mm]{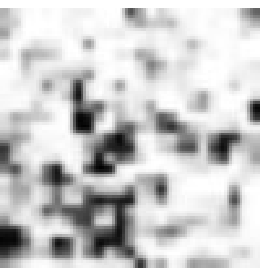}
      & 
			\includegraphics[width=22mm, height=22mm]{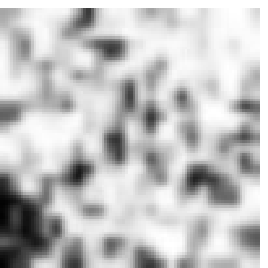}
      \\ 
			\hline
			\begin{sideways}
			{\scriptsize \macstt{}}
			\end{sideways}
			&
      \includegraphics[width=22mm, height=22mm]{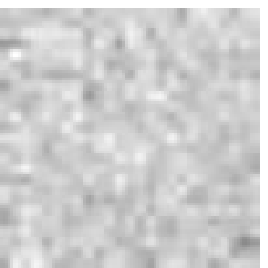}
      & 
			\includegraphics[width=22mm, height=22mm]{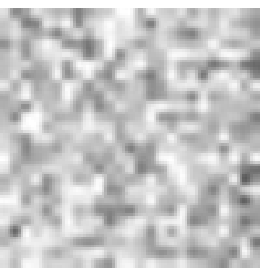}
			& 
			\includegraphics[width=22mm, height=22mm]{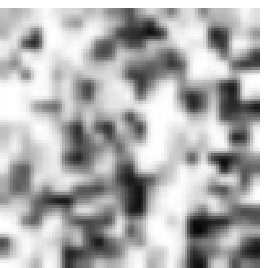}
      & 
			\includegraphics[width=22mm, height=22mm]{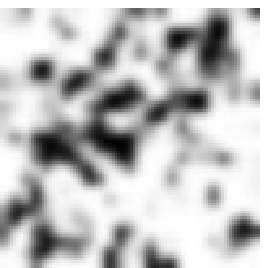}
      & 
			\includegraphics[width=22mm, height=22mm]{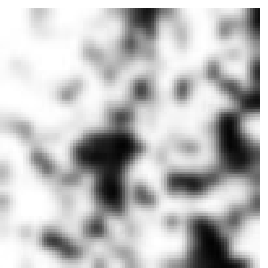}
      & 
			\includegraphics[width=22mm, height=22mm]{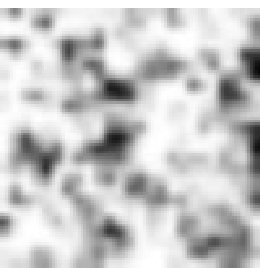}
      & 
			\includegraphics[width=22mm, height=22mm]{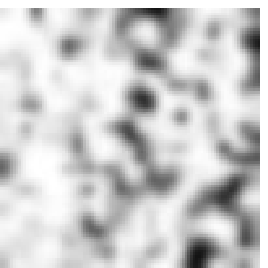}
      \\ 
			\hline
			\begin{sideways}
			{\scriptsize \MACS1931{}}
			\end{sideways}
			&
      \includegraphics[width=22mm, height=22mm]{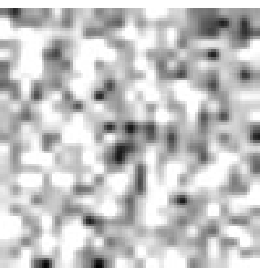}
      & 
			\includegraphics[width=22mm, height=22mm]{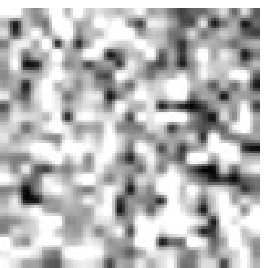}
			& 
			\includegraphics[width=22mm, height=22mm]{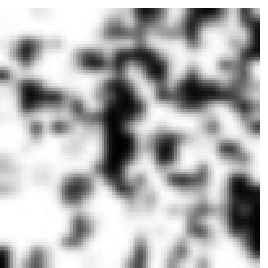}
      & 
			\includegraphics[width=22mm, height=22mm]{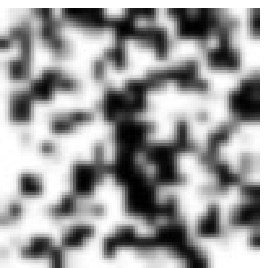}
      & 
			\includegraphics[width=22mm, height=22mm]{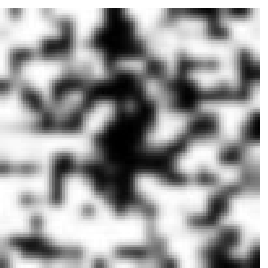}
      & 
			\includegraphics[width=22mm, height=22mm]{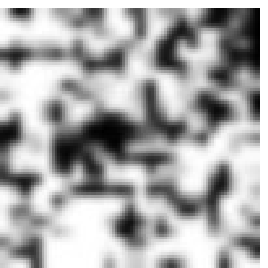}
      & 
			\includegraphics[width=22mm, height=22mm]{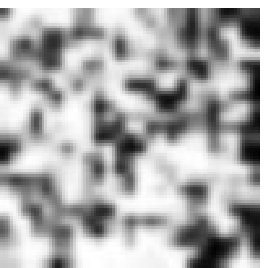}
      \\ 
			\hline
			\begin{sideways}
			{\scriptsize \RXJ1347{}}
			\end{sideways}
			&
      \includegraphics[width=22mm, height=22mm]{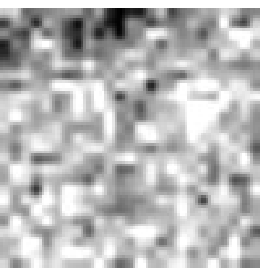}
      & 
			\includegraphics[width=22mm, height=22mm]{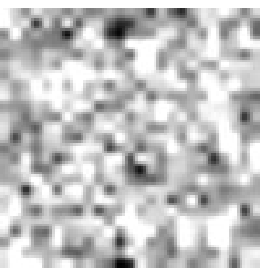}
			& 
			\includegraphics[width=22mm, height=22mm]{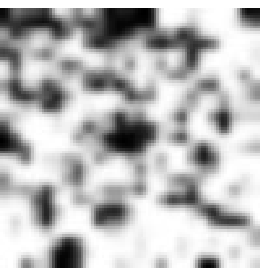}
      & 
			\includegraphics[width=22mm, height=22mm]{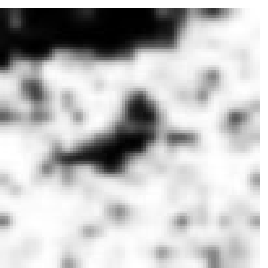}
      & 
			\includegraphics[width=22mm, height=22mm]{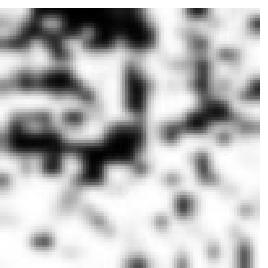}
      & 
			\includegraphics[width=22mm, height=22mm]{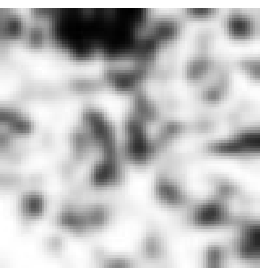}
      & 
			\includegraphics[width=22mm, height=22mm]{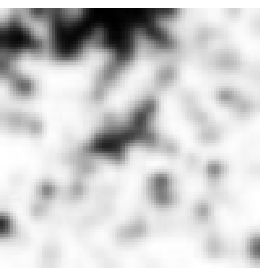}
      \\ 
			\hline
      \end{tabular}
      \caption{\footnotesize
1.6$''$ $\times$ 1.6$''$ Thumbnail images of the five Pop~III galaxy candidates. The rows contain the five objects, and the columns represent the seven filters with the longest wavelength coverage in the CLASH survey. The Ly$\alpha$-break is clearly observed in each of them. For most of them, the break appears between the filters F850LP and F105W. For the \macstt{} object, there is hardly any detection in F105W, so the Ly$\alpha$-break in this object is less clear and plausibly located between F105W and F110W instead. It is hard to discern any coherent morphology for \macstt{}, especially when comparing F110W and F125W, which raises doubts about the nature of the object (it could even be several smaller objects). The object \Abell209{} seems to have the clearest Ly$\alpha$-break of our five objects. Hints of flux can be seen in the F850LP filter for \macst{}, and in the F814W filter for \MACS1931{}. This could imply lower redshift objects, but could also be spurious detections in a few pixels.
}
      \label{fig:RF1}
      \end{center}
      \end{figure*}

The object we have discovered behind RXJ1347 possesses several interesting properties. In Figure~\ref{fig:colorcolordiagram}, we see that it has a very attractive combination of colors to correspond to a Pop~III galaxy. The galaxy has one very clear ($\mathrm{S/N} \approx 8.8$) detection in F110W, while the F125W and F160W falls short of 5$\sigma$ detections by a narrow margin. Even though vague, the galaxy also appears elongated from upper right to lower left in some of the thumbnails in Figure~\ref{fig:RF1}. This plausibly confirms the high magnification (as discussed in Section~\ref{sec:gravitationallensing}). The magnification of $\mu \approx 10$ corresponds to a brightening of the object by $\Delta \mathrm{m}_{\mathrm{AB}} \approx 2.5$. Since this means that all magnitudes would be $\gtrsim 30$ without magnification, we could not have detected the object without the flux boost due to gravitational lensing. In Figure~\ref{fig:RF2} we show the best fit models from each grid compared to the observations. As can be seen, the Pop~III model fits the data better, especially for the peculiarly low flux in F140W (compared to the surrounding filters). The P(z) graph in Figure~\ref{fig:RF2} displays a general preference in a quite broad range, $z \approx 6-9$, for Pop~III galaxy models even though the Yggdrasil $\mathrm{Z} > 0$ grid has competitive fits. However, there is a narrow peak at $z=8.0$ (see Section~\ref{sec:popiiigalaxycandidates}), where Pop~III galaxy templates have a significantly better quality of fit.

			\subsection{Dust attenuation}
\label{sec:dustextinction}

The content of dust in galaxies can significantly alter their spectra. Dust absorbs the radiation and generally re-emits it at a longer wavelength. This can be modeled as an attenuation curve, which is the modification applied to the spectrum of a galaxy as a function of wavelength. An attenuation curve has one parameter, E(B-V), regulating the amount of attenuation applied (E(B-V) is usually between 0 and 1). Pop~III galaxies are expected to be dust-free, as dust requires metals to form, thus we have not applied any attenuation curve to those models. To the CWW, Kinney and Gissel grids, we have applied three different dust models, \citep{2000ApJ...533..682C, 1984A&A...132..389P, 1979MNRAS.187P..73S}, with E(B-V) varying from 0 to 1 in 0.1 increments. The CWW, Kinney models are built on empirical spectra, hence some dust extinction is already included, and the Gissel grid contains some models with dust extinction already applied. I.e. the E(B-V) parameter in this Section is relative to the model grids regardless whether they already contain extinction. 

Since the attenuation generally is much higher at lower wavelengths, we would expect our models to turn more red by applying dust. The very blue nature of our objects would thus fit worse to dusty models. For the \citet{2000ApJ...533..682C} and \citet{1984A&A...132..389P} attenuation models, this is indeed what we find. The E(B-V)~=~0 solution is favored for all our objects and for both comparison grids over solutions with higher E(B-V).

\begin{figure*}[htp]
  \begin{center}
	  \subfigure[CWW, Kinney]{\label{fig:mwseatonextinctiongraphcwwkinnet}\includegraphics[scale=0.45]{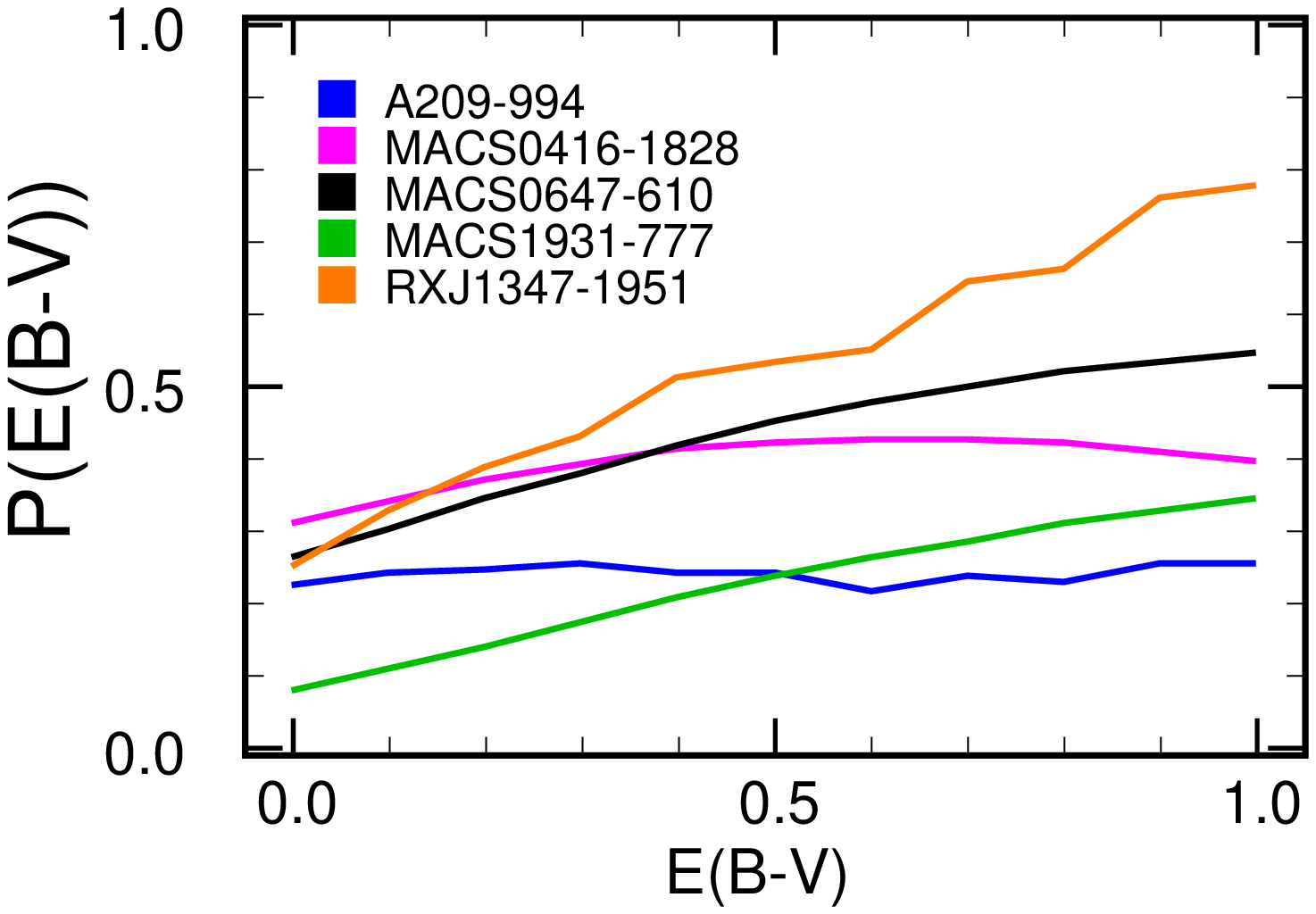}}
		\subfigure[Gissel]{\label{fig:mwseatonextinctiongraphgissel}\includegraphics[scale=0.45]{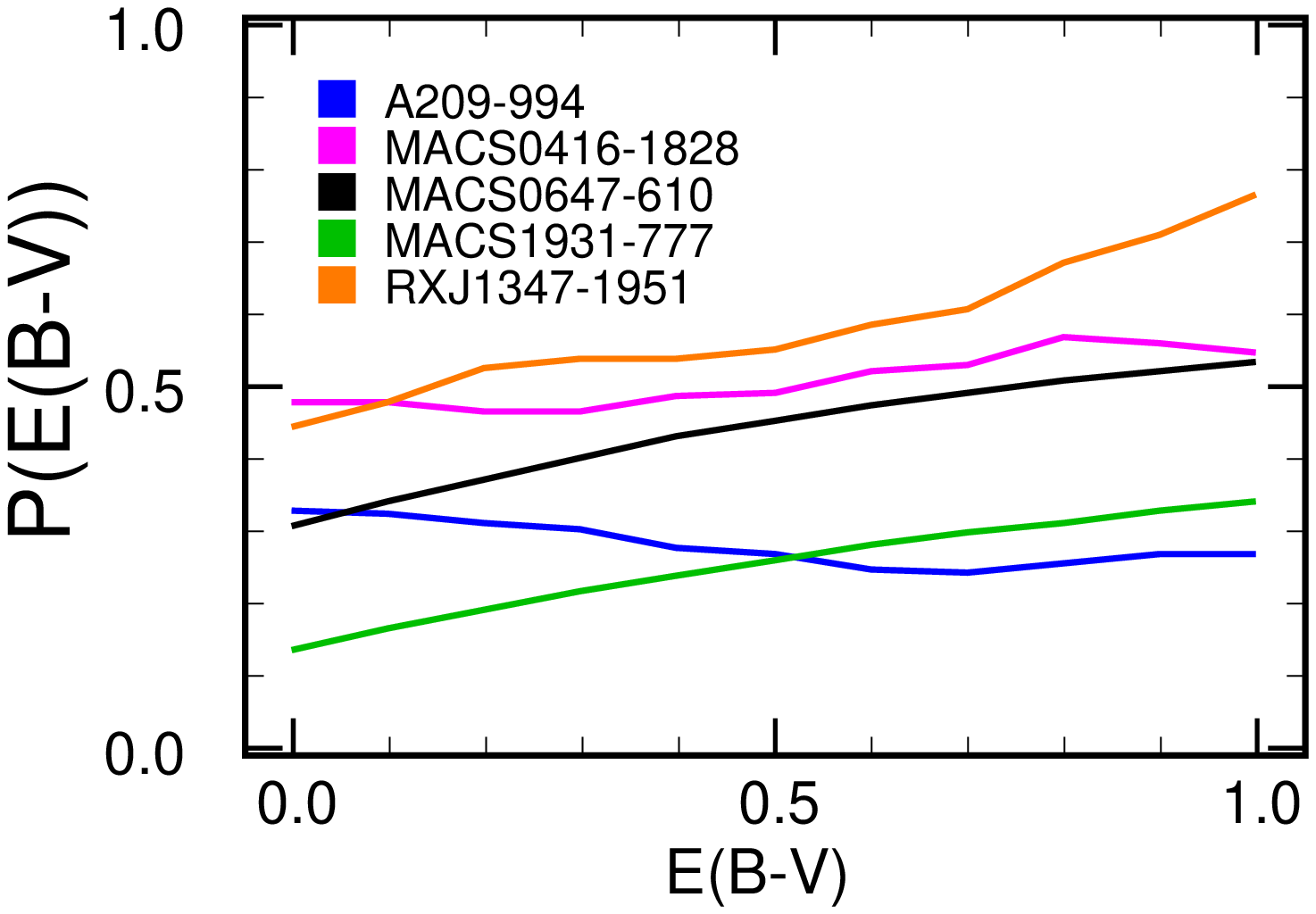}}
  \end{center}
  \caption{\footnotesize
     Quality of fit as a function of E(B-V) for the best-fitting models, using the \citet{1979MNRAS.187P..73S} attenuation model. The graph to the left uses the CWW, Kinney grid, while the graph to the right uses the Gissel grid. \Abell209{} and \macst{} do not improve their fits at all, or marginally, by adding dust attenuation. The other three objects improve their quality of fit significantly with increasing E(B-V). This follows from the ``bump" in the \citet{1979MNRAS.187P..73S} attenuation curve with locally increased obscuration around 2,175~$\mathrm{\AA}$. \macstt{} and \RXJ1347{} achieve fits competitive with their Pop~III fit while \MACS1931{} improves its fit significantly with non-zero E(B-V). However, this would assume very dusty objects (an E(B-V)~=~1.0 in the \citet{1979MNRAS.187P..73S} model corresponds to a dust attenuation in the rest-frame UV of $\Delta m_{\mathrm{AB}} \gtrsim 7.5$).
}
  \label{fig:mwseatonextinctiongraph}
\end{figure*}

The \citet{1979MNRAS.187P..73S} model has a ``bump''  with locally increased attenuation around 2,175~$\mathrm{\AA}$. This has the capacity, for high enough E(B-V), to actually turn the spectrum bluer, at least over a limited wavelength range. It turns out that the result of this effect is similar for our model grids CWW, Kinney and Gissel, but varies depending on object see Figure~\ref{fig:mwseatonextinctiongraph}. The \Abell209{} and \macst{} objects improve their fits marginally, or not at all, by adding \citet{1979MNRAS.187P..73S} dust attenuation, while the other three objects improve their fit significantly with E(B-V). \macstt{} and \RXJ1347{} achieve fits competitive with their Pop~III fit while \MACS1931{} improves its fit significantly with E(B-V). However, this would assume very dusty objects, an E(B-V) of 1.0 in the \citet{1979MNRAS.187P..73S} model corresponds to an attenuation of $\Delta m_{\mathrm{AB}} \approx 3$ in the rest-frame V-band. Since we are observing the redshifted UV-part of the spectrum, the attenuation is even higher. As a lower limit, we calculate the attenuation in the F225W and arrive at values $\Delta m_{\mathrm{AB}} \gtrsim 7.5$. In other words, our objects without dust would have apparent magnitudes $\sim 20$, which would imply a galaxy of implausible brightness at that redshift. Also, the \citet{1979MNRAS.187P..73S} model is derived for the Milky Way, while the \citet{2000ApJ...533..682C} and \citet{1984A&A...132..389P} models are derived for starburst galaxies and the Small Magellanic cloud, respectively. Our objects are more akin to the latter and at most a far-fetched caveat can be raised for the nature of the \macstt{}, \MACS1931{}, and \RXJ1347{} objects because of dust attenuation.

\section{Gravitational lensing}
\label{sec:gravitationallensing}

CLASH targets galaxy clusters at $z_\mathrm{L} \sim 0.2-0.9$, which act as gravitational lenses to improve S/N or even raise faint objects over the detection threshold. Magnified objects are distorted and often appear as extended arcs, typically extended along the circumference of the cluster. Therefore, an observed arc also confirms the lensed nature of an observed object. The magnification provided for a fixed lens redshift varies with the cosmological angular diameter distance of the source, and that from the lens to the source, and the object's position relative the lensing cluster. For a given source redshift, the magnification maps exhibits contours of infinite magnification -- the so-called critical curves. Background objects seen close to the critical curves for their redshift will thus have a high magnification, and are likely to be multiply imaged (specifically, if the source lies within the caustics, which are the critical curves when projected back to the source plane).

\begin{figure*}[htp]
  \begin{center}
	  \subfigure[Abell209]{\includegraphics[scale=0.12]{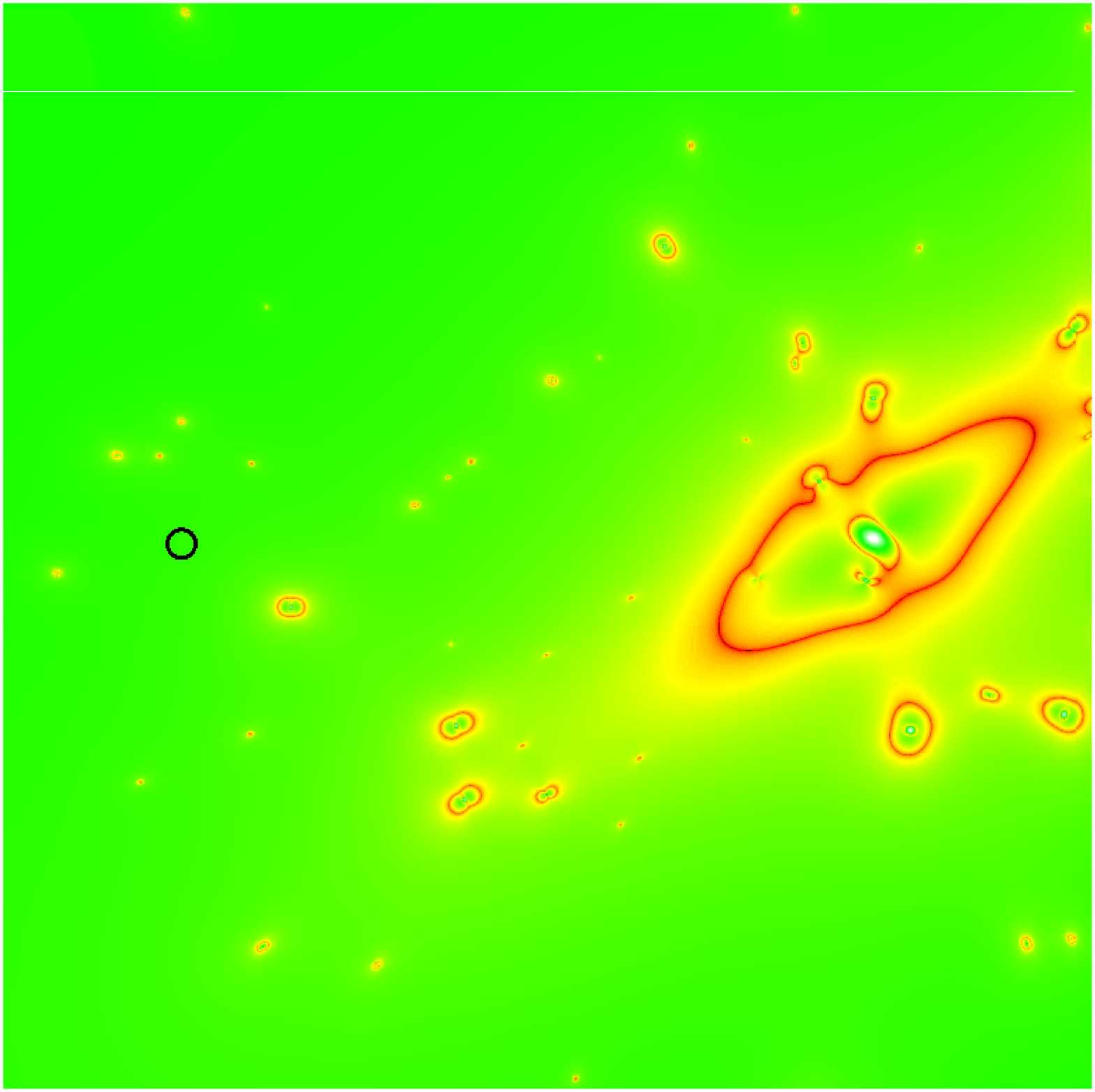}}
		\subfigure[MACS0416]{\includegraphics[scale=0.12]{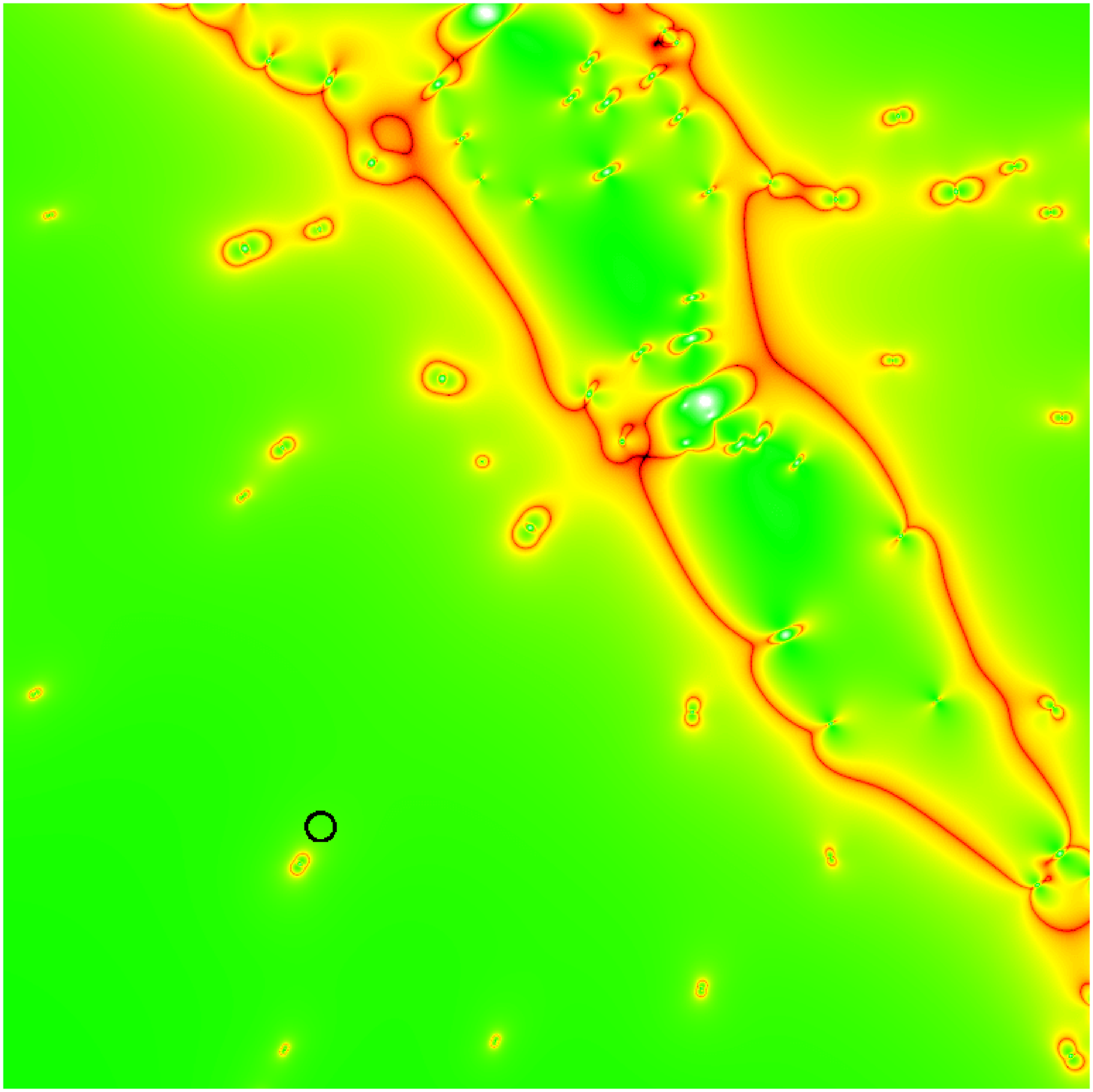}}
		\subfigure[MACS0647]{\includegraphics[scale=0.12]{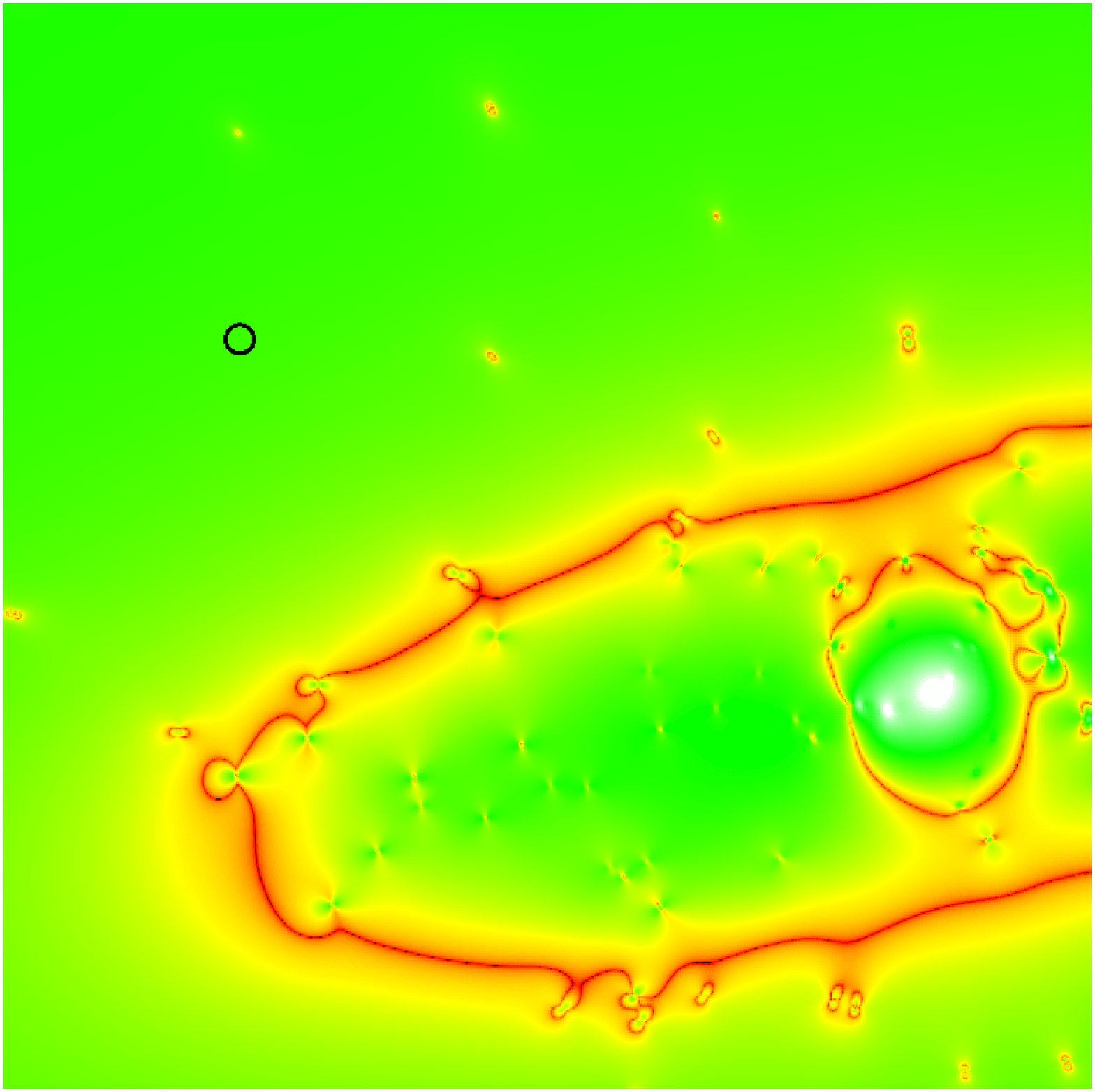}}
	  \subfigure[MACS1931]{\includegraphics[scale=0.12]{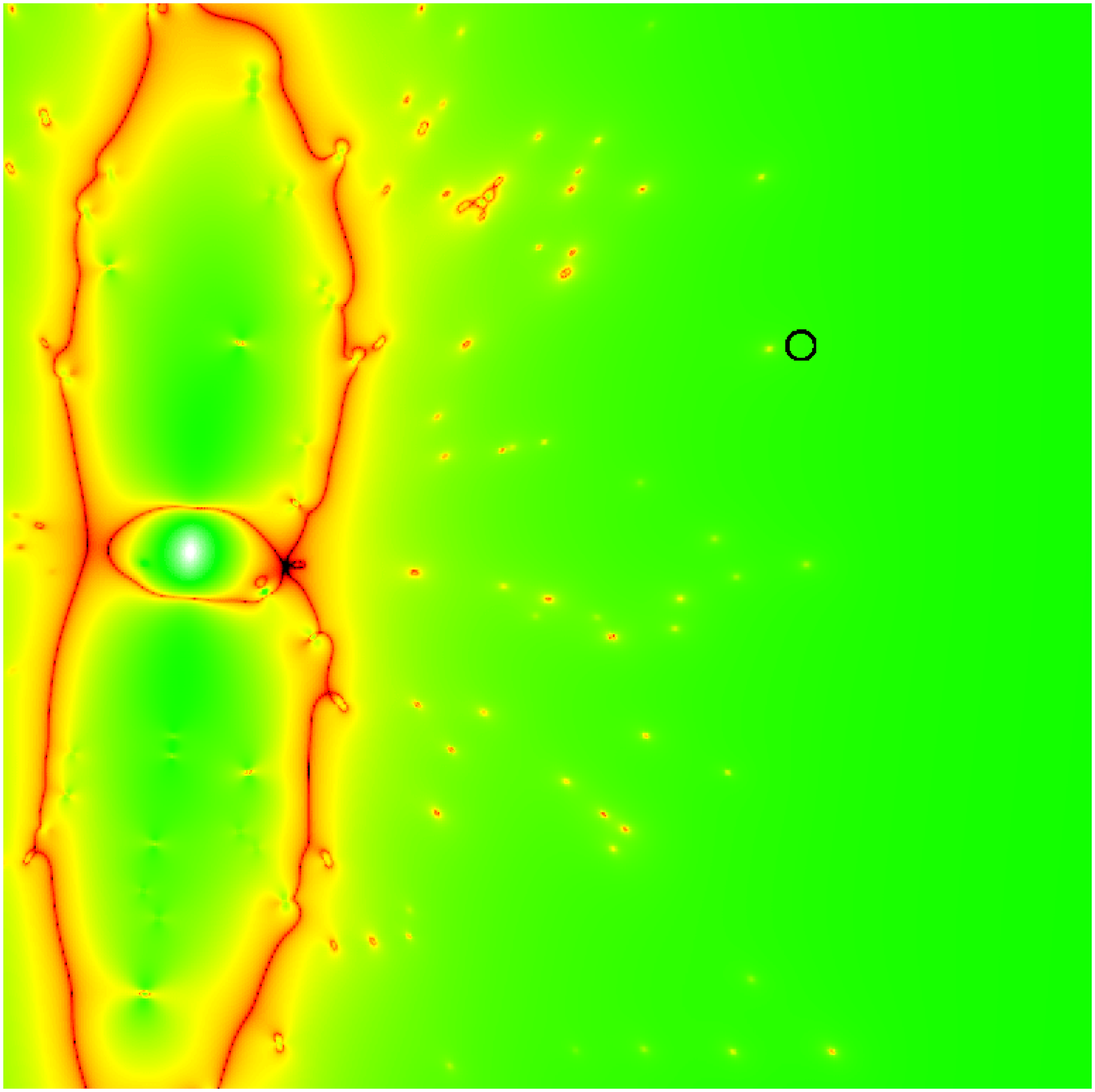}}
		\subfigure[RXJ1347]{\includegraphics[scale=0.12]{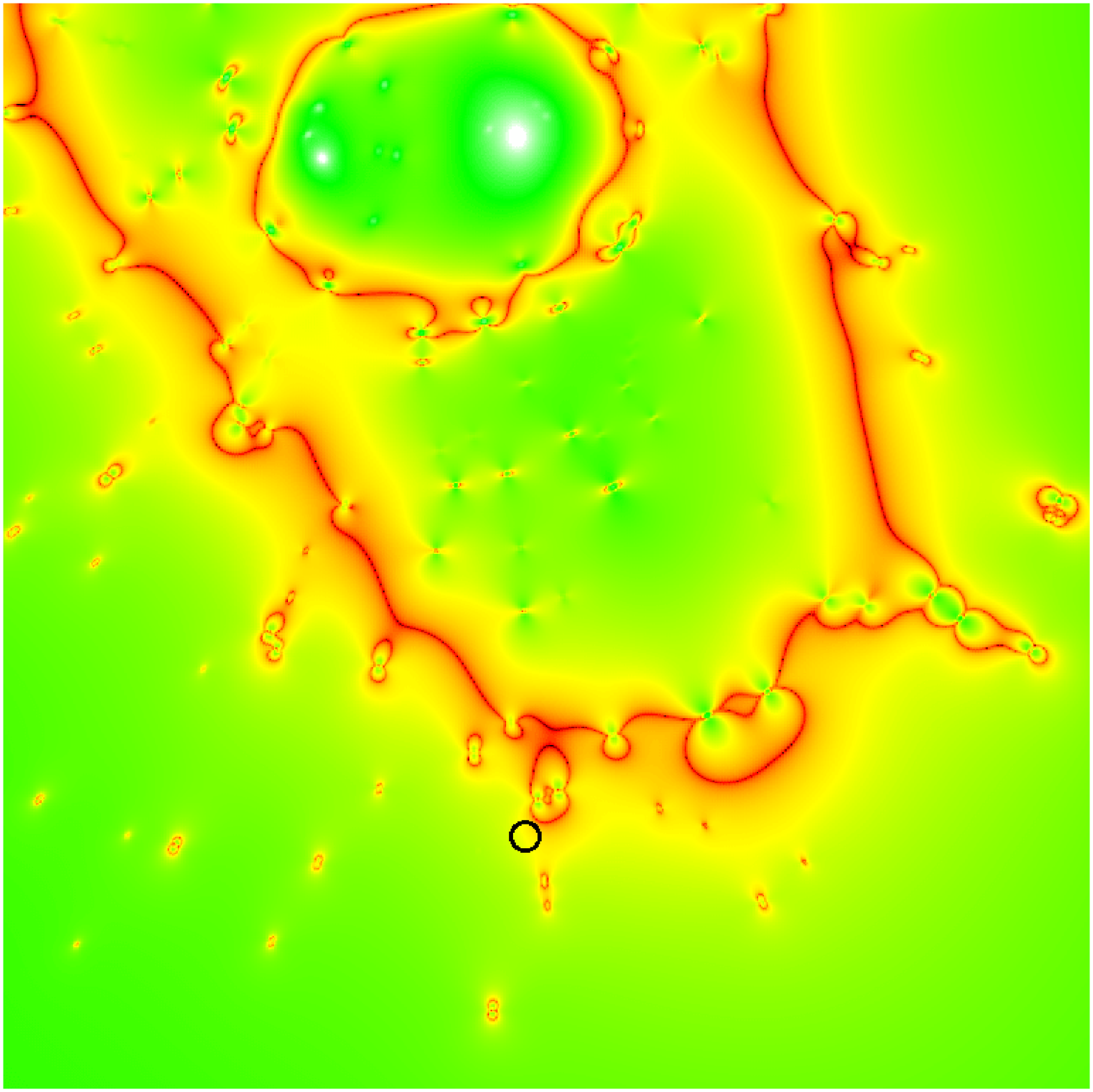}}
  \end{center}
  \caption{\footnotesize
Magnification maps for the five clusters \citep{2014arXiv1411.1414Z}, covering $100'' \times 100''$. The colors indicate the estimated magnification at the same source redshift as our best-fitting Pop~III galaxy model for the five objects discussed in this paper. White means demagnification by a factor $\sim 10$, green means no or at most a low magnification, yellow a magnification of $\sim 10$, orange a magnification of $\sim 10^2$, red a magnification of $\sim 10^3$, and black a magnification of $\sim 10^4$. The position of our objects are marked with black circles.
}
  \label{fig:magnificationmaps}
\end{figure*}

To obtain an estimate of the magnification of an observed object, mass models of the gravitational lens are required. We have used the models provided by the CLASH team \citep{2009ApJ...707L.102Z, 2013ApJ...762L..30Z, 2014arXiv1411.1414Z}, obtained through the Hubble Space Telescope Archive, as a high-end science product of the CLASH program \citep{2012ApJS..199...25P}. The models make the basic assumption that light traces mass for cluster galaxies, and adopt an analytic form for the dark matter. For the galaxies, a pseudo-isothermal elliptical density distribution \citep[PIEMD,][]{1993ApJ...417..450K, 2007NJPh....9..447J} is applied to each member galaxy, scaled by its luminosity. An elliptical Navarro Frenk White (eNFW) profile is applied to describe the dark matter halo of the cluster. The parameters describing the PIEMD scaling relations and the dark matter eNFW halo shape are minimized using a Markov chain Monte Carlo (MCMC) simulation with positions and redshifts of multiple images, as constraints \citep[see][for more details]{2013ApJ...762L..30Z, 2014arXiv1411.1414Z}. The resulting lens model can be used to obtain magnification estimates of individual objects, and also for finding new multiply lensed objects. Figure~\ref{fig:magnificationmaps} displays the magnification maps we have computed. The maps show the estimated magnification for a source redshift equal to our object behind each cluster.

\section{Physical properties}
\label{sec:physicalproperties}

		\begin{figure*}[h!]
     \begin{center}
     \begin{tabular}{ | l | c | c | c | c | }
     \hline
       & Z & $f_{\mathrm{Ly\alpha}}$ & Age (Myr) & Mass ($10^6 \mathrm{M}_{\odot}$) \\
			\hline
			\begin{sideways}
			\hspace{5mm}
			\Abell209{}
			\end{sideways}
			&
      \includegraphics[width=37mm, height=37mm]{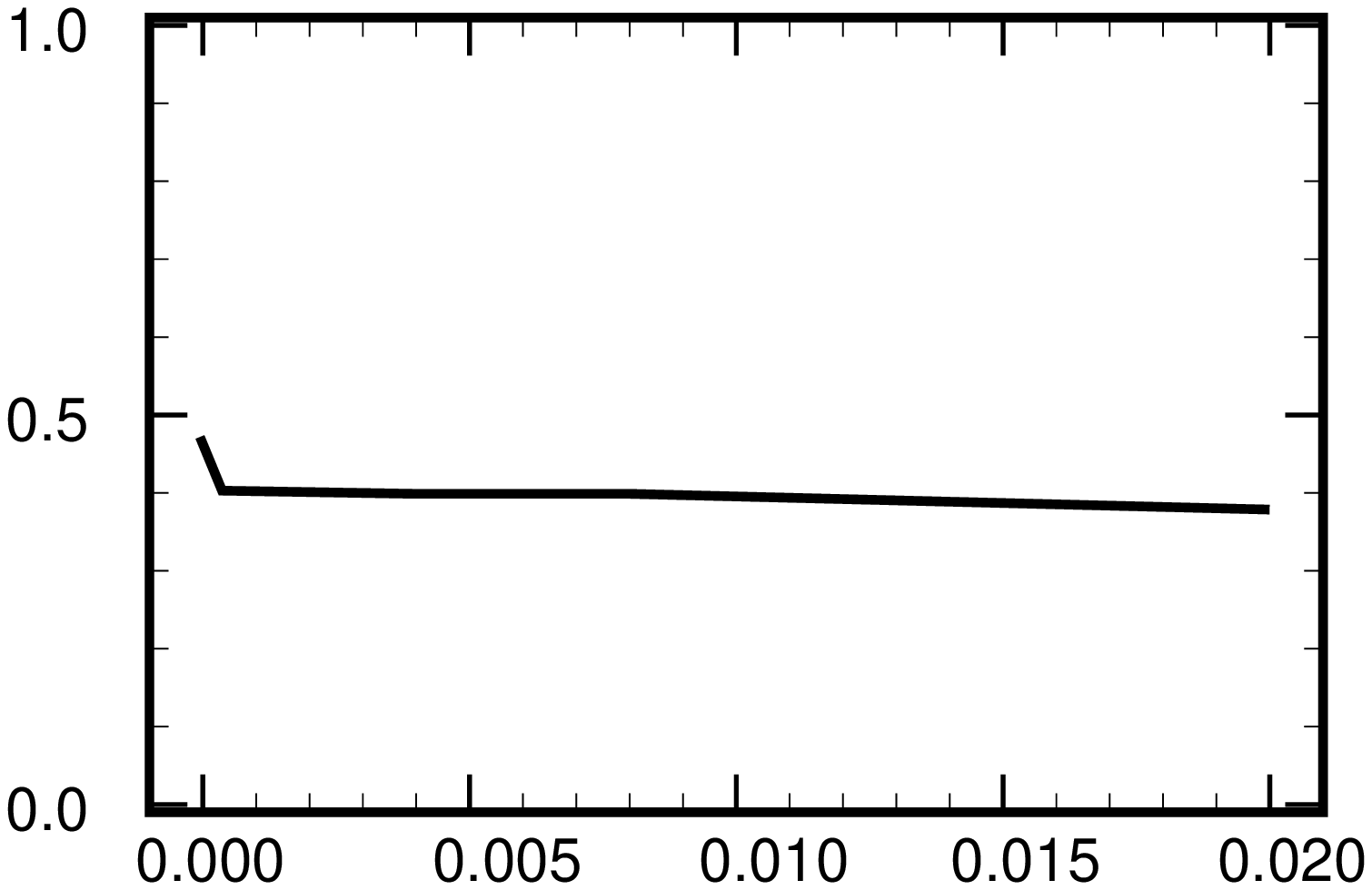}
      & 
			\includegraphics[width=37mm, height=37mm]{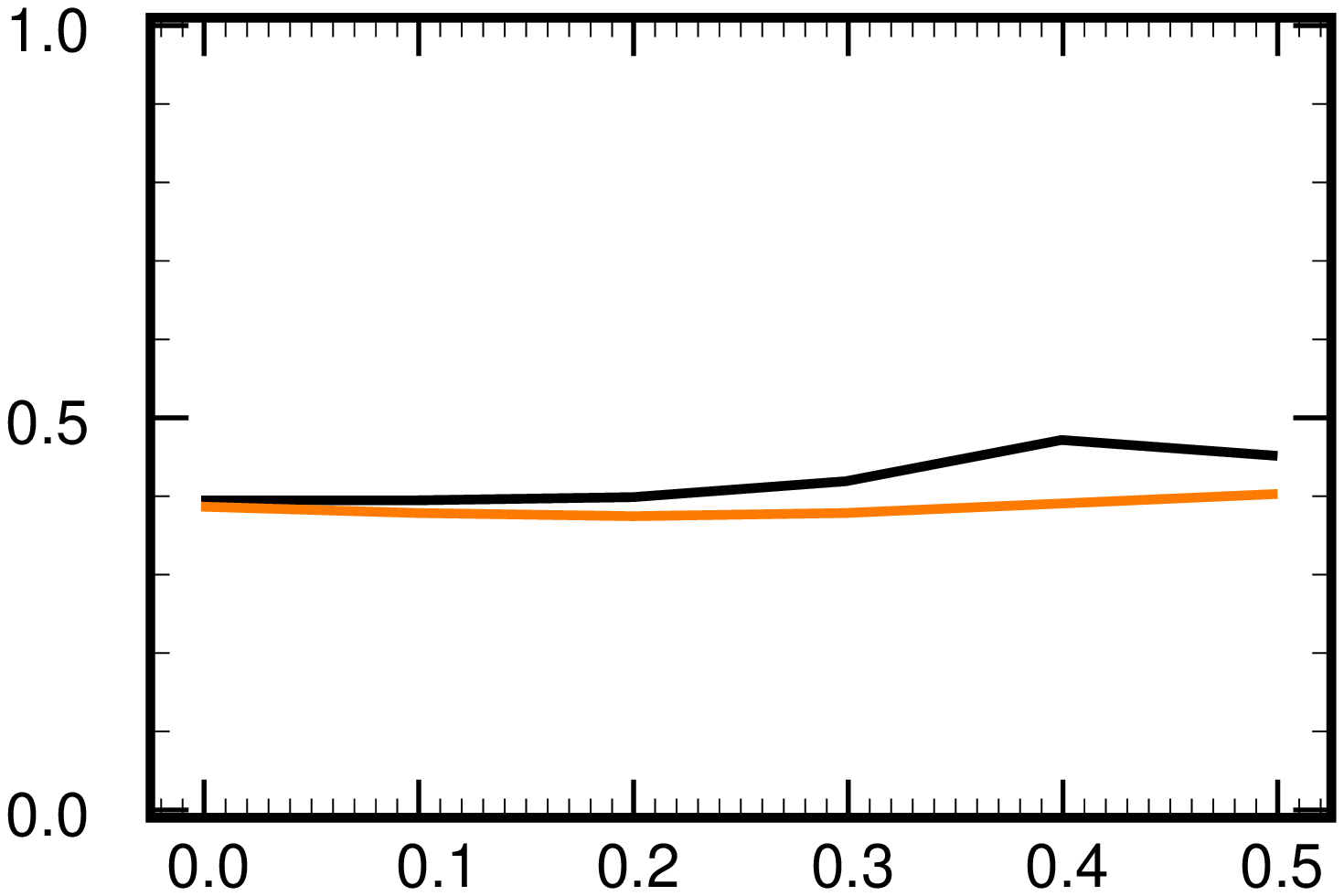}
			& 
			\includegraphics[width=37mm, height=37mm]{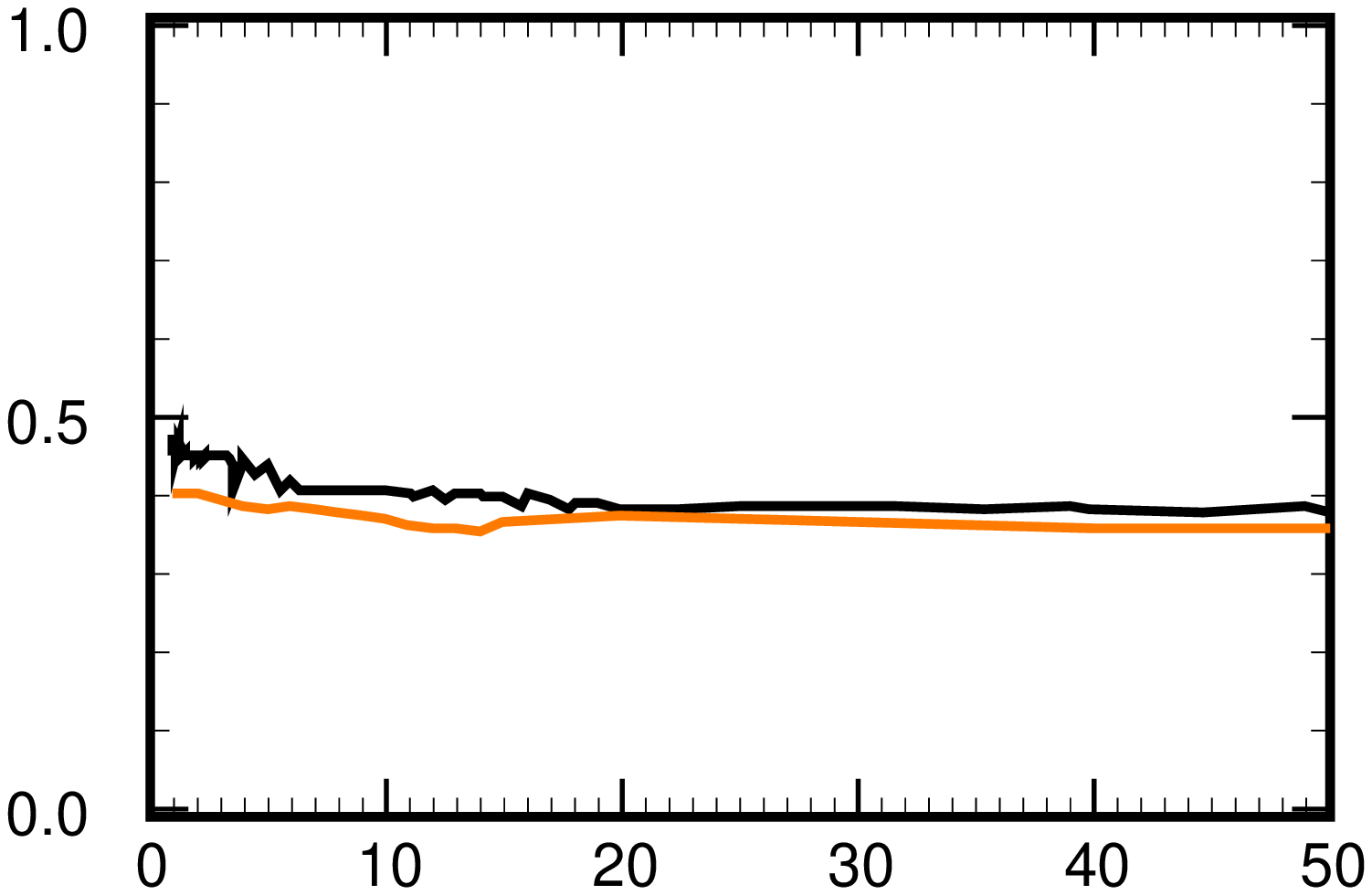}
			& 
			\includegraphics[width=37mm, height=37mm]{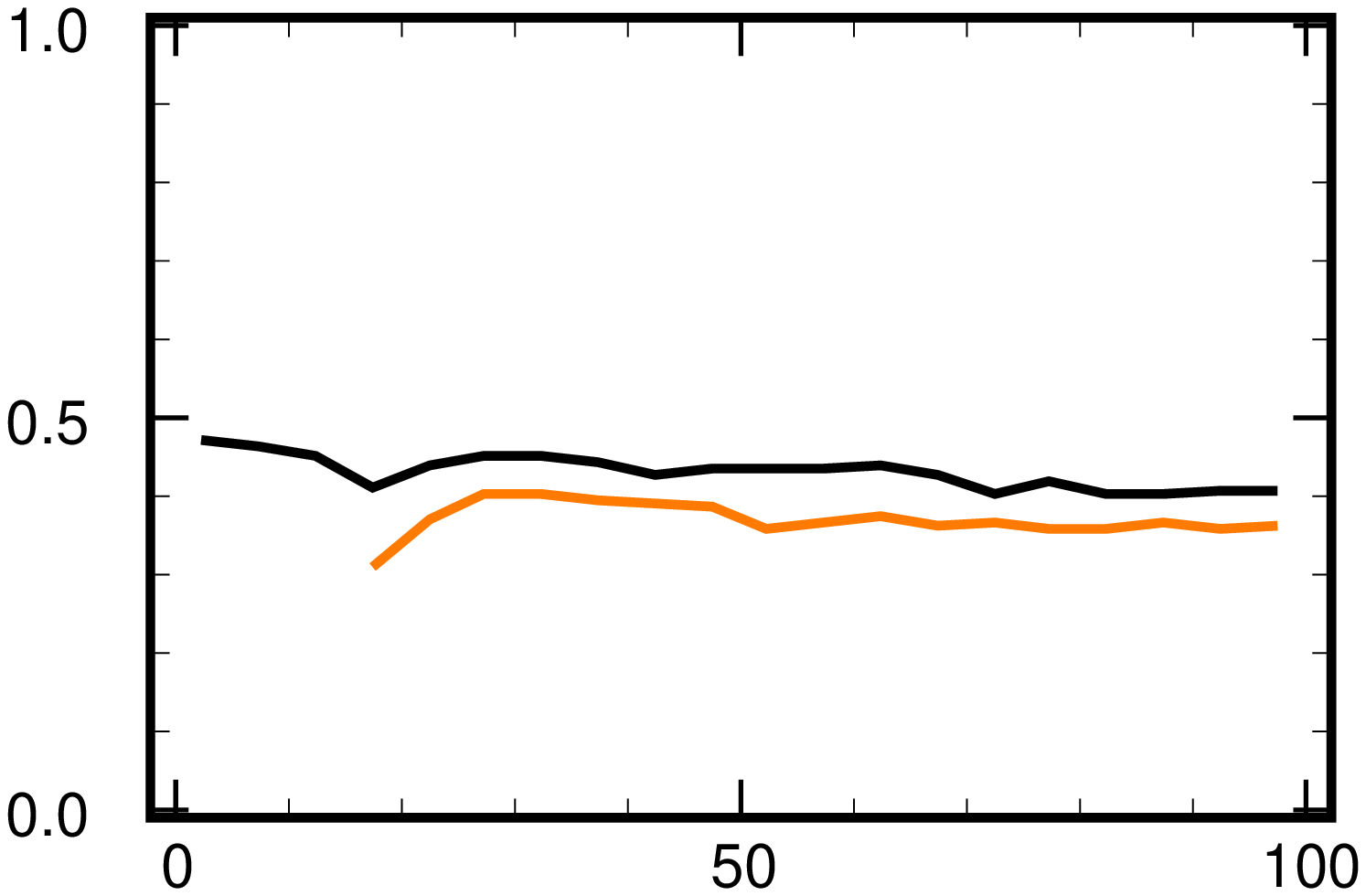}
      \\ 
			\hline
			\begin{sideways}
			\hspace{5mm}
			\macst{}
			\end{sideways}
			&
      \includegraphics[width=37mm, height=37mm]{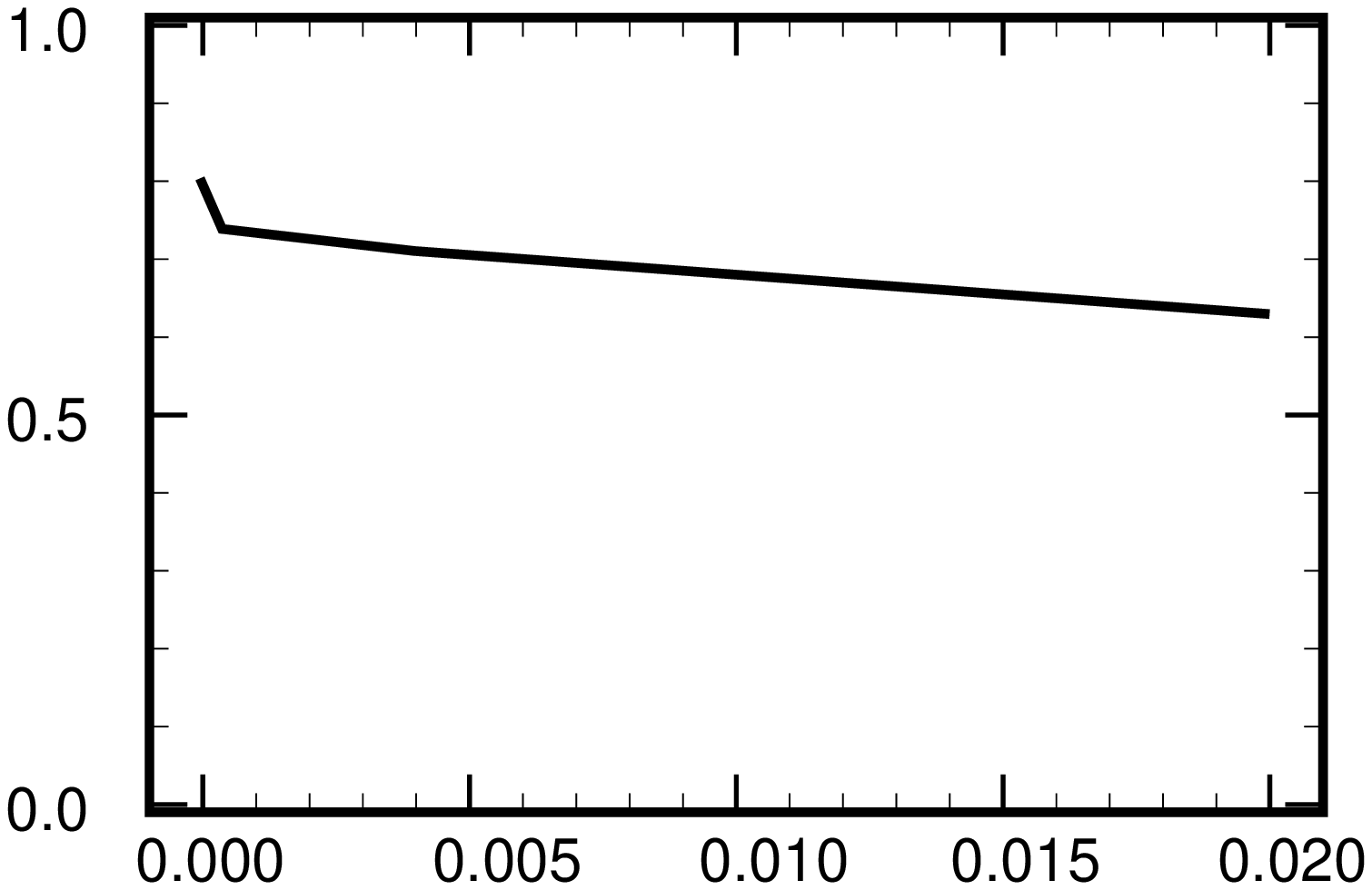}
      & 
			\includegraphics[width=37mm, height=37mm]{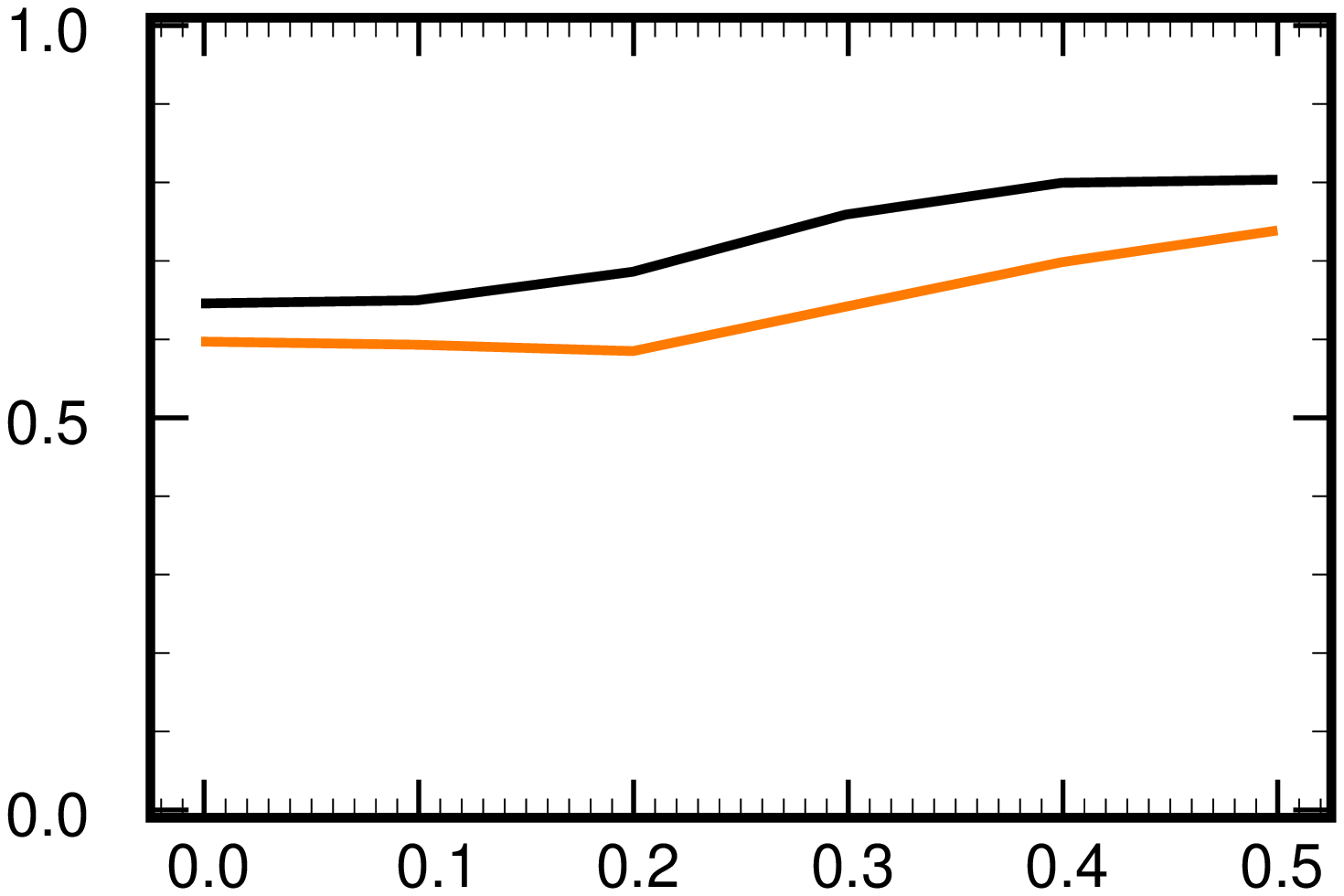}
			& 
			\includegraphics[width=37mm, height=37mm]{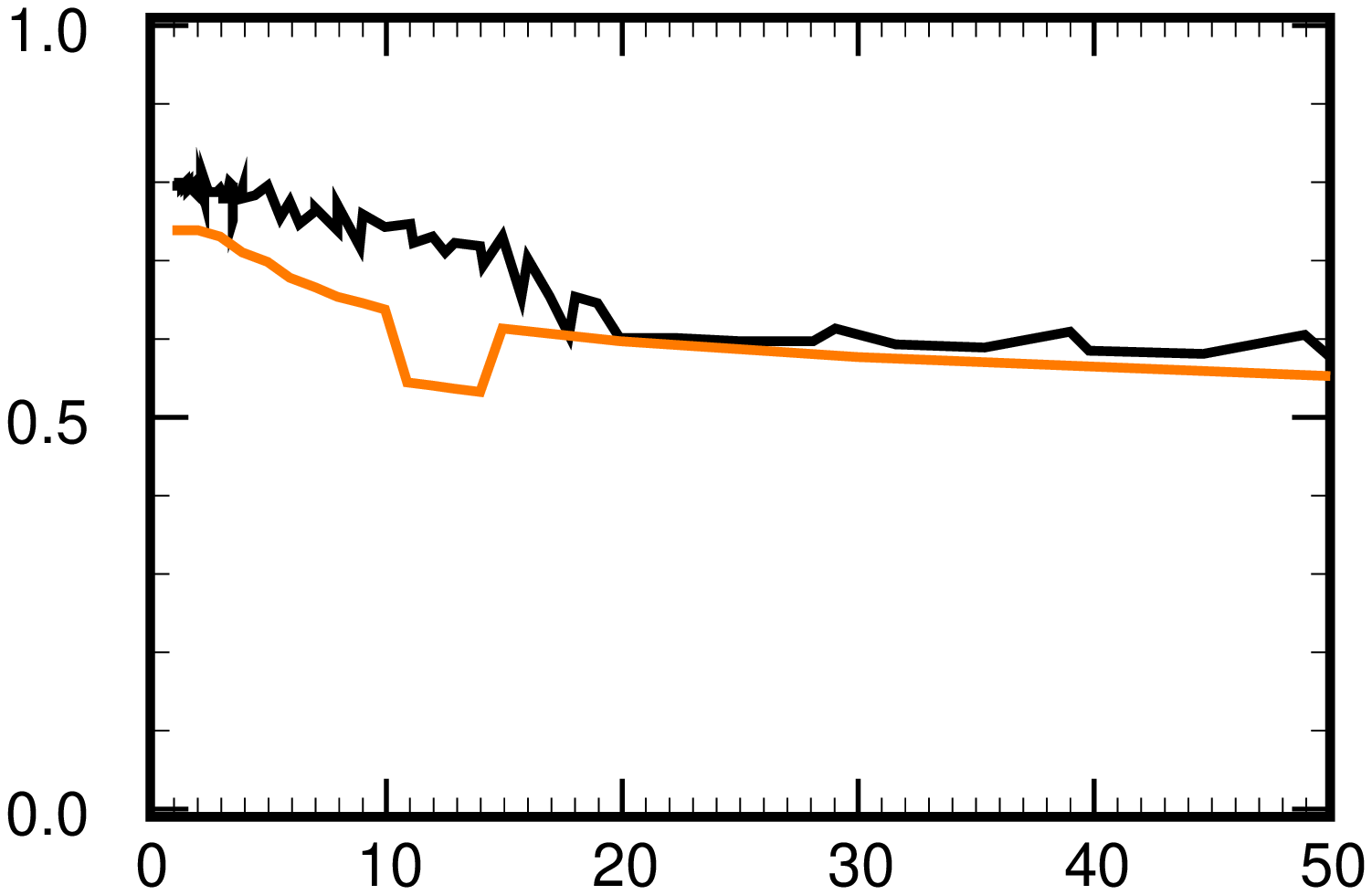}
			& 
			\includegraphics[width=37mm, height=37mm]{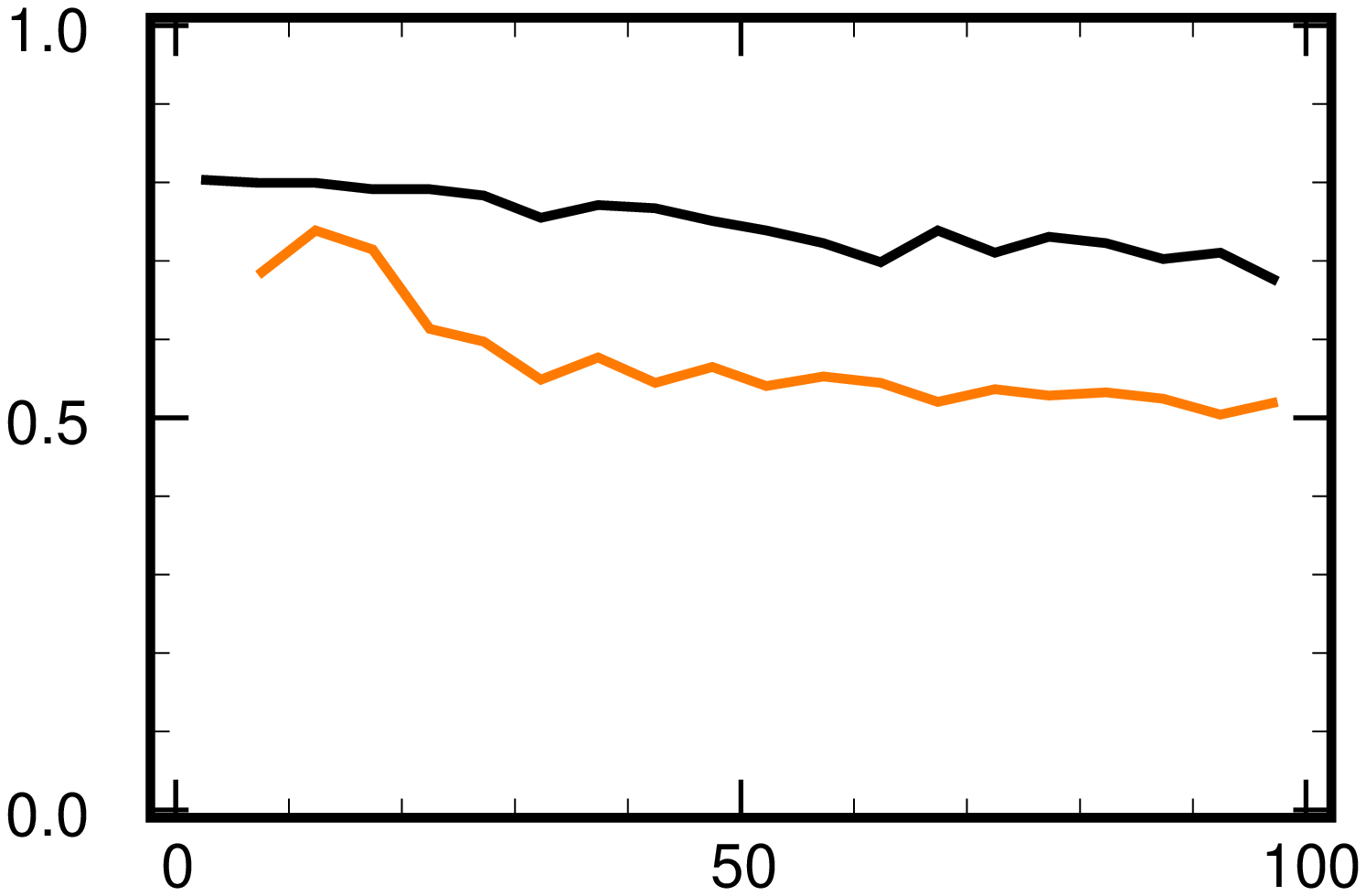}
      \\ 
			\hline
			\begin{sideways}
			\hspace{5mm}
			\macstt{}
			\end{sideways}
			&
      \includegraphics[width=37mm, height=37mm]{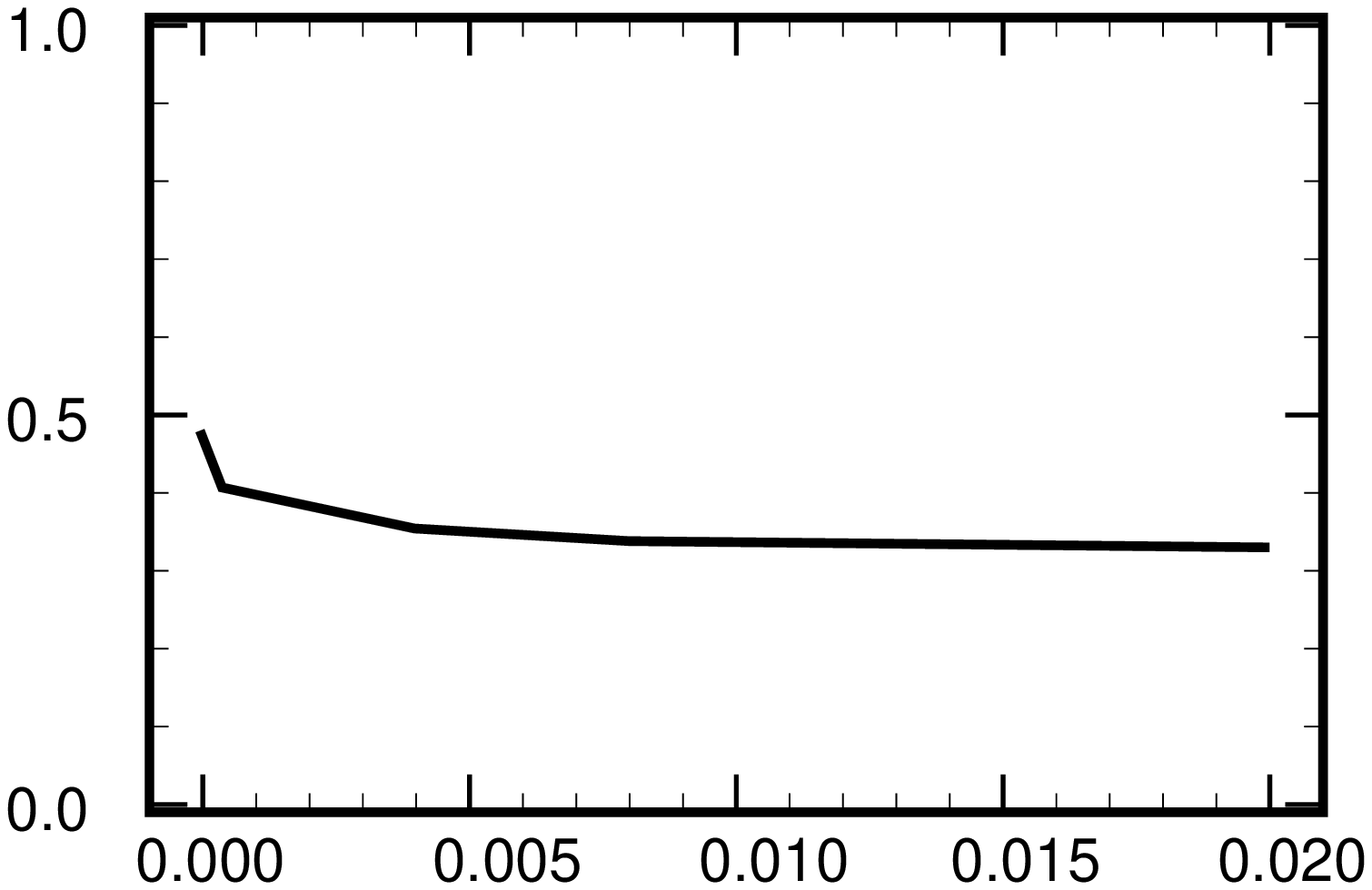}
      & 
			\includegraphics[width=37mm, height=37mm]{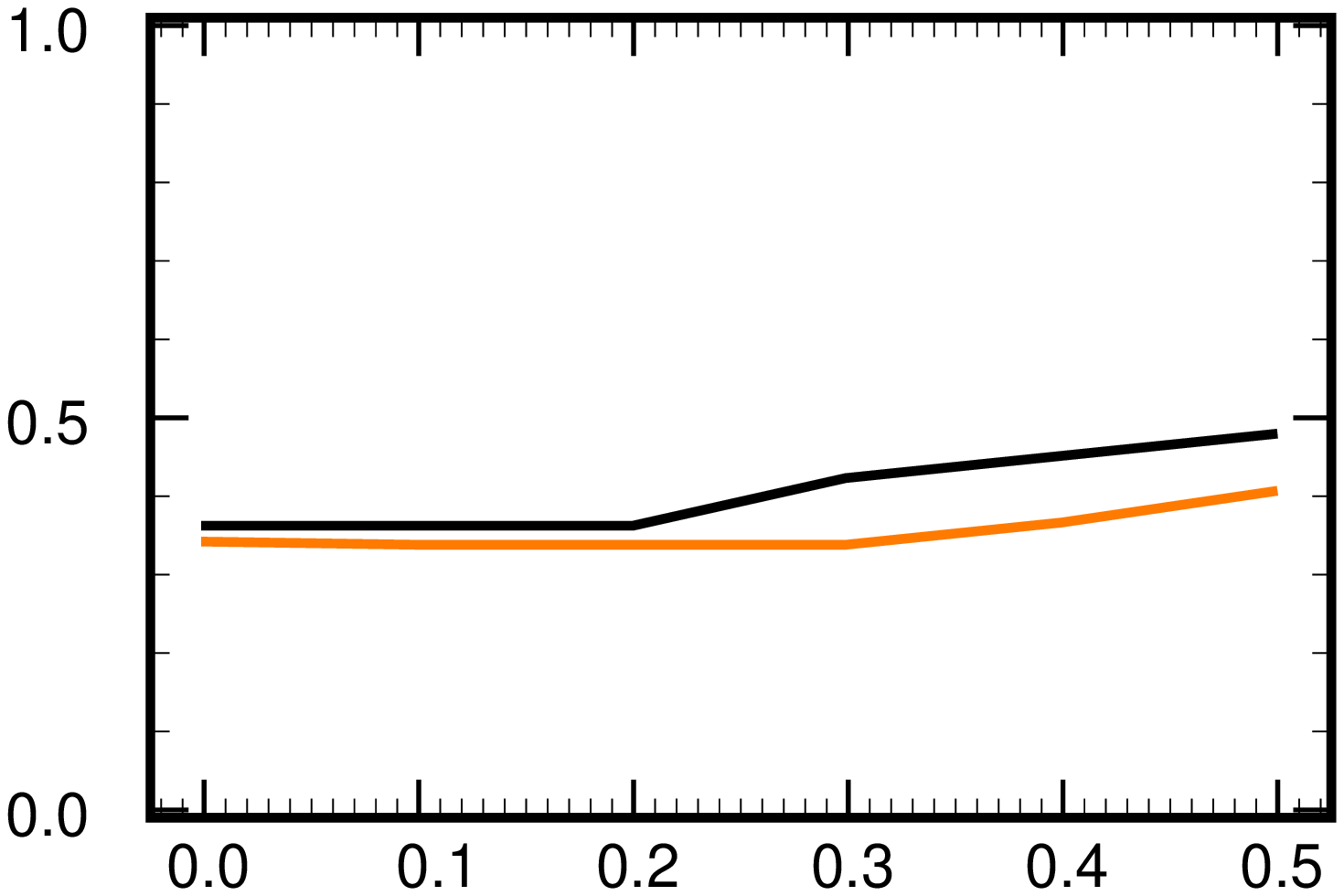}
			& 
			\includegraphics[width=37mm, height=37mm]{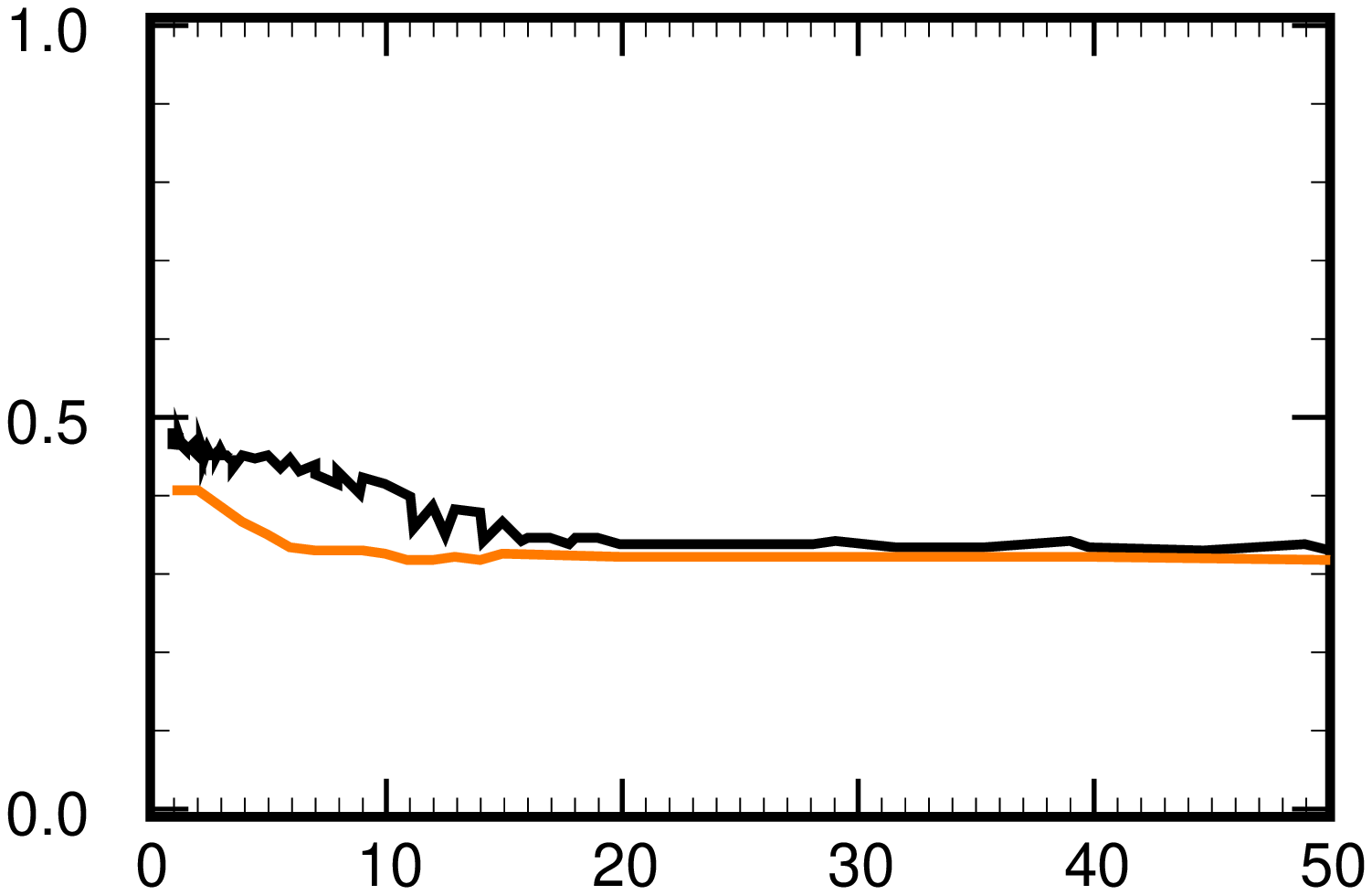}
			& 
			\includegraphics[width=37mm, height=37mm]{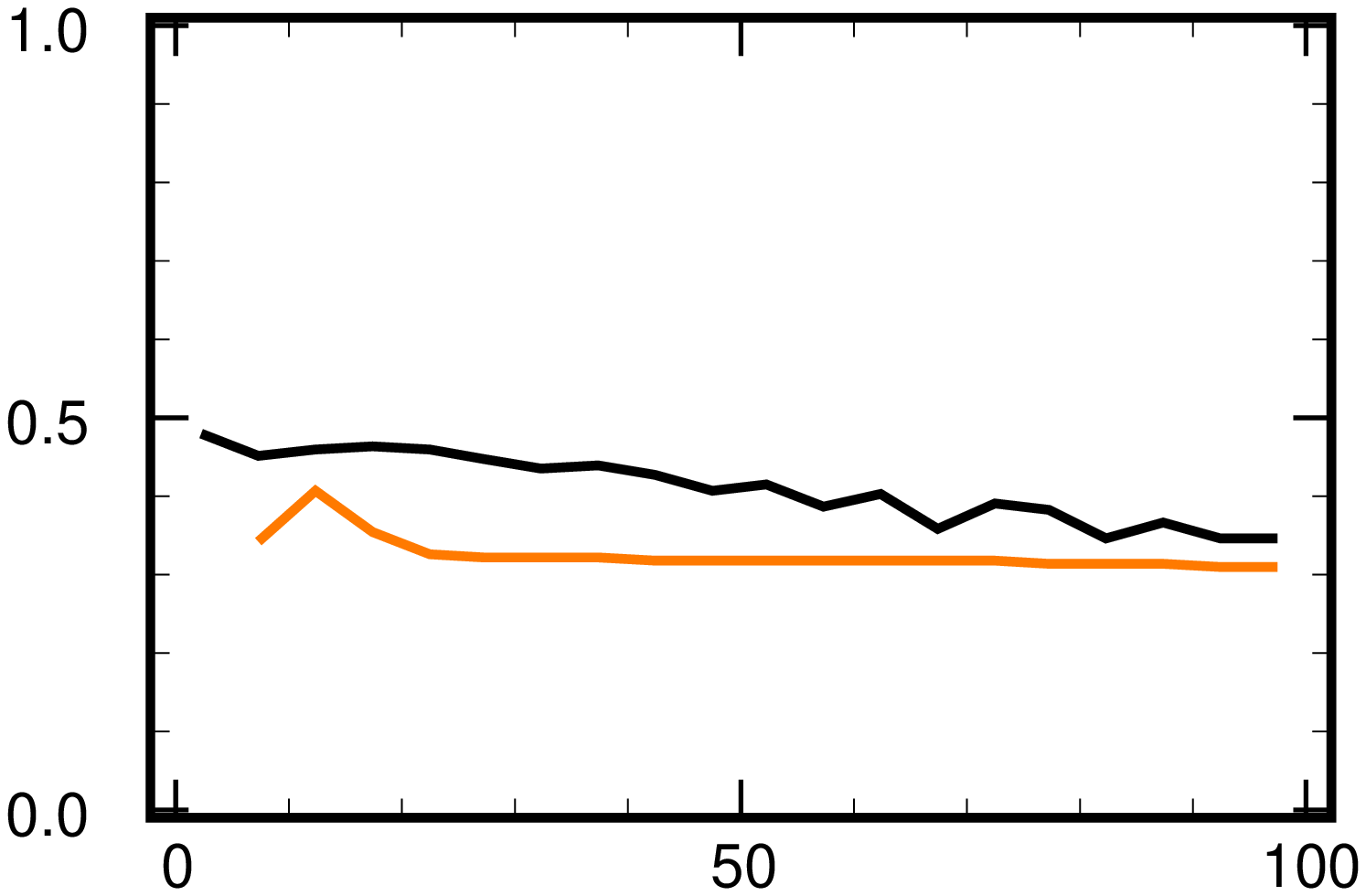}
      \\ 
			\hline
			\begin{sideways}
			\hspace{5mm}
			\MACS1931{}
			\end{sideways}
			&
      \includegraphics[width=37mm, height=37mm]{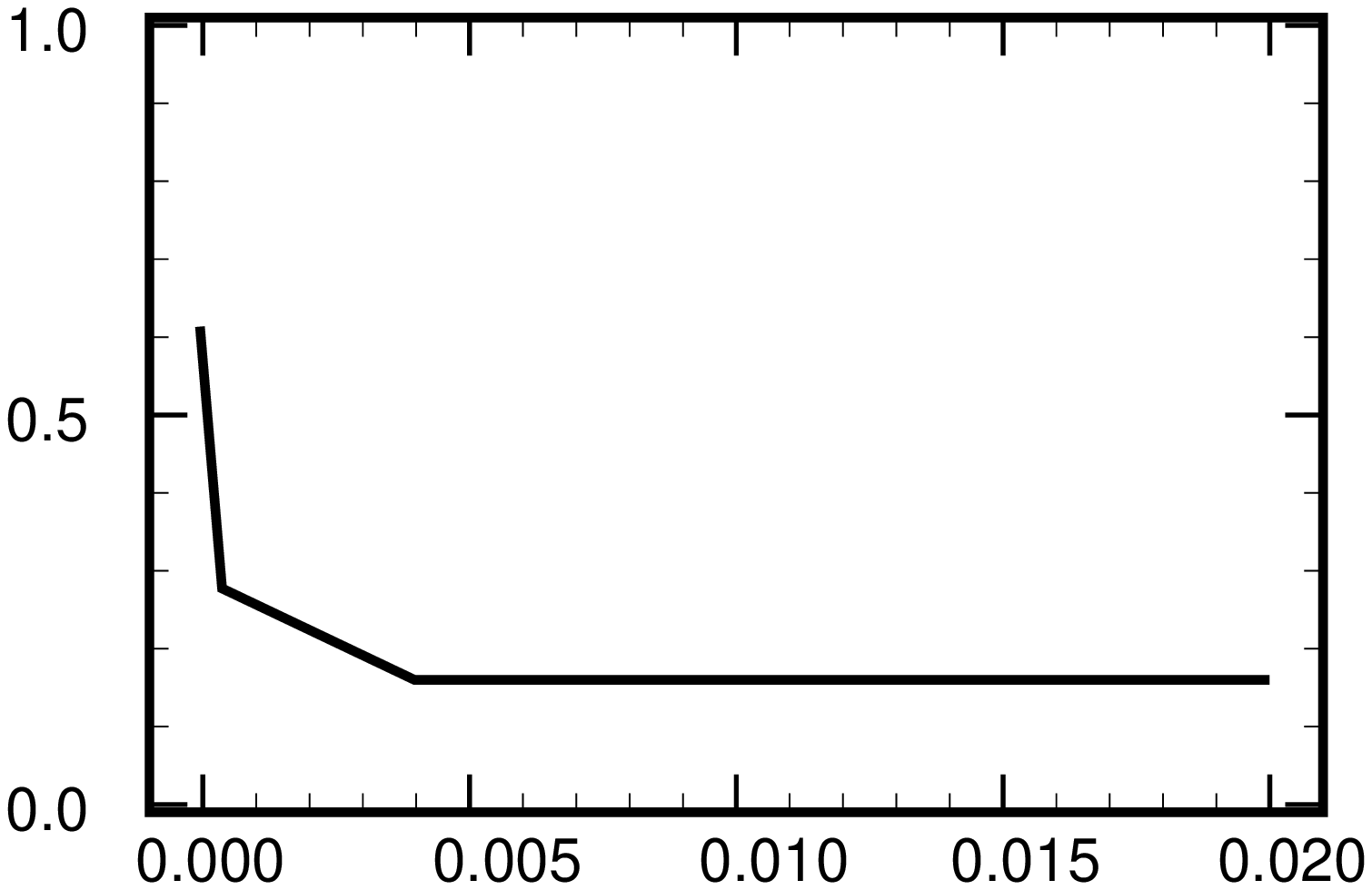}
      & 
			\includegraphics[width=37mm, height=37mm]{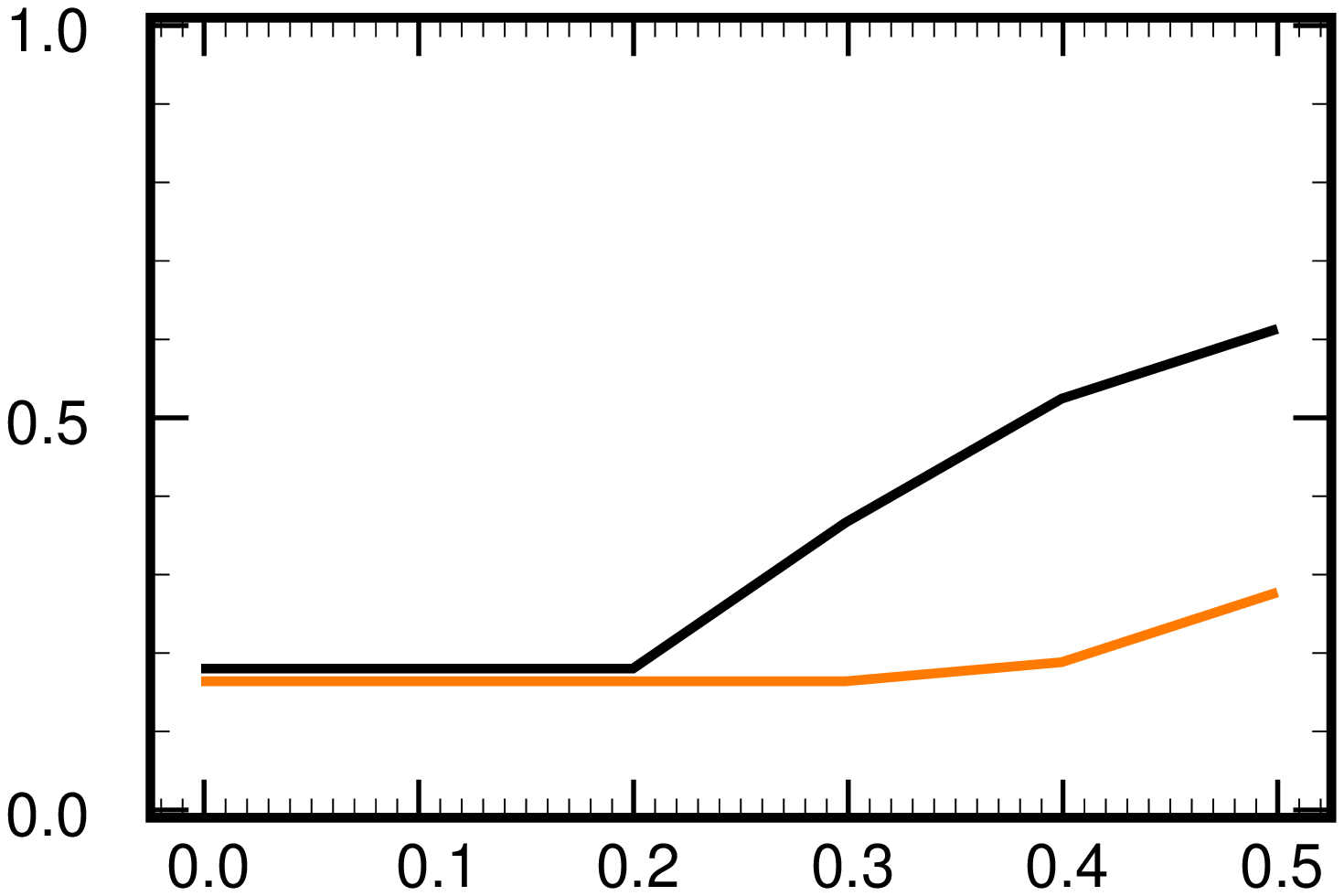}
			& 
			\includegraphics[width=37mm, height=37mm]{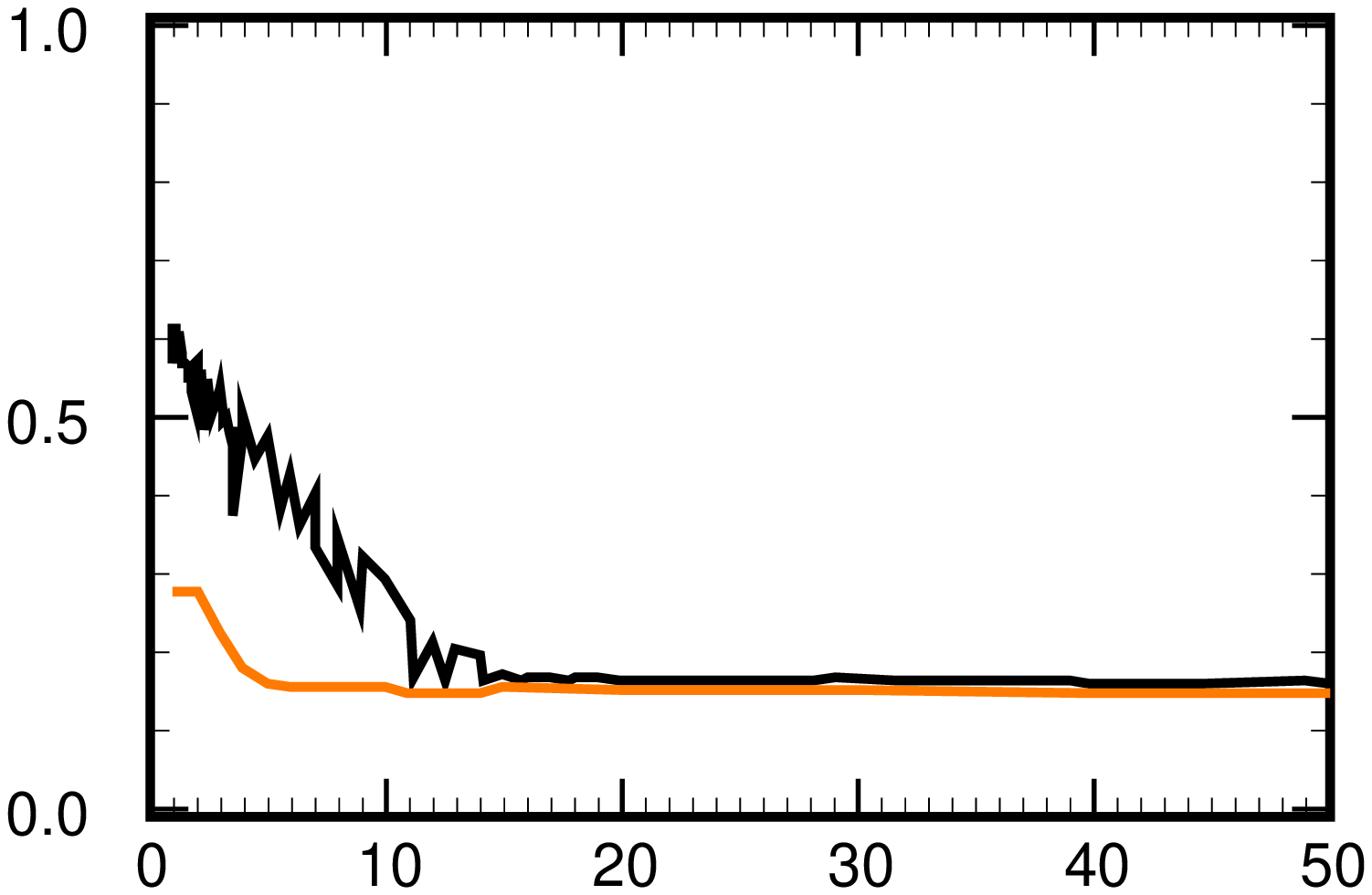}
			& 
			\includegraphics[width=37mm, height=37mm]{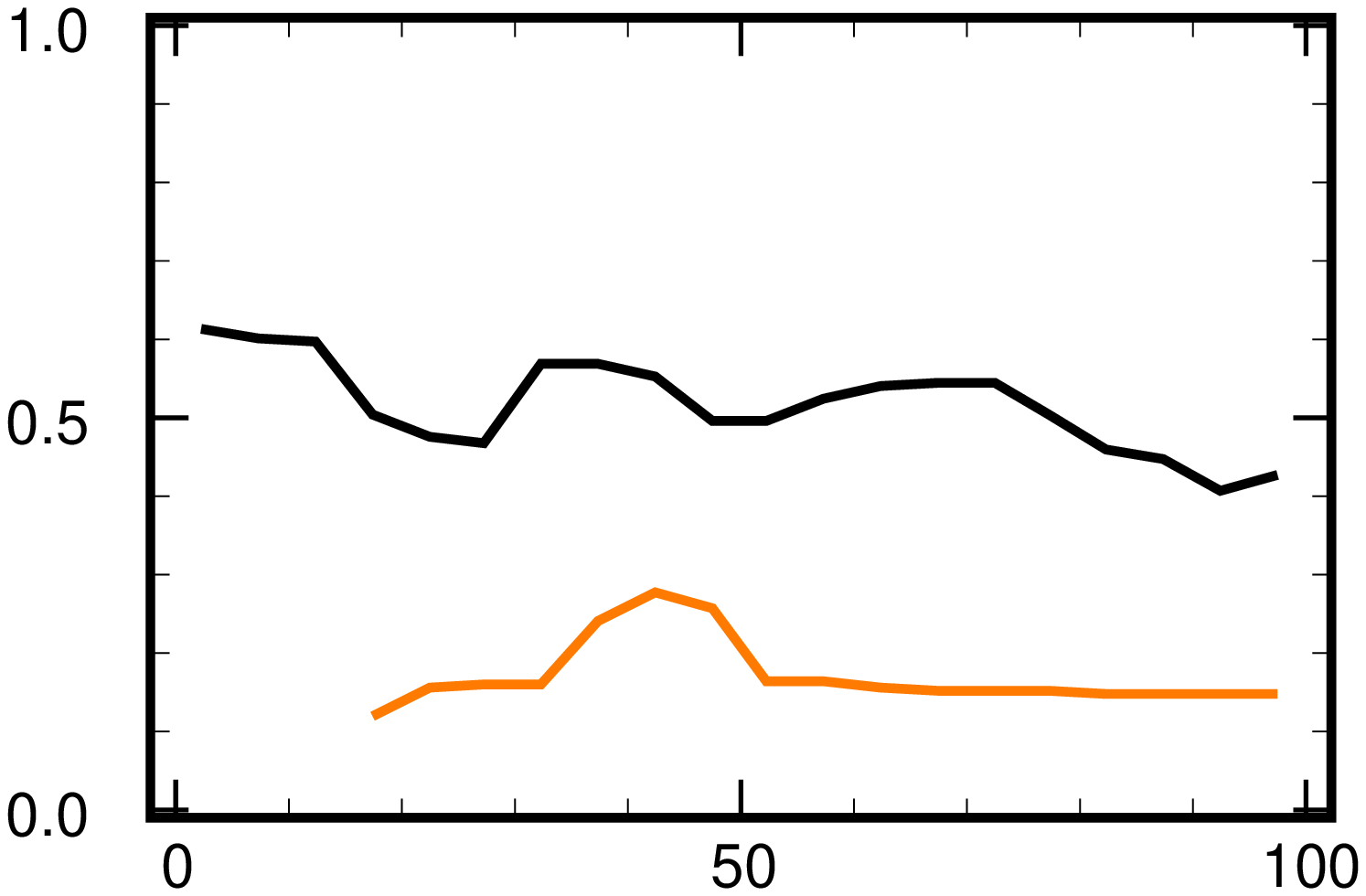}
      \\ 
			\hline
			\begin{sideways}
			\hspace{5mm}
			\RXJ1347{}
			\end{sideways}
			&
      \includegraphics[width=37mm, height=37mm]{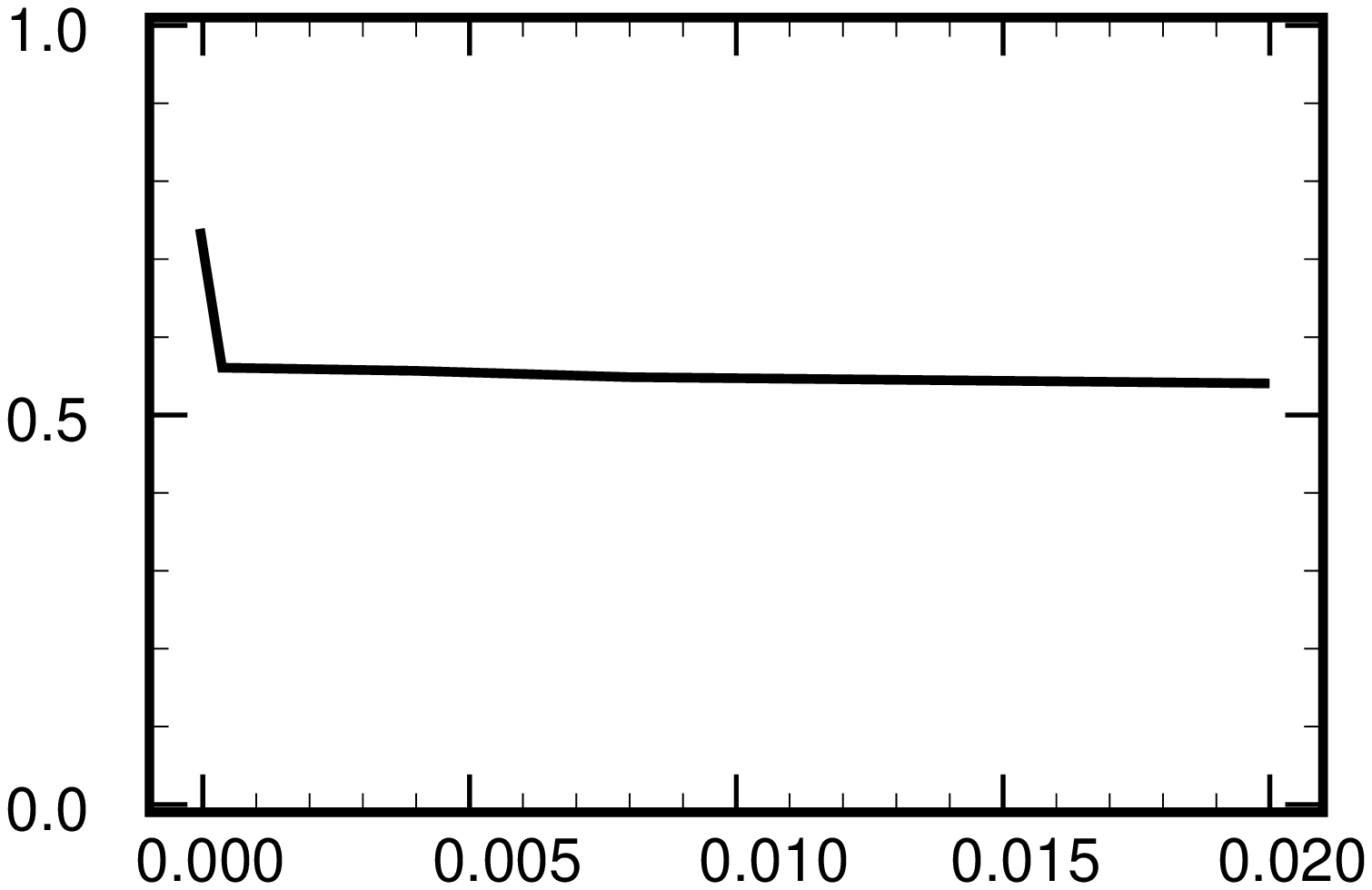}
      & 
			\includegraphics[width=37mm, height=37mm]{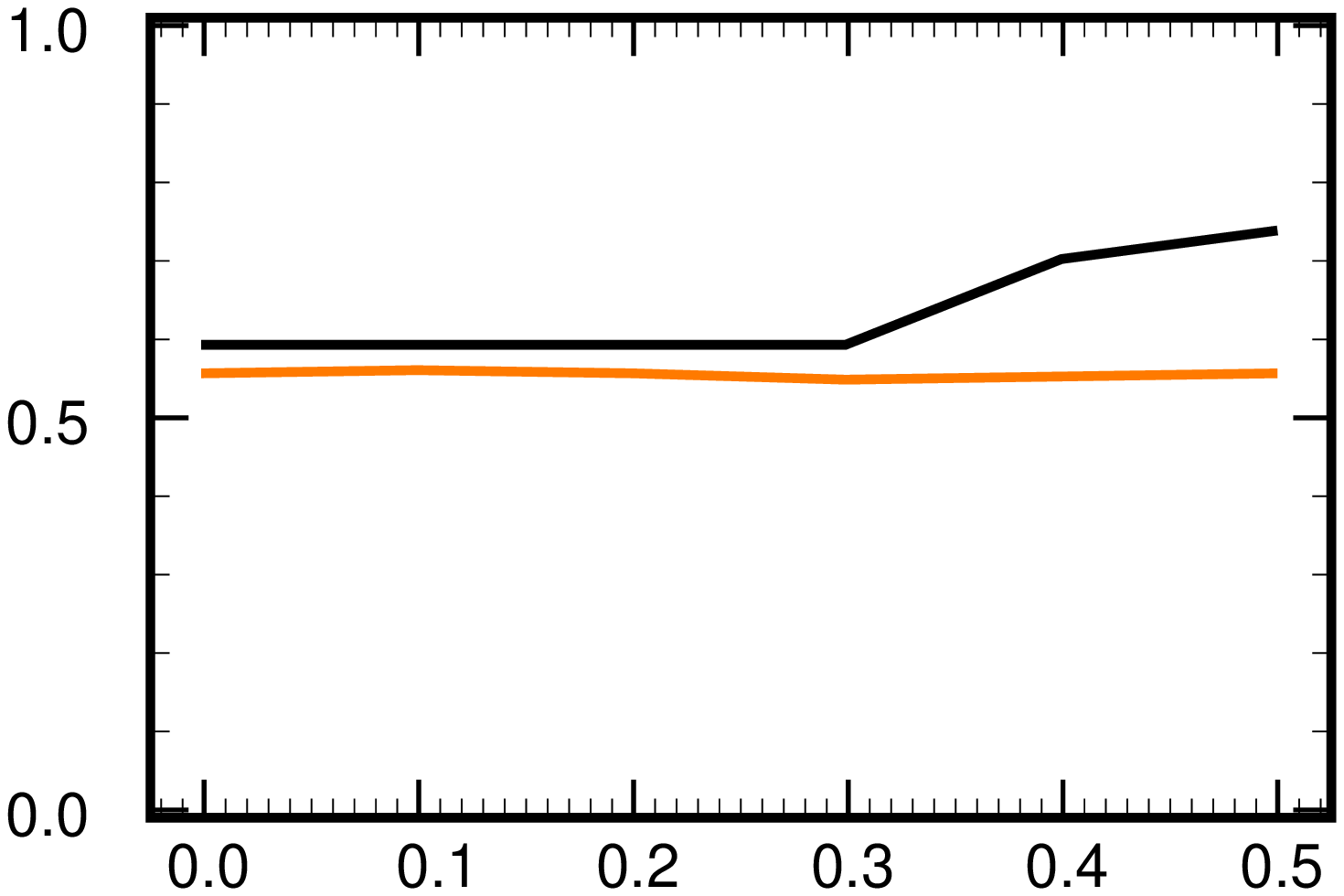}
			& 
			\includegraphics[width=37mm, height=37mm]{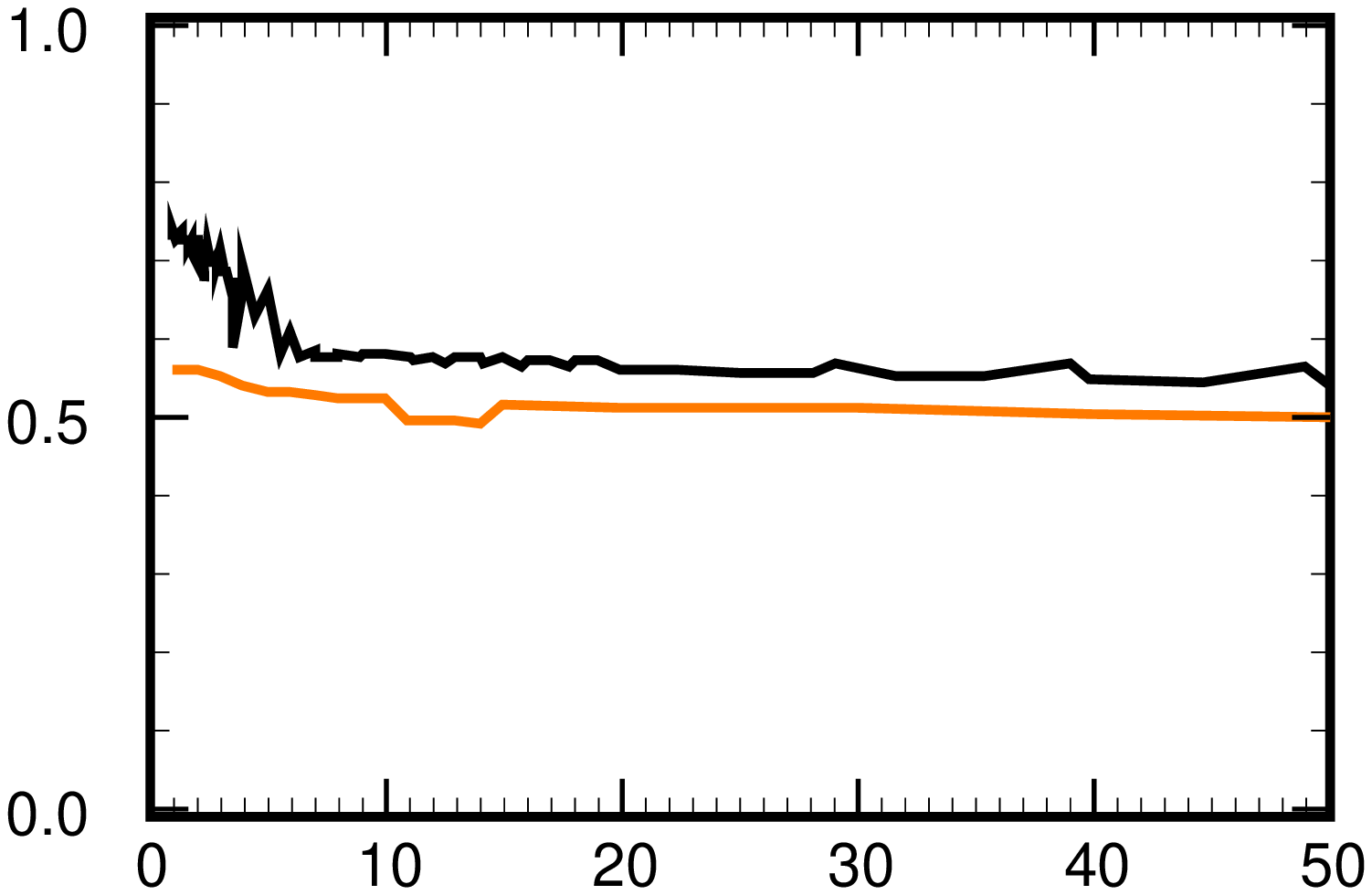}
			& 
			\includegraphics[width=37mm, height=37mm]{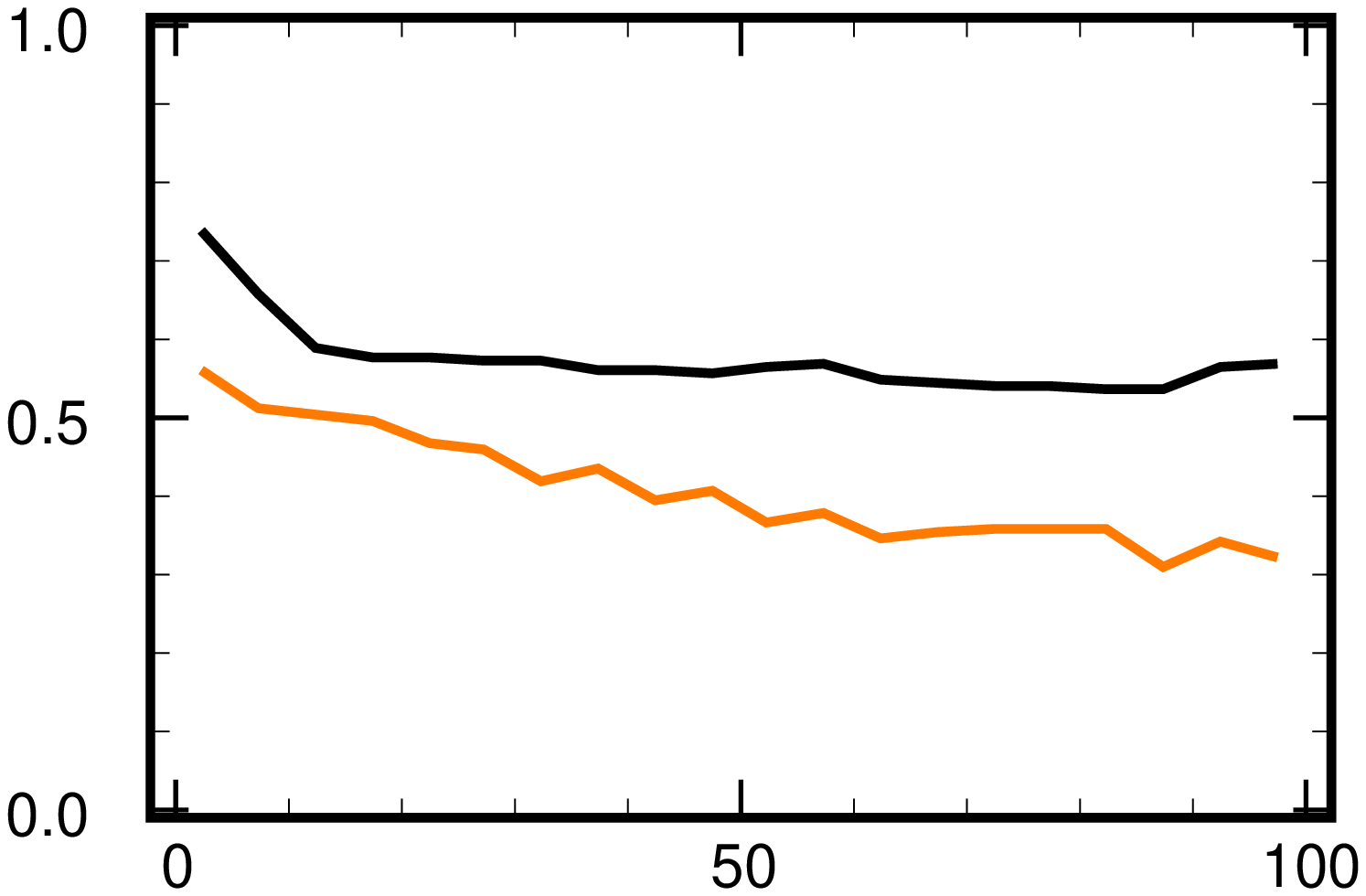}
      \\ 
			\hline
      \end{tabular}
      \caption{\footnotesize
Diagrams showing the best solution as a function of certain model parameters. Each row represents one of the five Pop~III galaxy candidates, and the columns explore one parameter each. The x-axis in each thumbnail corresponds to the parameter in the column header while the y-axis displays the quality of fit as the best cross-validation fit of any template as a function of the parameter. The black line represents the Pop~III model grid and the orange line represents the $\mathrm{Z}>0$ Yggdrasil grid. As can be seen, the quality of fit declines at least slightly with metallicity. There is a distinct but small difference between zero metallicity (Pop~III) and low metallicity ($\mathrm{Z}=0.0004$) in four of the objects. However, it is only in the case of \MACS1931{} that the difference is significant. In the column $f_\mathrm{Ly\alpha}$, we see an increase of the quality of fit with the escape fraction, again only significant for \MACS1931{} and only for Pop~III models. The exception is the $\mathrm{Z}>0$ Yggdrasil grid for the \RXJ1347{} object. For this object, the model with highest quality of fit lacks nebular emission ($f_\mathrm{cov} = 0$) and there is therefore no recombination Ly$\alpha$-emission. The third column shows the dependence of the fit to the age of the galaxy. For \MACS1931{} Pop~III models of young galaxies are strongly preferred while the other objects has a weak preference for younger galaxy solutions. The last column shows the dependence on the mass. The mass includes all mass that at any point has formed stars and should be regarded as an upper limit on the mass of stars that are still alive. There is a weak preference for low-mass galaxy models but there is a broad range of masses with good solutions.
}
      \label{fig:PF1}
      \end{center}
      \end{figure*}
			
We have investigated the physical properties of the objects, by examining how the optimal solution varies with the different parameters of our Yggdrasil models (the parameters are described in Section~\ref{sec:modelsyggdrasil}).

Different IMFs and \SFT{}s provide approximately equally good results. The fits slightly favor $f_{\mathrm{cov}}>0$, i.e. galaxies at least partially surrounded by photoionized gas. The exception is the object \RXJ1347{} when fitted to the $\mathrm{Z}>0$ Yggdrasil grid, which favors ``naked" stellar population spectra without nebular contributions. In Figure~\ref{fig:PF1}, we show the solution with highest quality of fit as a function of parameters Z and age, as well as of $f_{\mathrm{Ly\alpha}}$ and mass. In each row, there is one object while each parameter is represented by a column. Overall, the fit quality declines slightly with metallicity. There is also a distinct but small difference between zero metallicity (Pop~III) and low metallicity ($\mathrm{Z}=0.0004$) in four of the objects. In \MACS1931{} the difference is significant. \RXJ1347{} has a more or less flat dependence on Z, except for the difference between Pop~III models and models with non-zero metallicity. In the column $f_\mathrm{Ly\alpha}$, we see an increase of the quality of fit with the escape fraction. Low escape fraction generally results in a worse fit for both Pop~III models and models with non-zero metallicity. The quality of fit then increases with increasing escape fraction, with Pop~III models gaining more from the increase. However, the increase is only significant for \MACS1931{} and only for Pop~III models. The exception to the increase is the $\mathrm{Z}>0$ Yggdrasil grid for the \RXJ1347{} object, for which the fit quality is independent of $f_{\mathrm{Ly\alpha}}$. This follows since the model with best quality of fit for this object has no nebula ($f_\mathrm{cov} = 0$), and so there is no recombination Ly$\alpha$ emission. The age parameter favors young galaxies. This makes sense, since the galaxies are selected to be blue longward of the Ly$\alpha$-break. Young star forming galaxies are dominated by massive short-lived hot stars, providing them with a very blue signature. An extended starburst could provide the same signature during its \SFT{}, by continuously forming new massive hot stars. For \MACS1931{}, Pop~III models of young galaxies are strongly preferred while the other objects have a weak preference for young galaxy solutions. The last column shows the dependence on the mass. The mass graphs display the mass that has been converted to stars over the history of the galaxy as opposed to showing luminous mass, i.e. the graph contains the mass from all stars alive as well as stars that have formed and already died. This is because the Pop~III templates do not contain this information and the Yggdrasil $\mathrm{Z}>0$ models are implemented the same way to make the two grids comparable. For young templates this will not matter and for older it means the values are upper limits to the mass as they might really represent a galaxy containing less luminous mass. There is a weak preference for low-mass galaxy models but there is a broad range of masses with good solutions.

In \citet{2012MNRAS.427.2212Z} a parameter for typical star formation efficiency ($\epsilon$) is introduced. This is defined as the fraction of baryons in the dark matter halo converted to Pop~III stars throughout the \SFT{}. Assuming a halo mass of $10^8~\mathrm{M}_{\odot}$ \citep{2010ApJ...716L.190S} that contains an order of magnitude less baryons we can estimate the $\epsilon$ that the objects implies, assuming them to be Pop~III galaxies. The mass estimates of the objects (see Table~\ref{tab:RT1}) are of the order $\sim 10^6~\mathrm{M}_{\odot}$ resulting in $\epsilon \sim 0.1$. This is a very approximate number, and according to \citet{2012MNRAS.427.2212Z} $log_{10}(\epsilon) \gtrsim -1.2$ is excluded by observations, hence the derived masses are marginally consistent (given the errors involved).

The objects are very faint with large observational errors. This generally allows good quality of fits for a quite wide range of parameter values. Therefore, the dependence of the fits on different physical properties are often quite weak.

\section{Summary and conclusion}
\label{sec:summaryconclusion}

In this paper, we have reported on a search for $z \gtrsim 6$ Pop~III galaxy candidates behind the 25 galaxy clusters of the CLASH survey. Five galaxies \citep[three of which were  reported in a prior proceedings paper;][]{2014MmSAI..85..210R} that are consistent with the expectations of Pop~III galaxies with anomalously strong Ly$\alpha$ emission have been uncovered. Photometric data from the CLASH survey have been fitted to Pop~III and $\mathrm{Z}>0$ Yggdrasil model grids, as well as both empirical and synthetic models of mundane galaxies. The fitting was performed using $\chi^2$ minimization with the publicly available \textsc{Le Phare} code, as well as independent $\chi^2$ and cross-validation codes developed by our team. We have taken into account absorption by the IGM (the Gunn--Peterson trough), as well as absorption of Ly$\alpha$ photons in the form of an escape fraction varying between 0 (complete absorption) and 0.5. For the mundane models, we have included dust attenuation using the dust models of \citet{2000ApJ...533..682C}, \citet{1984A&A...132..389P} and \citet{1979MNRAS.187P..73S}. Three of our objects thereby improved their quality of fit. However, for the mundane models to display competitive fits in comparison to Pop~III models, an UV dust attenuation of $\Delta m_{\mathrm{AB}} \approx 7.5$ was needed, which seems implausibly high. We have also visually inspected each object and the surrounding fields in all available filters (including data in Spitzer filters, whenever available, to confirm their status as high-redshift objects) to rule out conspicuous optical artifacts and contamination from nearby sources.

To summarize our findings:

\begin{itemize}

\item	In the case of \RXJ1347{}, Pop~III galaxy models produce significantly better fits than the comparison templates in a narrow redshift interval around $z\approx 8.0$. This is due to the redshifted wavelength of the Pop~III galaxy models strong Ly$\alpha$-line causing only part of the transmission in the F125W filter to be effective for the line. This optimizes the quality of fit in a very narrow wavelength range. \RXJ1347{} also appears to be slightly extended like an arc. This indicates strong lensing, also confirmed by the high magnification estimate, $\mu=10$. \RXJ1347{} could not have been observed without magnification.
\item	\MACS1931{} has a quality of fit to Pop~III galaxy templates significantly better than our comparison model grids. However, \MACS1931{} has extreme colors, even for a Pop~III galaxy. Problems with the data in one filter, F125W, could for instance give rise to a spurious Pop~III signature.
\item \Abell209{}, \macst{}, and \macstt{} are not considered good Pop~III galaxy candidates, as they also have low-metallicity templates with competitive fits. \macst{} and \macstt{} both run the risk of being affected by diffraction spikes. \macstt{} also displays a morphology that is hard to discern when comparing the images in different filters.
\item The three objects \Abell209{}, \macst{}, and \RXJ1347{} were discovered at $z > 6$ despite violating drop-out criteria from color-color diagrams of the type commonly used in the literature. Hence, commonly used drop-out criteria from color-color diagrams possibly risk missing a number of $z > 6$ galaxies.
\item	The fits to the five objects discussed in this paper favor models with strong Ly$\alpha$ (albeit with a weak dependence in most cases). The rest-frame equivalent widths of the $\mathrm{Ly\alpha}$ line varies between 570--1,200~$\mathrm{\AA}$ for the best-fitting Pop~III models.
\item The model fits favor young objects with ages $\lesssim 20$~Myr. Mass estimates vary between $0.83-4.8 \times 10^6 \mathrm{M}_{\odot}$, but models implying masses of up to $\sim 10^8 \mathrm{M}_{\odot}$ provide reasonable fits for some objects.

\end{itemize}

If even just one of the five objects with Pop~III galaxy signatures we have found was to be confirmed through follow-up spectroscopy, it would strengthen the case for Pop~III stars constituting the ``missing link'' between Big Bang nucleosynthesis and the observed metal-enriched universe.

\section*{Acknowledgments}
\label{sec:acknowledgement}

CER acknowledges funding from the Swedish National Space Board and the Royal Swedish Academy of Sciences. EZ acknowledges funding from the Swedish National Space Board and the Swedish Research Council (Project 2011-5349). EZ and JG acknowledge funding from the Wenner-Gren foundations. Support for AZ was provided by NASA through Hubble Fellowship grant \#HST-HF2-51334.001-A awarded by STScI.\vspace{5mm}

\bibliographystyle{References}
\bibliography{References}

\appendix

We have generated two catalogs which have been used mainly for consistency checks. This appendix describes each of them.

\section{Catalog}
\label{sec:guaitacatalog}

To generate the catalogs, we used CLASH~v1 images with 0.065$''$/pixel after multi-drizzling. All the catalogs were generated using the SExtractor version 2 code in dual mode. Because we are looking for galaxies at $z\geq6$, we required detection in the WFC3/F105W ($z=7.7$ for Ly$\alpha$ at its pivot wavelength) and WFC3/F125W ($z=9.3$ for Ly$\alpha$ at its pivot wavelength) bands for selection. The flux  measurements were carried out on near-infrared (F105W, F110W, F125W, F140W, and F160W) images smoothed using a gaussian kernel to resemble the PSF of \#F160W filter image, while the detection was un-smoothed on the F105W and F125W images. We also measured fluxes in F814W and F850LP, in which $z\sim6$ sources could appear. In the $\sim5'\times5'$ wide CLASH fields, just a few not-saturated stars are observed. Therefore, we used the tabulated PSF as a reference\footnote{http://www.stsci.edu/hst/wfc3/documents/handbooks/currentIHB/\\c07\_ir07.html}.

\subsection{Detection} 

To be able to detect either faint or bright extended sources, such as the very high redshift ones lensed by the clusters, but also point-like sources at the border of the detection limits, we defined five sets of detection parameters in the SExtractor configuration file. The five cases are tabulated in Table~\ref{tab:detectioncases}. It contains the detection threshold, and the minimum area of continuous pixels fulfilling the detection threshold to define an object. Also the deblending parameter is displayed which is the ability for SExtractor to distinguish nearby groups of pixels (with detection $>$ threshold) separated by faint pixels. A low value tends to separate the groups of pixels into several objects, while a higher value tends to identify the whole group as one object.

\begin{table}
\caption{The five cases of detection parameters defined for our independent object catalogs.}
\label{tab:detectioncases}
\begin{center}
\begin{tabular}{lccc}
\hline
\\
Case & Threshold & Area$_\mathrm{min}$ & Deblend \\
\hline
\\
A & 2 & 3 & 0.005 \\
B & 0.5 & 10 & 0.1 \\
C & 0.5 & 20 & 1 \\
D & 0.3 & 100 & 0.01 \\
E & 0.5 & 10 & 0.0001 \\
\\
\hline
\end{tabular}
\end{center}
\end{table}

Case~A is a conservative detection method, in which the detection threshold was optimized to favor detection of real sources and limit the detection of the fake ones, due to background fluctuations. Case~B and Case~C reveal extended sources, even with signal-to-noise less than 1. Case~D allows to detect a few 100 continuous pixel sources, with signal 0.3 times the image background rms. It is optimized to detect faint (extended) arcs. Case~E is an extreme case. It is used to look for any possible small, either faint or bright signal.

\subsection{Photometry}

We used SExtractor AUTO and APER photometry. The AUTO photometry is defined as the most precise estimate of the total magnitude for galaxies. Based on SExtractor guide book, the updated version of AUTO photometry in SExtractor~v2 provides the magnitude value closest to the true one for faint sources. The AUTO photometry is expected to recover about 90\% of the flux and this behavior, for each band, is magnitude independent. We varied the Kron (Kr) and minimum radius (R$_{\mathrm{min}}$) parameters to allow SExtractor to retrieve flux also from the wings of extended objects and reduce the noise of point-like sources. However, AUTO is sensitive to crowdedness and it is not robust in the case of very close neighbors. This is the reason why we varied the deblending parameter in Case~B, C, D, and E to improve detection. 

We chose R$_{\mathrm{min}}$~=~3 to be able to measure flux down to $\sim 0.2''$ radius. Also, we adopted Kr~=~1, optimized to reduce contamination from neighbors; Kr~=~2, which allows a typical aperture size of 0.4$''$; Kr~=~2.5, the standard value expected to produce only 6--10\% of flux losses. APER photometry was applied because it can be optimized to give the highest signal to noise, S/N. We chose an aperture radius of 0.2$''$, which resembles the tabulated sharpness in WFC3/NIR images and 0.4$''$, which is expected to contain more than 80\% of the flux of WFC3/NIR sources. The model PSF gives the same encircled energy in F125W, F140W, and F160W at 0.4$''$ radius.  

For every source detected in our five catalogs (one for each case), we measured fluxes and flux errors with 3 AUTO and 2 APER settings.

\section{Catalog}
\label{sec:melindercatalog}

The fluxes of the detected sources were measured using SExtractor. To limit the amount of noise getting into the photometric aperture while still measuring most of the flux coming from the, sometimes extended and distorted, lensed galaxies we choose to use isophotal photometry (MAG\_ISO). The same aperture was used for all of the filters and was determined from the detection image for each object using the same threshold for a pixel to be included as in the detection step (DETECT\_THRES = ANALYSIS\_THRES = 1). On average this aperture has a size of $\sim 50$ pixels or $\sim 0.2$ sq. arcsec, but naturally varies quite a lot depending on the size of the object. 
 
As mentioned above, SExtractor was run in dual mode to use the same detection image in all of the filters.  In order to estimate the photometrical uncertainties we also used a separate rms image for each filter. These images were constructed by taking the square root of the inverted HST weight maps also released by the CLASH team. SExtractor was thus setup to calculate the photometrical errors from the rms maps rather than from the local background in the science images. Furthermore, the effect of correlated noise in drizzled output images was corrected for following the procedure outlined in \citet{2002PASP..114..144F}. The true background noise in the images after correcting for the correlation is $\sim 40\%$ higher than the noise measured in the frames.
 
For each of the detected sources we also estimated the limiting flux in the red optical filters (F814W and F850LP) by computing the total noise within an aperture of the same size as the isophotal area from the rms images. Also these limiting fluxes were corrected for the pixel correlation effect.

\end{document}